\documentclass[
 reprint,
 amsmath,amssymb,
 aps, prb,
]{revtex4}

\usepackage{amsfonts}
\usepackage{latexsym}
\usepackage{graphicx}
\usepackage[usenames]{color}
\usepackage[utf8]{inputenc}
\usepackage[english,bulgarian]{babel}
\usepackage{hyperref}

\expandafter\def\expandafter\UrlBreaks\expandafter{\UrlBreaks
  \do\a\do\b\do\c\do\d\do\e\do\f\do\g\do\h\do\i\do\j%
  \do\k\do\l\do\m\do\n\do\o\do\p\do\q\do\r\do\s\do\t%
  \do\u\do\v\do\w\do\x\do\y\do\z\do\A\do\B\do\C\do\D%
  \do\E\do\F\do\G\do\H\do\I\do\J\do\K\do\L\do\M\do\N%
  \do\O\do\P\do\Q\do\R\do\S\do\T\do\U\do\V\do\W\do\X%
  \do\Y\do\Z}

\addto\captionsbulgarian{%
}

\begin{document}
\title{Measuring the speed of light with electric and magnetic pendulum}

\author{Vasil~G.~Yordanov}
\email[E-mail: ]{vasil.yordanov@gmail.com}
\affiliation{Department of Atomic Physics, Faculty of Physics,\\
 St.~Clement of Ohrid University at Sofia,\\
5 James Bourchier Blvd., BG-1164 Sofia, Bulgaria}
\author{Vassil~N.~Gourev}
\email[E-mail: ]{gourev@phys.uni-sofia.bg}
\affiliation{Department of Atomic Physics, Faculty of Physics,\\
 St.~Clement of Ohrid University at Sofia,\\
5 James Bourchier Blvd., BG-1164 Sofia, Bulgaria}

\author{Stojan~G.~Manolev}
\email[E-mail: ]{manolest@yahoo.com}
\affiliation{Middle School Goce Delchev,\\
Purvomaiska str. 3, MKD-2460 Valandovo, R.~Macedonia}

\author{Albert~M.~Varonov}
\email[E-mail: ]{avaronov@phys.uni-sofia.bg}
\affiliation{Department of Theoretical Physics, Faculty of Physics,\\
St.~Clement of Ohrid University at Sofia,\\
5 James Bourchier Blvd., BG-1164 Sofia, Bulgaria}
\author{Todor~M.~Mishonov}
\email[E-mail: ]{mishonov@gmail.com}
\affiliation{Department of Theoretical Physics, Faculty of Physics,\\
St.~Clement of Ohrid University at Sofia,\\
5 James Bourchier Blvd., BG-1164 Sofia, Bulgaria}


\date{02.05.2016} 

\begin{abstract}
This problem was given at the Fourth Experimental Physics Olympiad Day of Light on 23 April 2016 in Sofia organized by the Sofia Branch  of the Union of Physicists in Bulgaria and the Society of Physicists of the Republic of Macedonia, Strumica.
\end{abstract}

\maketitle


\section{Introduction}

Over 110 children from Bulgaria, Macedonia, Bosnia and Herzegovina, and Russia took part and worked with inspiration more than 4 hours in the auditoriums of the Faculty of Physics of St.~Clement of Ohrid University at Sofia. The students measured the speed of light in honor of the passed UNESCO~Year~of~Light~2015~\cite{EPO_IYL2015}. This fundamental constant was measured with experimental setups, which remained for the high school physics laboratories after the Olympiad. The students work was shown on BTV and BNT (Bulgarian national television broadcast channels) evening news. The Olympiad absolute champion Dejan Maksimovski from Skopie measured the speed of light with accuracy better than 1~\%. This is an excellent result even for the best worldwide universities. The experimental setup directly reflects the educational program and could be reproduced in every high school. The Sofia Branch of the Union of Physicists in Bulgaria started this worldwide unique Olympiad with pilot competitions in 2011 and we have already achieved international recognition; see for example the tasks from the previous Olympiads: first,~\cite{EPO1} second,~\cite{EPO2} and third.~\cite{EPO3} Our Experimental Physics Olympiad (EPO) is open for children from the whole World and in this sense it is international. A teachers qualification course was held in parallel with the Olympiad, which 35 teachers took part in. The Olympiad was accomplished thanks to generous and noble sponsors, only deacon Ignatius worked with such kind people.

\section{Problem}

\begin{figure}[h]
\includegraphics[width=9cm]{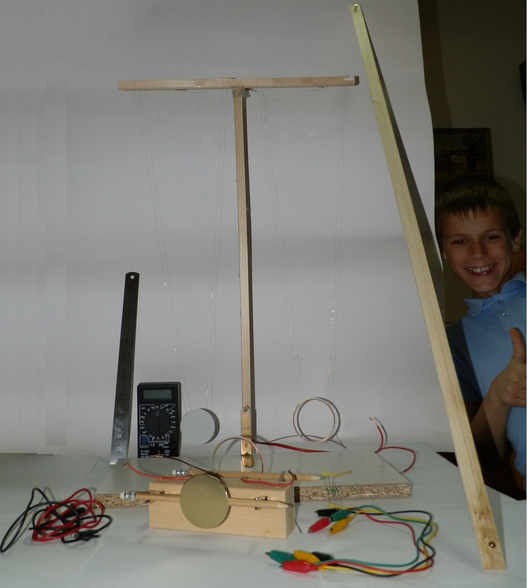}
\caption{Check whether you have the full setup of items shown in the picture: a stand with a metal plate and a ring of copper coils hanging on thin metal wires; a wooden block with another metal plate and ring of copper coils mounted; two battery size AA quadruple holders; 8~1.5~V AA size batteries; two voltage sources, placed in tubes with three metal screws (electrodes) each; a rheostat made of thin lath on top of which a thin resistive kanthal conductor is stretched; aluminum plate which is used as a slider by nipping it with a crocodile clip and held in hand; cables with crocodile clip connectors; 4~resistors; a metal ruler with 0.5~mm divisions and 1~multimeter. It is supposed that you have an additional multimeter with its connecting cables.}
\label{exp_setup_photo}
\end{figure}

Measure the speed of light $c$ using the given experimental setup shown in Figure~\ref{exp_setup_photo}.
You can follow the described instructions below.
The current text, as well as the experimental setup, remain for the participants after the Olympiad.

\begin{enumerate}

\section{Experimental problem.
\emph{Be careful not to tear the wires!}  
}
\subsection{Gravitational part. Measurement of Earth's gravitational acceleration $g$}

\begin{figure}[h]
\includegraphics[width=2.5cm]{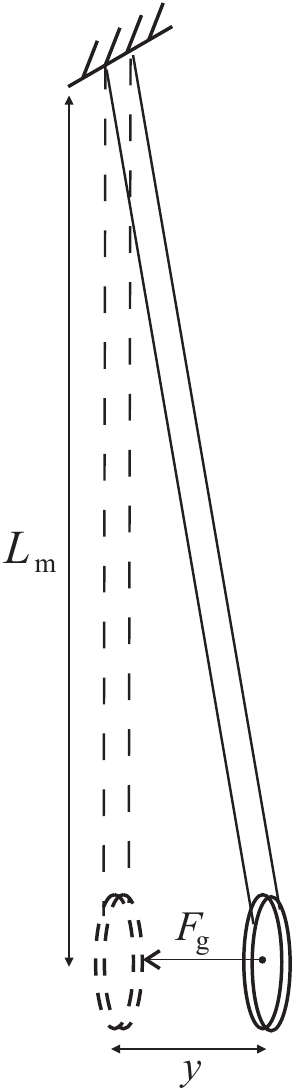}
\caption{
A ring with coiled wire hangs on two thin wires. The distance from the suspension point to the ring center is
$L_\mathrm{m}$. 
As the pendulum moves away from its equilibirum position by a little distance $y$, the restoring force is 
$F_g=-m gy/L_\mathrm{m}$. 
The minus sign means that the force is directed against the displacement.
}
\label{setup_g}
\end{figure}

\item Measure the period $T$ (in seconds) of oscillation of the suspended ring. 

Gently swing the ring in the direction of its axis
and count how many oscillations $N_\mathrm{osc}$ you can observe.
Swing again and measure the time of these oscillations $T_N$.
Find the period as the ratio of the full time and the number of oscillations $T=T_N/N_\mathrm{osc}$. 
Repeat the experiment several times and find the average value of the measured period.

\item Find the angular frequency $\omega=2\pi/T$.

\item Measure the distance $L_\mathrm{m}$ (in meters) from the suspension point to the hanging ring center (look in Figure~\ref{setup_g}). 

\item Find the Earth's gravitational acceleration $g=L_\mathrm{m}\omega^2$.

We remind you the well known formula for the period of a pendulum
\begin{equation}
T=2 \pi \sqrt \frac{L_\mathrm{m}}{g}.
\end{equation}
\item How many percent does you result for the Earth's gravitational acceleration $g$ differ with from the value you know?

\subsection{Measurement of the electromotive voltages $\mathcal{E}_1$, $\mathcal{E}_2$, $\mathcal{E}_3$ and $\mathcal{E}_4$  of the sources for the electric experiment with an ammeter and a voltmeter only}
\label{measure_varE}

The voltage source you are going to use for the electric experiment consists of many small 12~V batteries, size 23A, placed in two plastic tubes.
Each of the tubes has three metal screws (electrodes), one at each end and one in the middle.
There are labels $\mathcal{E}_1$, $\mathcal{E}_2$, $\mathcal{E}_3$ or $\mathcal{E}_4$ in the middle between any two electrodes.
There is a high-resistance resistance $r_i$ placed in series to the batteries in the corresponding sections between any two metal screws, as it is shown in the equivalent circuit of the voltage source in Figure~\ref{HV-Battery}.
That resistance is installed for protection purpose.
Do not disassemble the voltage source and do not remove the protection resistance!

An accurate measurement of the electromotive voltage with the given voltmeter is not possible because the source internal resistance $r_i$ is comparable in value to the voltmeter internal resistance.

\begin{figure}[h]
\includegraphics[width=12cm]{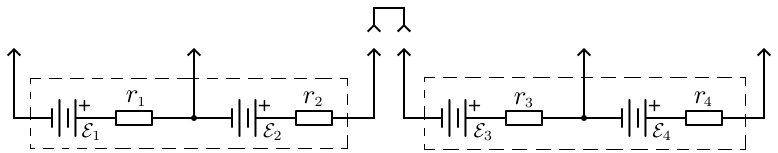}
\caption{
Equivalent circuit of the voltage source for the electric experiment. The plastic tubes containing the batteries and the resistance are designated with dashed lines. The screws coming out of the tubes (electrodes) are schematically drawn with arrows, while the connecting crocodile clips are shown with a complementary sign.
}
\label{HV-Battery}
\end{figure}

\item Measure the electromotive voltages $\mathcal{E}_1$, $\mathcal{E}_2$, $\mathcal{E}_3$ and $\mathcal{E}_4$
with an ammeter and a voltmeter only.

\begin{figure}[h]
\includegraphics[width=12cm]{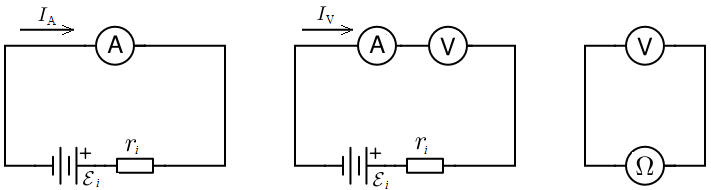}
\caption{Measurement of the voltage of a section of the voltage source $\mathcal{E}_i$ for the electrostatic experiment. 
A high resistance $r_i$ ($i=1,\,2,\,3,\,4$) that limits the current is hidden inside the tubes in between any two screws. 
Left figure: Connect an ammeter between two adjacent/neighboring screws (electrodes) and measure the flowing current is $I_\mathrm{A}$.
Middle figure: After connecting a voltmeter in series to the ammeter, the current $I_\mathrm{V}$ is smaller and the voltmeter shows voltage $U$.
Right Figure: Circuit for a direct measurement of the voltage internal resistance $R_\mathrm{V}$ using another multimeter switched as an ohmmeter.
}
\label{Measure-battery}
\end{figure}

Connect the ammeter with one of the voltage source sections as it is shown in the left circuit in Figure~\ref{Measure-battery} and measure the flowing current $I_\mathrm{A}$ through it.
There is a high-resistance resistance limiting the current in the voltagre source, so there is no danger of a large current flowing throught the ammeter.
Repeat the measurement for all four voltage source sections and write down the results in a table as it is shown in the exemplary table~\ref{tab:I12345}.

Connect the ammeter and the voltmeter to the voltage source in series, as it is shown in the middle circuit in Figure~\ref{Measure-battery}  and measure the current $I_\mathrm{V}$ flowing through the circuit and the voltage $U$ measured by the voltmeter.
Repeat the measurement for all four voltage source sections and write down the results in a table as it is shown in the exemplary Table~\ref{tab:I12345}.
 
\begin{table}[h]
\caption{Exemplary table for currents and voltages experimental results measured from the circuits in Figure~\ref{Measure-battery} as well as for the calculated value of the electromotive voltage of the different voltage source sections.}
\begin{tabular}{| r | r | r | r | r | r | r | }
\tableline
$i$& $I_\mathrm{A} \, [\mathrm{mA}]$ & $I_\mathrm{V} \,[\mathrm{mA}]$ & $U \,[\mathrm{V}]$ & $\mathcal{E} \,[\mathrm{V}]$ & $r \, [\Omega$] & $R_\mathrm{V} \, [\Omega$]\\
\tableline
1 & & & & & & \\
2 & & & & & & \\
3 & & & & & & \\
4 & & & & & & \\
\tableline
\end{tabular}
\label{tab:I12345}
\end{table}

For each voltage source section write down in exemplary Table~\ref{tab:I12345} the electromotive voltages calculated from the formula
\begin{equation}
\label{eq:E}
\mathcal{E}=\frac{U}{1-I_\mathrm{V}/I_\mathrm{A}}.
\end{equation}

\item What is the internal resistance of each voltage source section?

Using the data from Table~\ref{tab:I12345}, determine the internal resistance of each voltage source section 
\begin{equation}
\label{eq:r}
r=\frac{U}{I_\mathrm{A}-I_\mathrm{V}},
\end{equation}
for each of the four measurements.

\item Determine the voltmeter internal resistance.

Using the data from Table~\ref{tab:I12345}, determine the voltmeter internal resistance
\begin{equation}
\label{eq:R}
R_\mathrm{V}=\frac{U}{I_\mathrm{V}},
\end{equation}
for each of the four measurements and compute the average value from all measurements.

You can measure directly the ohmmeter internal resistance from the right circuit in Figure~\ref{Measure-battery}. 
Compare the calculated average value with the directly measured. How many percent is the difference? 

\item Derive formulas~\eqref{eq:E}, \eqref{eq:r}  and~\eqref{eq:R}.

\subsection{Measurement of the electromotive voltages $\mathcal{E}_1$, $\mathcal{E}_2$, $\mathcal{E}_3$ and $\mathcal{E}_4$ of the sources for the electric experiment with an ammeter and resistors}
\textit{This problem series (section C) is for the junior students competing in category S. The senior participants can return back to it after finishing the remaining problems.}

\item{Measure the resistances $R_1^*,$ $R_2^*,$ $R_3^*$ and $R_4^*$ of the four given resistors and represent the results tabulated.}

\item Connect the circuit in Figure~\ref{Measure-battery2} and measure the current $I$, when you have placed resistor $R_1=R_1^*,$ $R_2=R_1^*+R_2^*,$ $R_3=R_1^*+R_2^*+R_3^*$ and $R_4=R_1^*+R_2^*+R_3^*+R_4^*$.

\begin{figure}[h]
\includegraphics[width=10cm]{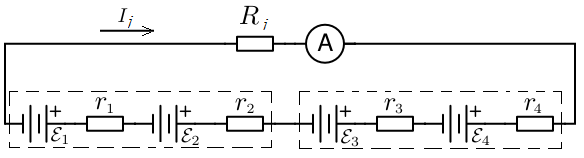}
\caption{Voltage measurement of a section of the voltage source for the electrostatic experiment.
The current through the circuit $I_j$ is measured
with different external resistances $R_j$
and thus the internal resistance of the source hidden in the tube is measured.
}
\label{Measure-battery2}
\end{figure}

Write down the measured values in a table, as it is shown in the exemplary Table~\ref{tab:E1234_2}.
Fill the last column with the reciprocal values of the current.

\begin{table}[h]
\caption{Exemplary table for experimental results for the currents and voltages measured with the circuits in Figure~\ref{Measure-battery2} for the different sections of the voltage sources.}
\begin{tabular}{| r | r | r | r | }
\tableline
i& $R_j \, [\mathrm{k}\Omega]$ & $I_j \,[\mu\mathrm{A}]$ &  $1/I_j \,[\mu\mathrm{A}^{-1}]$ \\
\tableline
1 &  &  &\\
2 & &  &\\
3 & & &\\
4 & &  &\\
\tableline
\end{tabular}
\label{tab:E1234_2}
\end{table}

\item Using the results from Table~\ref{tab:E1234_2}, represent graphically the relation of the resistance $R_j$ from the current reciprocal value $1/I_j$.

The resistance $R_j$ and the reciprocal value of the current $I_j$ are related as follows
\begin{equation}
\label{eq:R_j}
R_j=\mathcal{E}_\mathrm{tot} \frac{1}{I_j} - r_\mathrm{tot},
\end{equation}
where $\mathcal{E}_\mathrm{tot}$ is the summed electromotive voltage of all volatge sources connected in series and $r_\mathrm{tot}$ is their summed internal resistance.

\item Determine the electromotive voltage $\mathcal{E}_\mathrm{tot}=\mathcal{E}_1 + \mathcal{E}_2 + \mathcal{E}_3 + \mathcal{E}_4$ and the internal resistance $r_\mathrm{tot}=r_1+r_2+r_3+r_4$ of both voltage sources connected in series.

Draw a straight line passing closest to the experimental points. 
You can determine the battery electromotive voltage $\mathcal{E}_\mathrm{tot}$ from the angular coefficient and
determine the battery internal resistance $r_\mathrm{tot}$ from the intersection of the straight line with the ordinate.

\item Derive formula~\eqref{eq:R_j}.

\subsection{Electrostatic determination of $\varepsilon_0$}

\item Qualitative electric experiment

Using a crocodile clip cable, connect in series both voltage sources without connecting the voltmeter.
Be careful with the battery polarities. Connect (+) with (-) of both sources.
The total voltage
$U_4=\mathcal{E}_1+\mathcal{E}_2+\mathcal{E}_3+\mathcal{E}_4$ is calculated as a sum of the earlier determined electromotive voltages in section~\ref{measure_varE}.
Again using the crocodile clip connecting cables, apply that voltage to the round plates through their connecting conductors.
Let the plates be separated (farther than 1~cm) but parallel to each other and situated on a single axis lying on their centers.
Look at the diagram shown in Figure~\ref{setup_e}.
Gradually move the wooden block until the suspended plate suddenly sticks to the fixed to the wooden block plate.
Carefully repeat the experiment slowly, waiting for the oscillations to attenuate.
Move the wooden block in such a way that the plates remain parallel.
After the pendulum loses equilibrium and the plates stick together, cut their connection to the voltage source. 
Use a crocodile clip cable and short-circuit the wires coming out of both plates.
Then the plates separate from each other and after the oscillations attenuate, measure the distance $x_\mathrm{e}$ between the plates of this capacitor.

\item 
Carefully repeat the experiment measuring the distance $x_\mathrm{e}$ between the plates with 0.5~mm accuracy for different voltages:
$U_4=\mathcal{E}_1+\mathcal{E}_2+\mathcal{E}_3+\mathcal{E}_4,$
$U_3=\mathcal{E}_1+\mathcal{E}_2+\mathcal{E}_3,$
$U_2=\mathcal{E}_1+\mathcal{E}_2,$ and
$U_1=\mathcal{E}_1.$

Write down the measurement results in exemplary table~\ref{tab:xe3_vs_U2}.
Using a calculator, complete the table with two additional columns.
$x_\mathrm{e}^3$ and $U^2$.

\begin{table}[h]
\caption{A table with experimental results from the electric experiment}
\begin{tabular}{| r | r | r | r | r | r |}
\tableline
i& $U \,[\mathrm{V}]$ & $x_\mathrm{e} \,[\mathrm{mm}]$ & $U^2 \,[\mathrm{V}^2]$ & $x_\mathrm{e}^3 \,[\mathrm{mm}^3]$ \\
\tableline
1 &  &  &  & \\
2 &   & &  &  \\
3 &   &  &  &  \\
4 &  & &  &  \\
\tableline
\end{tabular}
\label{tab:xe3_vs_U2}
\end{table}

\item Graphically represent the relation $x_\mathrm{e}^3$ from $U^2$ on a millimeter paper and determine the slope $k_\mathrm{e}$ of the straight line passing closest to the experimental points.

\begin{figure}[h]
\includegraphics[width=7cm]{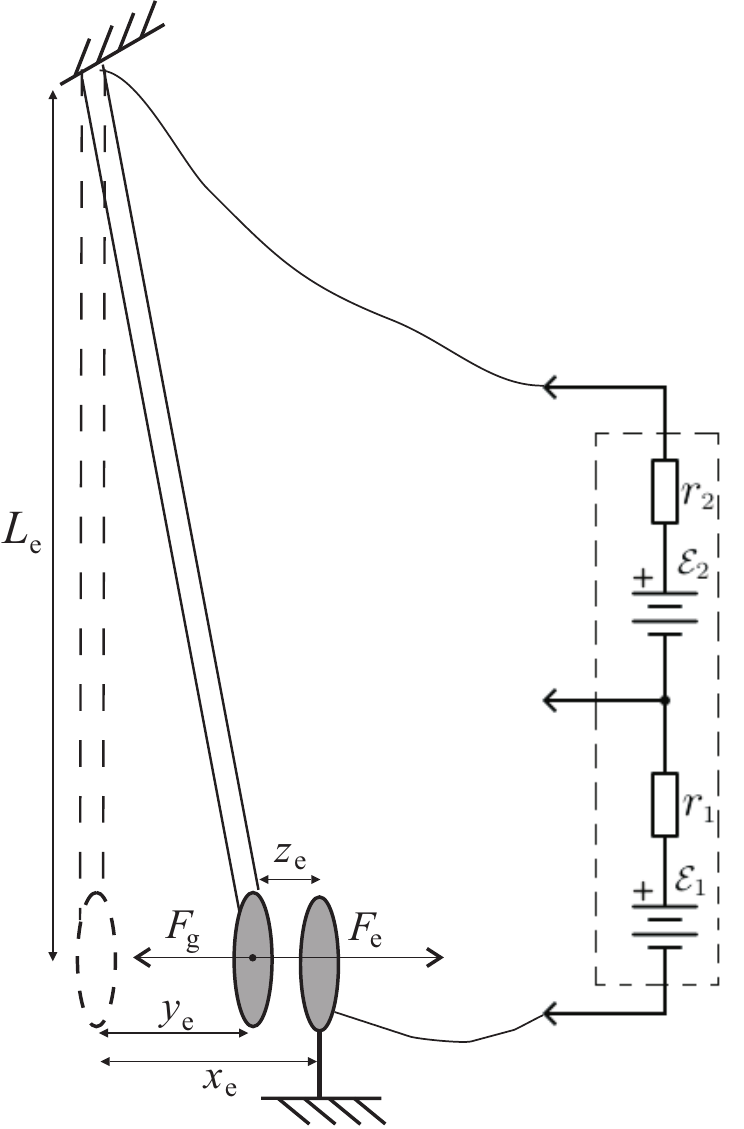}
\caption{
An electric pendulum for the vacuum dielectric permittivity $\varepsilon_0$.
The pendulum length is $L_\mathrm{e}$. 
The displacement from the pendulum equilibrium position shortly before sticking to the attractive fixed plate is $y_\mathrm{e}=x_\mathrm{e}-z_\mathrm{e}.$ Here 
$z_\mathrm{e}$ is the distance between the plates shortly before the pendulum swings towards the wooden block. The distance between  the pendulum equilibrium position and the fixed plate is $x_\mathrm{e}$.
The electric field between the plates is created by connected in series batteries with electromotive voltages
$\mathcal{E}_1$ and $\mathcal{E}_2$. 
The large resistances limiting the current are denoted with
$r_1$ and $r_2$. 
The resistances, as well as, the batteries are placed inside a plastic tube shown schematically with a dashed line.
The attractive electric force between the plates $F_\mathrm{e}$ is balanced with the gravitational force $F_\mathrm{g}$ in equilibrium.
}
\label{setup_e}
\end{figure}

\item Measure the mass $m_\mathrm{e}$ of the suspended plate. 

Use the electronic scale that is situated at the proctor.
Do not detach the suspended plate from the stand for the measurement.
Place the scale on the high pad that goes with the scale.
Make the measurement while the plate is suspended by the thin copper wire, being careful the wire not to imply additional forces on the plate during the measurement.

\item Measure the diameter $D_\mathrm{e}$ of both plates.

\item Measure the distance from the plate center to the horizontal suspension bar $L_\mathrm{e}$.

\item 
Determine the vacuum dielectric permittivity $\varepsilon_0$ via an approximate relation between $x_\mathrm{e}$ and $U$ 
derived assuming $x_\mathrm{e}\ll D_\mathrm{e}$
\begin{equation}
x_\mathrm{e}^3 =k_\mathrm{e}  U^2,
\end{equation}
where $k_\mathrm{e}$ is the linear coefficient
\begin{equation}
\label{eq:epsilon_0}
k_\mathrm{e} 
=\frac{27}{32} \pi  \varepsilon_0 \frac{L_\mathrm{e} D_\mathrm{e}^2}{m_\mathrm{e} g}.
\end{equation}

Using this formula and the experimentally measured value of
$k_\mathrm{e} $, determine the vacuum dielectric permittivity
\begin{equation}
\label{eq:k_e}
\varepsilon_0=
\frac{32}{27\pi}\frac{m_\mathrm{e} g}{L_\mathrm{e} D_\mathrm{e}^2}k_\mathrm{e}. 
\end{equation}

\item 
Add an additional column $x_\mathrm{e}^3 (1-\frac{4}{3 \pi} \frac{x_\mathrm{e}}{D_\mathrm{e}})$ to the table calculating the small correction
$\frac{4}{3 \pi} \frac{x_\mathrm{e}}{D_\mathrm{e}}$.
Repeat the described method above for a more accurate determination of the vacuum dielectric permittivity via the formula 
\begin{equation}
\label{Effect_of_ends}
x_\mathrm{e}^3 \left(1-\frac{4}{3 \pi} \frac{x_\mathrm{e}}{D_\mathrm{e}} \right) =
\frac{27}{32} \pi  \varepsilon_0 \frac{L_\mathrm{e} D_\mathrm{e}^2}{m_\mathrm{e} g}  U^2.
\end{equation}
This formula allows percent accuracy in using this experimental setup.

What is the difference in the determination of $\varepsilon_0$ between both methods?

\subsection{Magnetostatic determination of $\mu_0$}

\item Quality magnetic experiment

Place both rings parallel at a distance
$x_\mathrm{m}= \mathrm{20\;mm}$ between the middle of the coils.
Use crocodile clip cables and connect the circuit shown in Figure~\ref{setup_m}.
Let us follow the current path that starts from (+) of the battery.
First the current goes through the resistive kanthal wire stretched above the thin lath.
Such a variable resistance is also called a rheochord.
The circuit is closed by the aluminum slider clipped by a crocodile clip and held in hand.
Then the current flows through the ammeter, the suspended current coil, the fixed current coil with the same number of coils
$N=50$ and through (-) of the battery returns back to the current source.

Upon contact between the slider and the resistive wire the coils tremble.
If the coils repel each other, change the connection polarity of one of them.
The currents should be parallel and upon switching on the current, the coils attract each other.
Move the slider on the kanthal wire looking at the ammeter and the suspended coil at the same time.
Begin with small currents and moving the slider remember the situation where the pendulum loses stability and the suspended ring swings towards the fixed one.
Repeat the experiment slowly waiting for the oscillations to attenuate.
Write down the smallest (critical) current $I$ that causes the freely suspended ring suddenly to stick to the fixed one.

\begin{figure}[h]
\includegraphics[width=7cm]{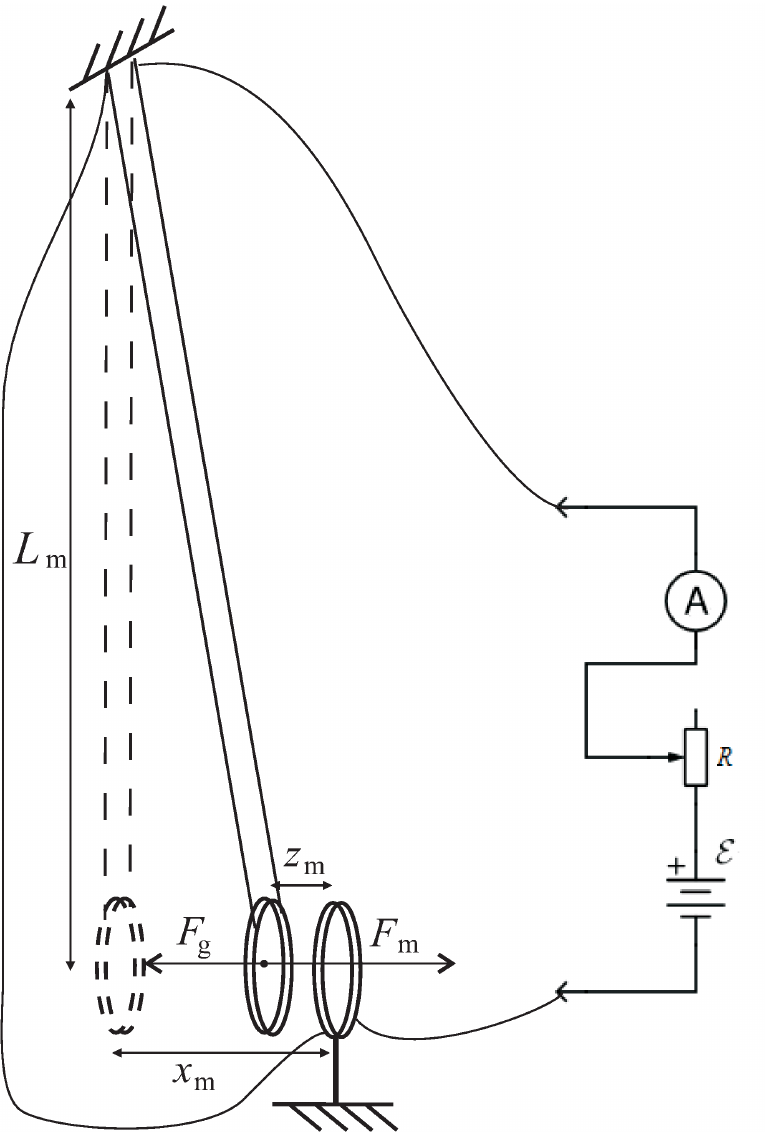}
\caption{
Magnetic pendulum for measurement of $\mu_0$, the vacuum magnetic permeability.
The pendulum length is $L_\mathrm{m}$. 
The distance from the fixed coil to the pendulum equilibrium position is $x_\mathrm{m}$.
The distance between the rings at the moment before the suspended one to swing towards the fixed one is $z_\mathrm{m}$.
The magnetic field between the coils is created by the current from 4 or 8 1.5~V batteries placed in holders and connected in series.
The current is measured from the ammeter and is regulated by the variable resistance $R$
via the slider sliding on the kanthal wire stretched above the lath.
The attractive magnetic force between the rings with parallel currents $F_\mathrm{m}$ is balanced with the gravitational force $F_\mathrm{g}$.
}
\label{setup_m}
\end{figure}

\item Repeat the experiment and measure the critical current $I$ for different values of $x_\mathrm{m}$, for instance 25, 20, 15, 10, 5~mm. 

Represent the results for the distance $x_\mathrm{m}$ and current $I$ in the first two columns of table~\ref{tab:mu0}.
Using a calculator, compute the additional columns for
$x_\mathrm{m}^2$ and $I^2$.
In other words, represent the tabulated relation $x_\mathrm{m}^2$ from $I^2$.

For the small distances use only one battery holder with 4 batteries.

\item Graphically represent the relation $x_\mathrm{m}^2$ from $I^2$ and determine the slope $k_\mathrm{m}$ of the straight line passing closest to the experimental points.
This is the method for experimental determination of the slope $k_\mathrm{m}$.

\item Measure the pendulum length, i.e. the distance $L_\mathrm{m}$ from the horizontal bar to the suspended coil center.

\item Measure the mass $m_\mathrm{m}$ of the suspended coil.

\item Measure the diameter $D_\mathrm{m}$ of both coils.

\item Determine the vacuum magnetic permeability $\mu_0$,
using the approximate formula
\begin{equation}
x_\mathrm{m}^2 =k_\mathrm{m}  I^2,
\end{equation}
where $k_\mathrm{m}$ is the linear coefficient
\begin{equation}
\label{eq:k_m}
k_\mathrm{m} =2 \mu_0 \frac{L_\mathrm{m} N^2 D_\mathrm{m}}{m g}.
\end{equation}
This formula is applicable when $x_\mathrm{m}\ll D_\mathrm{m}$
and we can accept that the conductors are straight.

This formula allows to express the vacuum magnetic permeability
 $\mu_0$ via the experimentally determined slope
\begin{equation}
\label{eq:k_m}
\mu_0=\frac{m gk_\mathrm{m}}{2 L_\mathrm{m} N^2 D_\mathrm{m}}.
\end{equation}

\item The attractive force between two current rings is a well known magneto-static problem with many applications.
The exact formula for $\mu_0$ determination with our experimental setup is
\begin{equation}
\label{Correction}
x_\mathrm{m}^2 \left[1+f \left(\frac{x_\mathrm{m}}{D_\mathrm{m}}\right) \right] =k_\mathrm{m}  I^2,
\end{equation}
where the function $f(x_\mathrm{m}/D_\mathrm{m})$ 
is shown tabulated below and graphically in Figure~\ref{f_delta}.

With the accuracy we work, the correction can also be calculated with the approximate formula
\begin{equation}
\label{function_2_rings}
f \left(\frac{x_\mathrm{m}}{D_\mathrm{m}}\right)\approx
\frac{1}{16}\left(-5+6 \log{ \frac{8 D_\mathrm{m}}{x_\mathrm{m}} } \right)
 \frac{x_\mathrm{m}^2}{D_\mathrm{m}^2}.
\end{equation}

An example for the experimental results and correction representation is given in Table~\ref{tab:mu0}.
Fill in your table.

\begin{table}[h]
\caption{An exemplary table for experimental results and correction function representation for the experiment with the current coils.  There is a linear relation between $I^2$ and $x_\mathrm{m}^2(1+f)$ and the proportionality coefficient determines $\mu_0$.}
\begin{tabular}{| r | r | r | r | r | r | r | r| }
\tableline
$i$& $x_\mathrm{m}[\mathrm{mm}]$& $I [\mathrm{mA}]$  & $x_\mathrm{m}^2 [\mathrm{mm^2}]$ & $I^2 [\mathrm{mA^2}]$ &  $x_\mathrm{m}/D_\mathrm{m}$ & $1+f(x_\mathrm{m}/D_\mathrm{m})$ & $x_\mathrm{m}^2(1+f) [\mathrm{mm^2}] $ 
\\
\tableline
1 &  5 & && & & & \\
2 & 10& && & & & \\
3 & 15& && & & & \\
4 & 20& & && & & \\
5 & 25& && & & & \\
\tableline
\end{tabular}
\label{tab:mu0}
\end{table}

\begin{center}
\begin{tabular}{| c | c || c | c || c | c || c |  c || c | c |}
		\hline
			\multicolumn{1}{| c |}{ } & \multicolumn{1}{ c ||}{ } & \multicolumn{1}{ c |}{ } & \multicolumn{1}{ c ||}{ } & \multicolumn{1}{ c |}{ } & \multicolumn{1}{ c ||}{ } 
			& \multicolumn{1}{ c |}{ } & \multicolumn{1}{ c ||}{ } & \multicolumn{1}{ c |}{ } & \multicolumn{1}{ c |}{ } \\
			\boldmath$x/D$ & \boldmath$f(x/D)$ & \boldmath$x/D$ &\boldmath$f(x/D)$ & \boldmath$x/D$ & \boldmath$f(x/D)$ & \boldmath$x/D$ & \boldmath$f(x/D)$  & \boldmath$x/D$ & \boldmath$f(x/D)$ \\[15pt] \hline 
			0.005	&	0.0001	&	0.105	&	0.0148	&	0.205	&	0.0480	&	0.305	&	0.0976	&	0.405	&	0.1644	\\	\hline
0.010	&	0.0002	&	0.110	&	0.0162	&	0.210	&	0.0502	&	0.310	&	0.1005	&	0.410	&	0.1682	\\	\hline
0.015	&	0.0005	&	0.115	&	0.0175	&	0.215	&	0.0522	&	0.315	&	0.1034	&	0.415	&	0.1721	\\	\hline
0.020	&	0.0008	&	0.120	&	0.0188	&	0.220	&	0.0545	&	0.320	&	0.1066	&	0.420	&	0.1760	\\	\hline
0.025	&	0.0012	&	0.125	&	0.0202	&	0.225	&	0.0566	&	0.325	&	0.1096	&	0.425	&	0.1800	\\	\hline
0.030	&	0.0016	&	0.130	&	0.0216	&	0.230	&	0.0589	&	0.330	&	0.1126	&	0.430	&	0.1840	\\	\hline
0.035	&	0.0021	&	0.135	&	0.0231	&	0.235	&	0.0613	&	0.335	&	0.1156	&	0.435	&	0.1880	\\	\hline
0.040	&	0.0027	&	0.140	&	0.0245	&	0.240	&	0.0634	&	0.340	&	0.1190	&	0.440	&	0.1920	\\	\hline
0.045	&	0.0034	&	0.145	&	0.0262	&	0.245	&	0.0659	&	0.345	&	0.1221	&	0.445	&	0.1965	\\	\hline
0.050	&	0.0040	&	0.150	&	0.0278	&	0.250	&	0.0684	&	0.350	&	0.1252	&	0.450	&	0.2006	\\	\hline
0.055	&	0.0048	&	0.155	&	0.0294	&	0.255	&	0.0709	&	0.355	&	0.1287	&	0.455	&	0.2047	\\	\hline
0.060	&	0.0056	&	0.160	&	0.0312	&	0.260	&	0.0732	&	0.360	&	0.1319	&	0.460	&	0.2092	\\	\hline
0.065	&	0.0064	&	0.165	&	0.0328	&	0.265	&	0.0757	&	0.365	&	0.1355	&	0.465	&	0.2134	\\	\hline
0.070	&	0.0073	&	0.170	&	0.0345	&	0.270	&	0.0783	&	0.370	&	0.1390	&	0.470	&	0.2181	\\	\hline
0.075	&	0.0083	&	0.175	&	0.0365	&	0.275	&	0.0810	&	0.375	&	0.1423	&	0.475	&	0.2223	\\	\hline
0.080	&	0.0092	&	0.180	&	0.0382	&	0.280	&	0.0837	&	0.380	&	0.1460	&	0.480	&	0.2270	\\	\hline
0.085	&	0.0103	&	0.185	&	0.0401	&	0.285	&	0.0864	&	0.385	&	0.1497	&	0.485	&	0.2317	\\	\hline
0.090	&	0.0114	&	0.190	&	0.0421	&	0.290	&	0.0891	&	0.390	&	0.1530	&	0.490	&	0.2361	\\	\hline
0.095	&	0.0125	&	0.195	&	0.0440	&	0.295	&	0.0919	&	0.395	&	0.1568	&	0.495	&	0.2408	\\	\hline
0.100	&	0.0136	&	0.200	&	0.0461	&	0.300	&	0.0947	&	0.400	&	0.1606	&	0.500	&	0.2456	\\	\hline

		\hline
\end{tabular}
\end{center}

\begin{figure}[h]
\includegraphics[width=18cm]{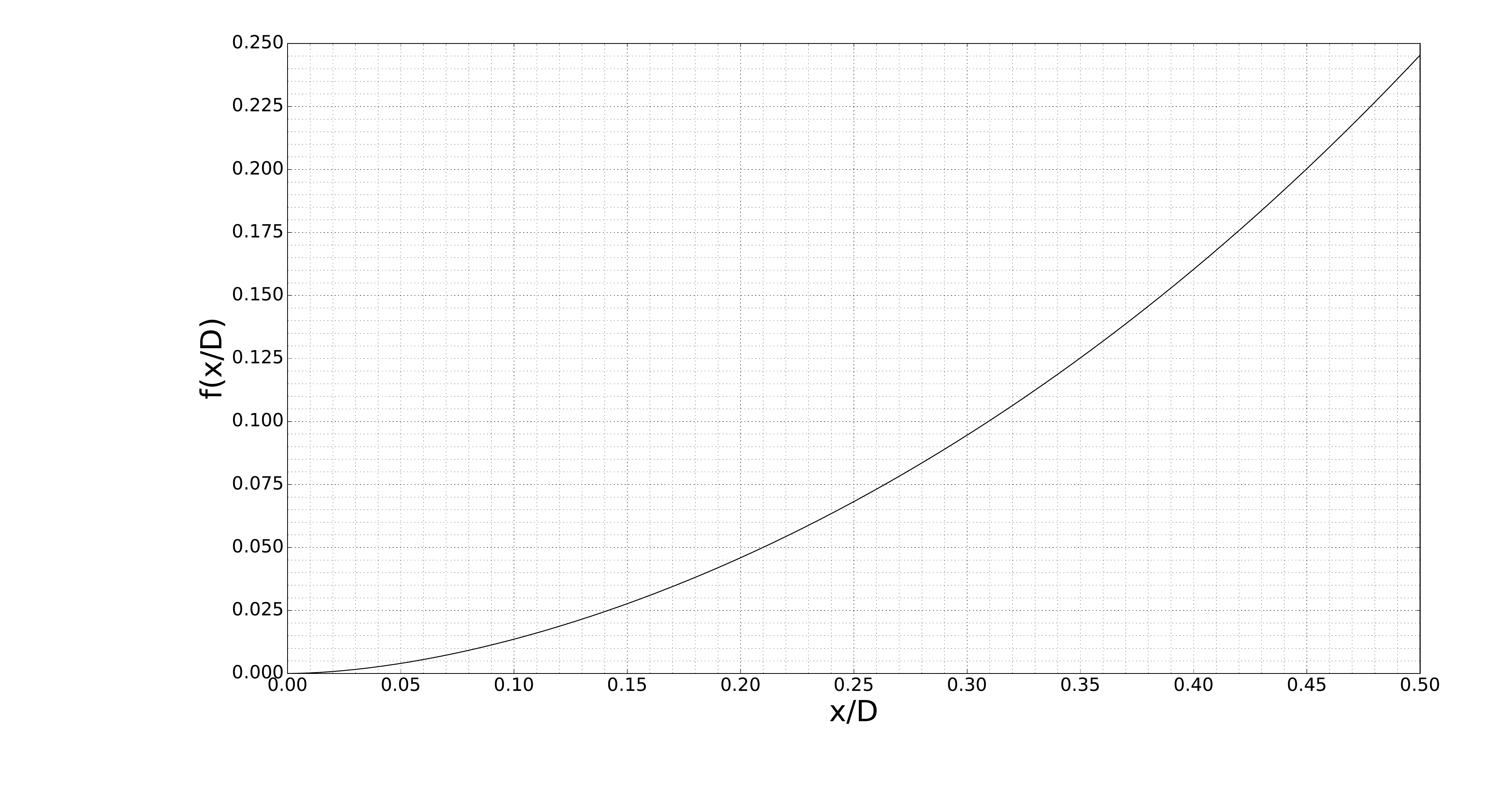}
\caption{The correction $f(x_\mathrm{m}/D_\mathrm{m})$ from equation (\ref{Correction})
as a function of the dimensionless ratio of the equilibrium displacement $x_\mathrm{m}$ and the coils diameter $D_\mathrm{m}$.
}
\label{f_delta}
\end{figure}

\item 
Using the exact formula \eqref{Correction}
and the depicted correction function 
in Fig.~\ref{f_delta} tabulated above,
graphically represent the experimental data:
in ordinate
$(1+f)x_\mathrm{m}^2$, 
in abscissa $I^2$
and determine the vacuum magnetic permeability $\mu_0$ from the slope.
How many percent is the difference between the results from the approximate and exact formula?
\end{enumerate}

\section{Determination of the speed of light}

Using the determined $\varepsilon_0$ and
$\mu_0$ in the previous subsections, calculate the speed of light
\begin{equation}
c=\frac{1}{\sqrt{\varepsilon_0 \mu_0}}.
\end{equation}
Do not confuse yourself if your result is different from the value you know, this is your first measurement of a fundamental constant.

\section{Theoretical problems}

Using the formulas for the attractive force between the plates of an infinite parallel plane capacitor, derive formula 
(\ref{eq:epsilon_0}) that you used to determine $\varepsilon_0$.

Using the formulas for the attractive forces between infinite parallel currents, derive the formula
 (\ref{eq:epsilon_0}) that you used for approximate determination of $\mu_0.$

\section{Homework problem. Sommerfeld prize 137 leva}

Look in electrodynamics textbooks, encyclopedias or Internet for the formula for the capacity of a capacitor that takes into account the effect of ends and derive the exact formula (\ref{Effect_of_ends}) for determination of
$\varepsilon_0$ with our experimental setup.

Similarly, look in textbooks for the formulas for mutual inductance and interaction forces between current coils and derive the exact formula
(\ref{Correction}).
The correction function
$f(x_\mathrm{m}/D_\mathrm{m})$
can be calculated numerically or only the first approximation of equation~(\ref{function_2_rings}) derived in power series of  $x_\mathrm{m}/D_\mathrm{m}$.

Send the solution to epo@bgphysics.eu from the address you registered until 07:00 on 24 April 2016.

You can work in teams and consult professors in theoretical physics, electrodynamics or electrotechnics. The prize is given personally to the participant at the day of the results announcement only.

\acknowledgments{}

The Olympiad was held with the cooperation of Faculty of Physics of University of Sofia and Regional inspectorate of education in Sofia city. Special thanks to the dean prof.~A.~Dreischuh and the chief inspector P.~Ivanova. 

The Olympiad opening ceremony was visited by the president of Balkan Physical Union acad.~A.~Petrov and the president of Macedonian Physical Society assoc.~prof.~B.~Mitrevski.

At the opening was read the address from Presidency of Bulgaria and  
senior expert in physics from Bulgarian Ministry of Education and Science 
read the address from the minister M.~Kuneva.

The Olympiad organizers highly appreciate the indispensable support of Theoretical and Computational Physics and Astrophysics foundation, Programista; Ltd, New Bulgarian University (NBU), VMware, Mobius Ads, Kamen Mirchev, and other private anonymous sponsors.

The organizers are extremely grateful to family members, friends and colleagues for their overnight support and patience: A.~Petkov, S.~Velkov, A.~Bozhankova, Z.~Dimitrov, P.~Beckyarov, D.~Todorov, M.~Stoev, L.~V.~Georgiev,  P.~Bobeva, D.~Kocheva, A.~Stefanov, N.~Yordanova, Ts.~Metodieva, K.~Koleva, I.~Dimitrov, K.~Gourev, B.~Nathan, D.~Tsvetanov, M.~Angelova, L.~Angelov, L.~Atanasova, D.~Damyanov, A.~Vassileva, assoc.~prof.~E.~Peneva, assoc.~prof.~S.~Kolev, G.~Parkov, R.~Simeonov, N.~Panaiotov, K.~Angelov. In the critical moments the Olympiad was saved by the University Rescue Team (\url{http://www.uaso.org/index.php?&Lang=2}) led by V.~N.~Gourev.

Next Olympiad EPO5 will be held in the autumn of 2017. \textit{Hasta luego!}


\newpage

\section{Original Bulgarian Text: Условие на задачата}
\section{Задача}

\begin{figure}[h]
\includegraphics[width=9cm]{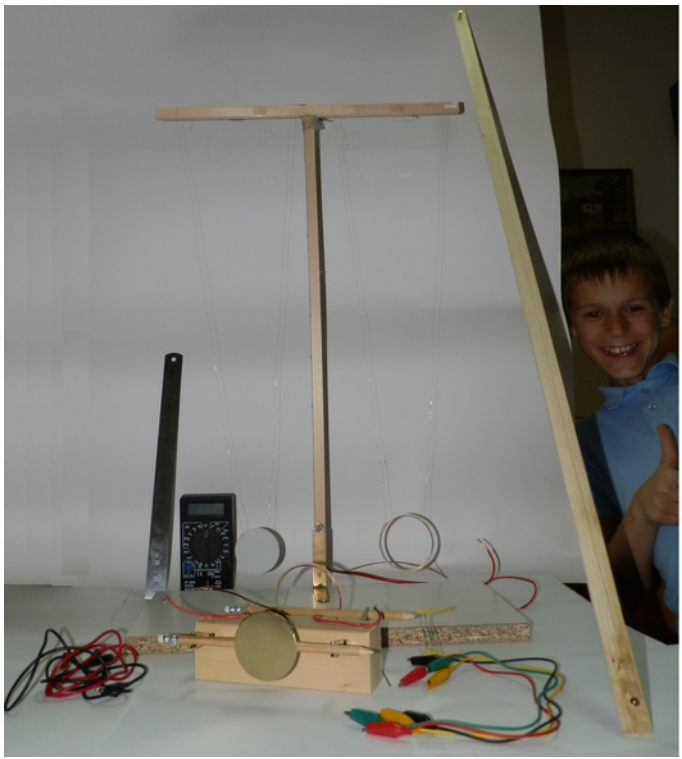}
\caption{Проверете дали имате пълния набор елементи, показан на снимката: стойка, на която са закачени метална пластина и пръстен от медни навивки, висящи на тънки медни проводници; дървено трупче, на което са закрепени друга метална пластинка и друг пръстен с медни намотки; два четворни държача за батерии с размер AA; 8 батерии от 1.5 V размер AA; два източника на напрежение, поставени в тръби с по три метални извода (електрода); реостат, направен от тънка летва, върху която е опънат тънък съпротивителен проводник от кантал; алуминиева пластинка, която се захапва с крокодил и като се държи с ръка и се използва за плъзгач; кабели с накрайници крокодил-крокодил; 4 съпротивления; метална линия с деления 0.5~mm и 1 мултиметър. Предполага се, че Вие носите още един мултиметър със съединяващите го кабели.}
\label{BG:exp_setup_photo}
\end{figure}

Измерете скоростта на светлината $c$, като използвате дадения експериментален набор, показан на Фигура~\ref{BG:exp_setup_photo}.
Може да следвате описаните по-долу инструкции. 
След олимпиадата, настоящият текст, както и постановката остават за участниците.

\begin{enumerate}

\section{Експериментална задача. 
\large{Внимавайте да не скъсате жичките!}  
}
\subsection{Гравитационна част. Измерване на земното ускорение $g$}

\begin{figure}[h]
\includegraphics[width=2.5cm]{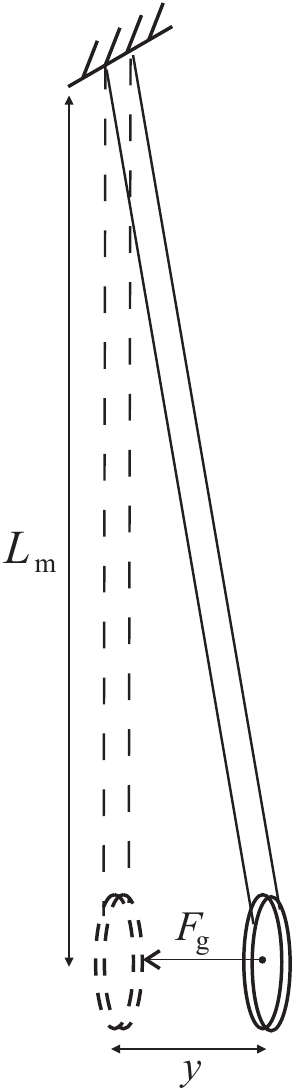}
\caption{
Пръстен с навита жичка виси на две тънки жички.
Разстоянието от точката на окачване до центъра на пръстена е
$L_\mathrm{m}$. 
Когато махалото се отклони от равновесното си положение на малко разстояние $y$, връщащата сила е 
$F_g=-m gy/L_\mathrm{m}$. 
Знакът минус означава, че силата е насочена срещу отместването.
}
\label{BG:setup_g}
\end{figure}

\item Измерете периода $T$ (в секунди) на люлеене на висящия пръстен. 

Заклатете леко пръстена в посока на оста му
и пребройте колко трептения $N_\mathrm{osc}$ може да наблюдавате.
Заклатете още веднъж и измерете времето за тези трептения $T_N$.
Определете периода като отношението на пълното време и броя на трептенията $T=T_N/N_\mathrm{osc}$. 
Повторете опита няколко пъти и намерeте средната стойност на измерения период.

\item Определете кръговата честота $\omega=2\pi/T$.  

\item Измерете разстоянието $L_\mathrm{m}$ (в метри) от точката на окачване до центъра на висящия пръстен (виж Фигура~\ref{BG:setup_g}). 

\item Определете земното ускорение $g=L_\mathrm{m}\omega^2$. 

Напомняме известната формула за периода на махало
\begin{equation}
T=2 \pi \sqrt \frac{L_\mathrm{m}}{g}.
\end{equation}
\item С колко процента вашият резултат за земното ускорение $g$ се различава от известната ви стойност?

\subsection{Измерване на електродвижещите напрежения $\mathcal{E}_1$, $\mathcal{E}_2$, $\mathcal{E}_3$ и $\mathcal{E}_4$ на източниците за електричния експеримент с помощта само на амперметър и волтметър}
\label{BG:measure_varE}

Източникът на напрежение, който ще използвате за електричния експеримент, е съставен от много на брой малки батерии от 12~V, размер 23A, поставени в две пластмасови тръби.
Всяка от тръбите има три метални извода (електрода) - по един във всеки край и един в средата.
По средата между два от електродите е залепен етикет с надпис
$\mathcal{E}_1$, $\mathcal{E}_2$, $\mathcal{E}_3$ или $\mathcal{E}_4$.
В съответните участъци последователно на батериите има поставено високоомно съпротивление $r_i$ между всеки два от металните изводи, както е показано на еквивалентната схема на източника на напрежение на Фигура~\ref{BG:HV-Battery}.
Това съпротивление е поставено с предпазна цел. 
Не разглобявайте източника на напрежение и не махайте предпазното съпротивление!

Точното измерване на електродвижещото напрежение само с предоставения ви волтметър не е възможно, тъй като вътрешното съпротивление $r_i$ на източника е сравнимо по стойност с вътрешното съпротивление на волтметъра.

\begin{figure}[h]
\includegraphics[width=12cm]{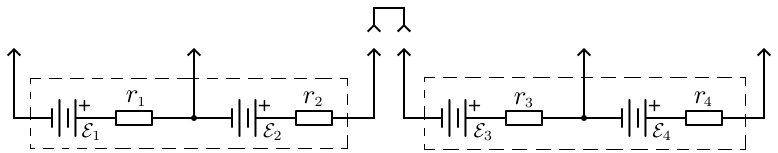}
\caption{
Еквивалентна схема на източника на напрежение за електричния експеримент. Пластмасовите тръби, в които са поставени батериите и съпротивлението, са означени с прекъснати линии.
Винтовете (електродите), които излизат от тръбите, са означени схематично със стрелки, а захапващите ги крокодили - с допълващ знак.
}
\label{BG:HV-Battery}
\end{figure}

\item Измерете електродвижещите напрежения $\mathcal{E}_1$, $\mathcal{E}_2$, $\mathcal{E}_3$ и $\mathcal{E}_4$
само с помощта на амперметър и волтметър. 

\begin{figure}[h]
\includegraphics[width=12cm]{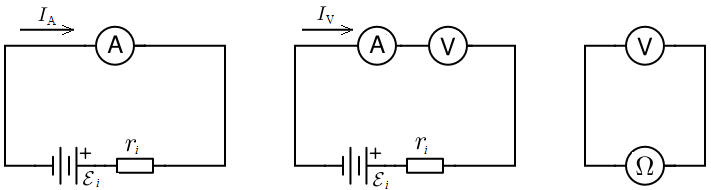}
\caption{Измерване на напрежението на един дял от източника на напрежение $\mathcal{E}_i$ за електростатичния експеримент. 
Вътре в тръбите между всеки два болта е скрито голямо съпротивление $r_i$ ($i=1,\,2,\,3,\,4$), ограничаващо тока.
Лява фигура: Между два съседни болта (електрода) 
включвате амперметър и токът, който протича е $I_\mathrm{A}$.
Средна фигура: Когато към амперметъра е включен последователно и волтметър, 
токът $I_\mathrm{V}$ е по-малък, a волтметърът показва напрежение $U.$
Дясна фигура: Схема за пряко измерване на вътрешното съпротивление на волтметъра $R_\mathrm{V}$, използвайки друг мултиметър, превключен като омметър.
}
\label{BG:Measure-battery}
\end{figure}

Свържете амперметъра с един от дяловете от източника на напрежение, както е показано в лявата схема от Фигура~\ref{BG:Measure-battery} и измерете тока $I_\mathrm{A}$, който протича през него.
В източника на напрежение има високоомно съпротивление, което ограничава тока и няма опасност да протече голям ток през амперметъра.
Повторете измерването за всички четири дяла от източниците на напрежение и нанесете резултата в таблица, както е показано в примерната Таблица~\ref{BG:tab:I12345}.

Свържете амперметъра и волтметъра последователно към източника на напрежение, както е показано в средната схема от Фигура~\ref{BG:Measure-battery}, и измерете тока $I_\mathrm{V}$, който протича през веригата и напрежението $U$, което показва волтметъра.
Повторете измерването за всички четири дяла от източниците на напрежение и нанесете резултата в таблица, както е показано в примерната Таблица~\ref{BG:tab:I12345}.
 
\begin{table}[h]
\caption{Примерна таблица за експериментални резултати за токовете и напрежението, измерени по схемите от Фигура~\ref{BG:Measure-battery}, както и за изчислената стойност на електродвижещото напрежение за различните дялове на източниците на напрежение.}
\begin{tabular}{| r | r | r | r | r | r | r | }
\tableline
$i$& $I_\mathrm{A} \, [\mathrm{mA}]$ & $I_\mathrm{V} \,[\mathrm{mA}]$ & $U \,[\mathrm{V}]$ & $\mathcal{E} \,[\mathrm{V}]$ & $r \, [\Omega$] & $R_\mathrm{V} \, [\Omega$]\\
\tableline
1 & & & & & & \\
2 & & & & & & \\
3 & & & & & & \\
4 & & & & & & \\
\tableline
\end{tabular}
\label{BG:tab:I12345}
\end{table}

За всеки един дял на източника на напрежение запишете в примерната Таблица~\ref{BG:tab:I12345} електродвижещите напрежения, изчислени по формулата
\begin{equation}
\label{BG:eq:E}
\mathcal{E}=\frac{U}{1-I_\mathrm{V}/I_\mathrm{A}}.
\end{equation}

\item Какво е вътрешното съпротивление на всеки един дял на източниците на напрежение?

Като използвате данните от Таблица~\ref{BG:tab:I12345} определете вътрешното съпротивление за всеки един дял на източника на напрежение
\begin{equation}
\label{BG:eq:R}
r=\frac{U}{I_\mathrm{A}-I_\mathrm{V}},
\end{equation}
за всяко от четирите измервания.

\item Определете вътрешното съпротивление на волтметъра.

Като използвате данните от Таблица~\ref{BG:tab:I12345}, определете вътрешното съпротивление на волтметъра
\begin{equation}
\label{BG:eq:R}
R_\mathrm{V}=\frac{U}{I_\mathrm{V}},
\end{equation}
за всяко от четирите измервания и пресметнете средната стойност от всички измервания.

Вътрешното съпротивление на омметъра може да измерите пряко по дясната схема от Фигура~\ref{BG:Measure-battery}. 
Сравнете изчислената средна стойност с пряко измерената. Колко процента е разликата? 

\item Изведете формули~\eqref{BG:eq:E}, \eqref{BG:eq:R}  и~\eqref{BG:eq:R}.

\subsection{Измерване на електродвижещите напрежения $\mathcal{E}_1$, $\mathcal{E}_2$, $\mathcal{E}_3$ и $\mathcal{E}_4$ на източниците за електричния експеримент с помощта на амперметър и резистори}
\textit{Тази серия от задачи (секция В) е за по-малките ученици, състезаващи се в категория S. По-големите ученици могат да се върнат към нея, след като приключат с останалите задачи.}

\item{Измерете съпротивленията $R_1^*,$ $R_2^*,$ $R_3^*$ и $R_4^*$ на предоставените ви четири резистора и представете резултатите таблично.}

\item Свържете схемата, дадена на Фигура~\ref{BG:Measure-battery2} и измерете тока $I$, когато сте поставили резистор $R_1=R_1^*,$ $R_2=R_1^*+R_2^*,$ $R_3=R_1^*+R_2^*+R_3^*$ и $R_4=R_1^*+R_2^*+R_3^*+R_4^*$.

\begin{figure}[h]
\includegraphics[width=10cm]{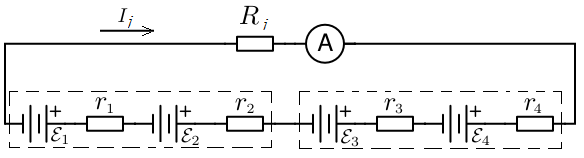}
\caption{Измерване на напрежението на един дял от източника на напрежение за електростатичния експеримент.
Токът през веригата $I_j$ се измерва 
при различни външни съпротивления $R_j$
и така се определя вътрешното съпротивление на 
източника, скрито в тръбата.  
}
\label{BG:Measure-battery2}
\end{figure}

Нанесете измерените стойности в таблица, както е показано на примерната Таблица~\ref{tab:E1234_2}.
В последната колона на таблицата запишете реципрочната стойност на тока.

\begin{table}[h]
\caption{Примерна таблица за експериментални резултати за токовете и напрежението, измерени по схемите от Фигура~\ref{BG:Measure-battery2} за различните дялове на източниците на напрежение}
\begin{tabular}{| r | r | r | r | }
\tableline
i& $R_j \, [\mathrm{k}\Omega]$ & $I_j \,[\mu\mathrm{A}]$ &  $1/I_j \,[\mu\mathrm{A}^{-1}]$ \\
\tableline
1 &  &  &\\
2 & &  &\\
3 & & &\\
4 & &  &\\
\tableline
\end{tabular}
\label{tab:E1234_2}
\end{table}

\item Използвайки резултатите от Таблица~\ref{tab:E1234_2} представете графично зависимостта на съпротивлението $R_j$ от реципрочната стойност на тока $1/I_j$.

Съпротивлението $R_j$ и реципрочната стойност на тока $I_j$ са свързани със следната зависимост
\begin{equation}
\label{BG:eq:R_j}
R_j=\mathcal{E}_\mathrm{tot} \frac{1}{I_j} - r_\mathrm{tot},
\end{equation}
където $\mathcal{E}_\mathrm{tot}$ е сумарното електродвижещо напрежение на всички последователно свързани източници на напрежение, а $r_\mathrm{tot}$ тяхното сумарно вътрешно съпротивление.

\item Определете електродвижещото напрежение $\mathcal{E}_\mathrm{tot}=\mathcal{E}_1 + \mathcal{E}_2 + \mathcal{E}_3 + \mathcal{E}_4$ и вътрешното съпротивление $r_\mathrm{tot}=r_1+r_2+r_3+r_4$ на двата последователно свързани източници на напрежение.

Прекарайте права, която минава най-близко до експерименталните точки.
От ъгловия коефициент може да определите електродвижещото напрежение $\mathcal{E}_\mathrm{tot}$ на батерията, а от пресечната на правата и ординатата може да определите вътрешното съпротивление на батерията $r_\mathrm{tot}$.

\item Изведете формула~\eqref{BG:eq:R_j}.

\subsection{Електростатично определяне на $\varepsilon_0$}

\item Качествен електричен експеримент

С помощта на кабел с крокодили свържете последователно двата източника на напрежение, без да свързвате към тях волтметъра.
Внимавайте с полярността на батериите. Свържете (+) с (-) на двата източника.
Сумарното напрежение
$U_4=\mathcal{E}_1+\mathcal{E}_2+\mathcal{E}_3+\mathcal{E}_4$ се пресмята като сума от определените по-рано в секция ~\ref{BG:measure_varE} електродвижещи напрежения.
Пак използвайки свързващите кабели с крокодили, подайте това напрежение върху кръглите пластинки през излизащите от тях проводници.
Нека пъвоначално пластинките са далеч една от друга (по далече от 1~cm), но са успоредни и разположени на една ос, минаваща през центъра им.
Вижте още и схемата показана на Фигура~\ref{BG:setup_e}.
Постепенно придвижвайте трупчето, докато висящата пластина внезапно се залепи за неподвижната пластинка, закрепената към трупчето.
Внимателно повторете експеримента по-бавно така, че да изчаквате трептенията да затихнат. 
Придвижвайте трупчето така, че винаги пластинките да са успоредни.
След като махалото загуби равновесие и пластинките се залепят, прекъснете връзката им към източника на напрежението.
Използвайте кабел с крокодили и свържете на късо жиците излизащи от двете пластинки.
Тогава пластинките се отделят една от друга и след като трептенията затихнат, измерете разстоянието $x_\mathrm{e}$ между пластинките на този плосък кондензатор.

\item 
Повторете внимателно този експеримент като измервате разстоянието  $x_\mathrm{e}$ между пластинките, с точност 0.5~mm, за различни напрежения:
$U_4=\mathcal{E}_1+\mathcal{E}_2+\mathcal{E}_3+\mathcal{E}_4,$
$U_3=\mathcal{E}_1+\mathcal{E}_2+\mathcal{E}_3,$
$U_2=\mathcal{E}_1+\mathcal{E}_2,$ и
$U_1=\mathcal{E}_1.$

Запишете резултатите от измерванията в примерната таблица~\ref{BG:tab:xe3_vs_U2}.
С помощта на калкулатор допълнете таблицата с две допълнителни колони
$x_\mathrm{e}^3$ и $U^2$.

\begin{table}[h]
\caption{Таблица с експериментални резултати от електричния експеримент}
\begin{tabular}{| r | r | r | r | r | r |}
\tableline
i& $U \,[\mathrm{V}]$ & $x_\mathrm{e} \,[\mathrm{mm}]$ & $U^2 \,[\mathrm{V}^2]$ & $x_\mathrm{e}^3 \,[\mathrm{mm}^3]$ \\
\tableline
1 &  &  &  & \\
2 &   & &  &  \\
3 &   &  &  &  \\
4 &  & &  &  \\
\tableline
\end{tabular}
\label{BG:tab:xe3_vs_U2}
\end{table}

\item Представете графично върху милиметрова хартия зависимостта $x_\mathrm{e}^3$ от $U^2$, и определете наклона $k_\mathrm{e}$ на правата, минаваща най-близо до експерименталните точки.

\begin{figure}[h]
\includegraphics[width=7cm]{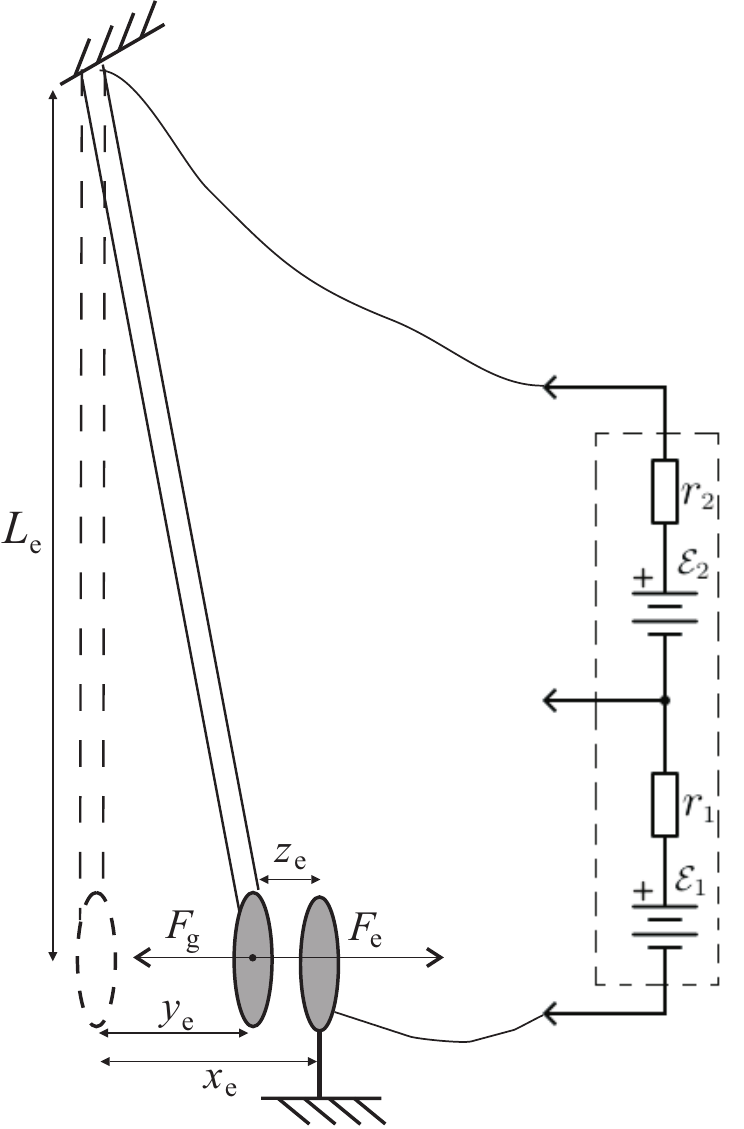}
\caption{
Електрично махало за измерване на диелектричната проницаемост  $\varepsilon_0$ на вакуума.
Дължината на махалото е $L_\mathrm{e}$. 
Отместването от равновесното положение на махалото
малко преди да се залепи към привличащата го неподвижна пластинка е $y_\mathrm{e}=x_\mathrm{e}-z_\mathrm{e}.$ Тук 
$z_\mathrm{e}$ е разстоянието между пластинките малко преди 
махалото да се устреми към трупчето. А $x_\mathrm{e}$ е разстоянието между равновесното положение на махалото и неподвижната пластинка.
Eлектричното поле между пластинките се създава от последователно свързани батерии с електродвижещи напрежения 
$\mathcal{E}_1$ и $\mathcal{E}_2$. 
Големите съпротивления, които ограничават тока са означени с 
$r_1$ и $r_2$. 
Съпротивленията, както и батериите, са скрити в пластмасова тръба,
означена символично с прекъсната линия.
В равновесие електричната сила на привличане между пластинките $F_\mathrm{e}$ се уравновесява с гравитационната сила $F_\mathrm{g}$.
}
\label{BG:setup_e}
\end{figure}

\item Измерете масата $m_\mathrm{e}$ на висящата пластинка.

За целта използвайте електронната везна, която се намира при квестора.
За измерването не откачайте висящата пластинката от стойката. 
Поставете везната върху високата подложка, която върви заедно с нея. 
Направете измерването докато пластинката е закачена за тънката медна нишка, като внимавате нишката да не оказва допълнителни сили върху пластинката по време на измерването.

\item Измерете диаметъра $D_\mathrm{e}$ на двете пластинки.

\item Измерете разстоянието от центъра на пластинката до летвата на окачване $L_\mathrm{e}$.

\item 
Определете диелектричната проницаемост на вакуума $\varepsilon_0$ чрез приближена зависимост между $x_\mathrm{e}$ и $U$, 
която е изведена при предположение, че $x_\mathrm{e}\ll D_\mathrm{e}$
\begin{equation}
x_\mathrm{e}^3 =k_\mathrm{e}  U^2,
\end{equation}
където $k_\mathrm{e}$ е линейният коефициент
\begin{equation}
\label{BG:eq:Epsilon_0}
k_\mathrm{e} 
=\frac{27}{32} \pi  \varepsilon_0 \frac{L_\mathrm{e} D_\mathrm{e}^2}{m_\mathrm{e} g}.
\end{equation}

Използвайки тази формула и експериментално измерената стойност на 
$k_\mathrm{e} $, определете диелектричната проницаемост на вакуума
\begin{equation}
\label{BG:eq:k_e}
\varepsilon_0=
\frac{32}{27\pi}\frac{m_\mathrm{e} g}{L_\mathrm{e} D_\mathrm{e}^2}k_\mathrm{e}. 
\end{equation}

\item 
Към таблицата добавете още един стълб $x_\mathrm{e}^3 (1-\frac{4}{3 \pi} \frac{x_\mathrm{e}}{D_\mathrm{e}})$, като пресметнете малката поправка
$\frac{4}{3 \pi} \frac{x_\mathrm{e}}{D_\mathrm{e}}$.
Повторете описания по-горе метод за едно по-точно определяне на диелектричната проницаемост на вакуума чрез формулата 
\begin{equation}
\label{BG:Effect_of_ends}
x_\mathrm{e}^3 \left(1-\frac{4}{3 \pi} \frac{x_\mathrm{e}}{D_\mathrm{e}} \right) =
\frac{27}{32} \pi  \varepsilon_0 \frac{L_\mathrm{e} D_\mathrm{e}^2}{m_\mathrm{e} g}  U^2.
\end{equation}
Тази формула позволява достигане на процентна точност при използването на дадената постановка.

Каква е разликата в определянето на $\varepsilon_0$ по двата метода?

\subsection{Магнитостатично определяне на $\mu_0$}

\item Качествен магнитен експеримент

Поставете двата пръстена успоредно на разстояние 
$x_\mathrm{m}= \mathrm{20\;mm}$, отчитано между средата на навивките.
Използвайте кабели с крокодили и сглобете схемата показана на Фигура~\ref{BG:setup_m}.
Нека проследим пътя на тока, който тръгва от (+) на батерията.
Първо преминава през съпротивителната канталова жица, опъната върху летвата.
Такова променливо съпротивление се нарича още реохорд.
Веригата се съединява от алуминиевия плъзгач защипан с крокодил, който се държи с ръка.
После токът преминава през амперметъра, висящия токов пръстен,
неподвижния токов пръстен със същия брой навивки $N=50$
и през (-) на батерията се завръща в захранващия източник.

Когато допрете плъзгача до съпротивителната жица
намотките потрепват. 
Ако бобините се отблъскват, сменете полярността на свързването на една от тях.
Токовете трябва да са успоредни и при включване на тока бобините да се привличат.
Движете плъзгача по жицата, като гледате едновременно амперметъра и висящата навивка.
Започнете с малки токове и като движите плъзгача запомнете положението, при което махалото губи устойчивост и висящият пръстен се устремява към неподвижния.
Повторете експеримента бавно като изчаквате трептенията да затихнат.
Запишете най-малкия (критичния) ток $I$, при който спокойно висящият пръстен се залепва внезапно за неподвижния.

\begin{figure}[h]
\includegraphics[width=7cm]{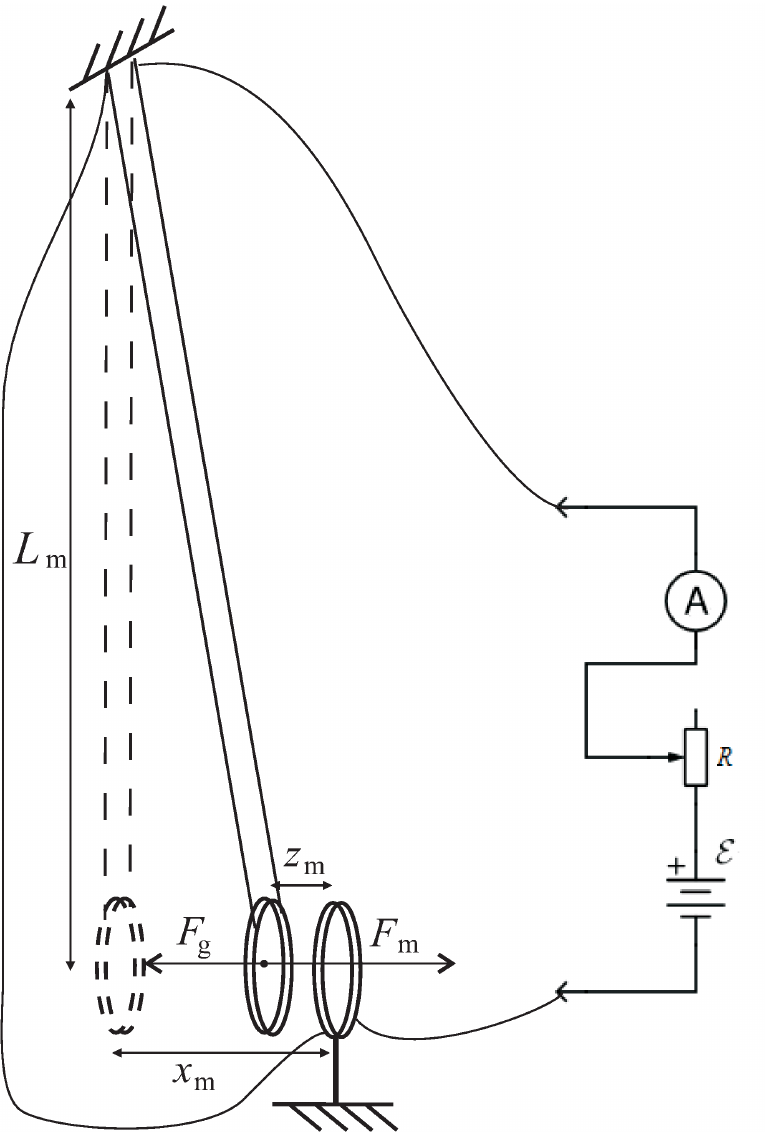}
\caption{
Магнитно махало за измерване на $\mu_0$, магнитната проницаемост на вакуума.
Дължината на махалото е $L_\mathrm{m}$. 
Разстоянието от неподвижната намотка до равновесното положение на махалото е $x_\mathrm{m}$.
А разстоянието между токовите пръстени в момента преди висящият да се устреми към неподвижния е $z_\mathrm{m}$.
Магнитното  поле между намотките се създава от тока на 4 или 8 последователно свързани батерии от 1.5~V, поставени в държачи.
Токът се измерва от амперметъра 
и се регулира с променливото съпротивление $R$
с помощта на плъзгача, плъзгащ се по
канталовата жица, опъната върху летвата.
В равновесие магнитната сила на привличане между пръстените с успоредни токове $F_\mathrm{m}$ се уравновесява с гравитационната сила $F_\mathrm{g}$.
}
\label{BG:setup_m}
\end{figure}

\item Повторете експеримента и измерете критичния ток $I$ за различни стойности на $x_\mathrm{m}$, например 25, 20, 15, 10, 5~mm. 

Резултатите за разстоянието $x_\mathrm{m}$ и тока $I$ представете таблично в първите две колони на таблица~\ref{BG:tab:mu0}.
С помощта на калкулатор пресметнете допълнителни колони за 
$x_\mathrm{m}^2$ и $I^2$. 
С други думи, представете таблично зависимостта  $x_\mathrm{m}^2$ от $I^2$. 

За малките разстояния ползвайте само един държач с четири батерии.

\item Представете графично зависимостта $x_\mathrm{m}^2$ от $I^2$ и определете наклонa $k_\mathrm{m}$ на правата, минаваща най-близо до експерименталните точки.
Това е методика за експериментално определяне на наклона $k_\mathrm{m}$.

\item Измерете дължината на махалото, т.е. разстоянието $L_\mathrm{m}$ от летвата до центъра на висящата бобинка.

\item Измерете масата $m_\mathrm{m}$ на висящата бобинка.

\item Измерете диаметъра $D_\mathrm{m}$ на двете бобинки.

\item Определете магнитната проницаемост на вакуума $\mu_0$,
като използвате приближената формула
\begin{equation}
x_\mathrm{m}^2 =k_\mathrm{m}  I^2,
\end{equation}
където $k_\mathrm{m}$ е линейният коефициент
\begin{equation}
\label{BG:eq:k_m}
k_\mathrm{m} =2 \mu_0 \frac{L_\mathrm{m} N^2 D_\mathrm{m}}{m g}.
\end{equation}
Тази формула е приложима, когато $x_\mathrm{m}\ll D_\mathrm{м}$
и може да приемем, че проводниците са прави.

Тази формула дава възможност за изразяването на 
магнитната проницаемост $\mu_0$ на вакуума
чрез експериментално определения наклон
\begin{equation}
\label{BG:eq:k_m}
\mu_0=\frac{m gk_\mathrm{m}}{2 L_\mathrm{m} N^2 D_\mathrm{m}}.
\end{equation}

\item Силата на привличане между два токови пръстена е известна магнетостатична задача, която има много приложения.
Точната формула за определянето на $\mu_0$ чрез нашата постановка се дава с формулата 
\begin{equation}
\label{BG:Correction}
x_\mathrm{m}^2 \left[1+f \left(\frac{x_\mathrm{m}}{D_\mathrm{m}}\right) \right] =k_\mathrm{m}  I^2,
\end{equation}
където функцията $f(x_\mathrm{m}/D_\mathrm{m})$ 
е представена таблично по-долу и графично на Фигура~\ref{BG:f_delta}.

При точността, с която работим, корекционната поправка може да се пресметне и чрез приближената формула
\begin{equation}
\label{BG:function_2_rings}
f \left(\frac{x_\mathrm{m}}{D_\mathrm{m}}\right)\approx
\frac{1}{16}\left(-5+6 \log{ \frac{8 D_\mathrm{m}}{x_\mathrm{m}} } \right)
 \frac{x_\mathrm{m}^2}{D_\mathrm{m}^2}.
\end{equation}

Пример за подреждането на експерименталните данни и корекционния множител е даден в Таблица~\ref{BG:tab:mu0}.
Попълнете Вашата таблица.

\begin{table}[h]
\caption{Примерна таблица за подреждане на експериментални резултати и корекционната функция за експеримента с токовите пръстени. Между $I^2$ и $x_\mathrm{m}^2(1+f)$ има линейна зависимост и коефициента на пропорционалност определя $\mu_0.$}
\begin{tabular}{| r | r | r | r | r | r | r | r| }
\tableline
$i$& $x_\mathrm{m}[\mathrm{mm}]$& $I [\mathrm{mA}]$  & $x_\mathrm{m}^2 [\mathrm{mm^2}]$ & $I^2 [\mathrm{mA^2}]$ &  $x_\mathrm{m}/D_\mathrm{m}$ & $1+f(x_\mathrm{m}/D_\mathrm{m})$ & $x_\mathrm{m}^2(1+f) [\mathrm{mm^2}] $ 
\\
\tableline
1 &  5 & && & & & \\
2 & 10& && & & & \\
3 & 15& && & & & \\
4 & 20& & && & & \\
5 & 25& && & & & \\
\tableline
\end{tabular}
\label{BG:tab:mu0}
\end{table}

\begin{center}
\begin{tabular}{| c | c || c | c || c | c || c |  c || c | c |}
		\hline
			\multicolumn{1}{| c |}{ } & \multicolumn{1}{ c ||}{ } & \multicolumn{1}{ c |}{ } & \multicolumn{1}{ c ||}{ } & \multicolumn{1}{ c |}{ } & \multicolumn{1}{ c ||}{ } 
			& \multicolumn{1}{ c |}{ } & \multicolumn{1}{ c ||}{ } & \multicolumn{1}{ c |}{ } & \multicolumn{1}{ c |}{ } \\
			\boldmath$x/D$ & \boldmath$f(x/D)$ & \boldmath$x/D$ &\boldmath$f(x/D)$ & \boldmath$x/D$ & \boldmath$f(x/D)$ & \boldmath$x/D$ & \boldmath$f(x/D)$  & \boldmath$x/D$ & \boldmath$f(x/D)$ \\[15pt] \hline 
			0.005	&	0.0001	&	0.105	&	0.0148	&	0.205	&	0.0480	&	0.305	&	0.0976	&	0.405	&	0.1644	\\	\hline
0.010	&	0.0002	&	0.110	&	0.0162	&	0.210	&	0.0502	&	0.310	&	0.1005	&	0.410	&	0.1682	\\	\hline
0.015	&	0.0005	&	0.115	&	0.0175	&	0.215	&	0.0522	&	0.315	&	0.1034	&	0.415	&	0.1721	\\	\hline
0.020	&	0.0008	&	0.120	&	0.0188	&	0.220	&	0.0545	&	0.320	&	0.1066	&	0.420	&	0.1760	\\	\hline
0.025	&	0.0012	&	0.125	&	0.0202	&	0.225	&	0.0566	&	0.325	&	0.1096	&	0.425	&	0.1800	\\	\hline
0.030	&	0.0016	&	0.130	&	0.0216	&	0.230	&	0.0589	&	0.330	&	0.1126	&	0.430	&	0.1840	\\	\hline
0.035	&	0.0021	&	0.135	&	0.0231	&	0.235	&	0.0613	&	0.335	&	0.1156	&	0.435	&	0.1880	\\	\hline
0.040	&	0.0027	&	0.140	&	0.0245	&	0.240	&	0.0634	&	0.340	&	0.1190	&	0.440	&	0.1920	\\	\hline
0.045	&	0.0034	&	0.145	&	0.0262	&	0.245	&	0.0659	&	0.345	&	0.1221	&	0.445	&	0.1965	\\	\hline
0.050	&	0.0040	&	0.150	&	0.0278	&	0.250	&	0.0684	&	0.350	&	0.1252	&	0.450	&	0.2006	\\	\hline
0.055	&	0.0048	&	0.155	&	0.0294	&	0.255	&	0.0709	&	0.355	&	0.1287	&	0.455	&	0.2047	\\	\hline
0.060	&	0.0056	&	0.160	&	0.0312	&	0.260	&	0.0732	&	0.360	&	0.1319	&	0.460	&	0.2092	\\	\hline
0.065	&	0.0064	&	0.165	&	0.0328	&	0.265	&	0.0757	&	0.365	&	0.1355	&	0.465	&	0.2134	\\	\hline
0.070	&	0.0073	&	0.170	&	0.0345	&	0.270	&	0.0783	&	0.370	&	0.1390	&	0.470	&	0.2181	\\	\hline
0.075	&	0.0083	&	0.175	&	0.0365	&	0.275	&	0.0810	&	0.375	&	0.1423	&	0.475	&	0.2223	\\	\hline
0.080	&	0.0092	&	0.180	&	0.0382	&	0.280	&	0.0837	&	0.380	&	0.1460	&	0.480	&	0.2270	\\	\hline
0.085	&	0.0103	&	0.185	&	0.0401	&	0.285	&	0.0864	&	0.385	&	0.1497	&	0.485	&	0.2317	\\	\hline
0.090	&	0.0114	&	0.190	&	0.0421	&	0.290	&	0.0891	&	0.390	&	0.1530	&	0.490	&	0.2361	\\	\hline
0.095	&	0.0125	&	0.195	&	0.0440	&	0.295	&	0.0919	&	0.395	&	0.1568	&	0.495	&	0.2408	\\	\hline
0.100	&	0.0136	&	0.200	&	0.0461	&	0.300	&	0.0947	&	0.400	&	0.1606	&	0.500	&	0.2456	\\	\hline

		\hline
\end{tabular}
\end{center}

\begin{figure}[h]
\includegraphics[width=18cm]{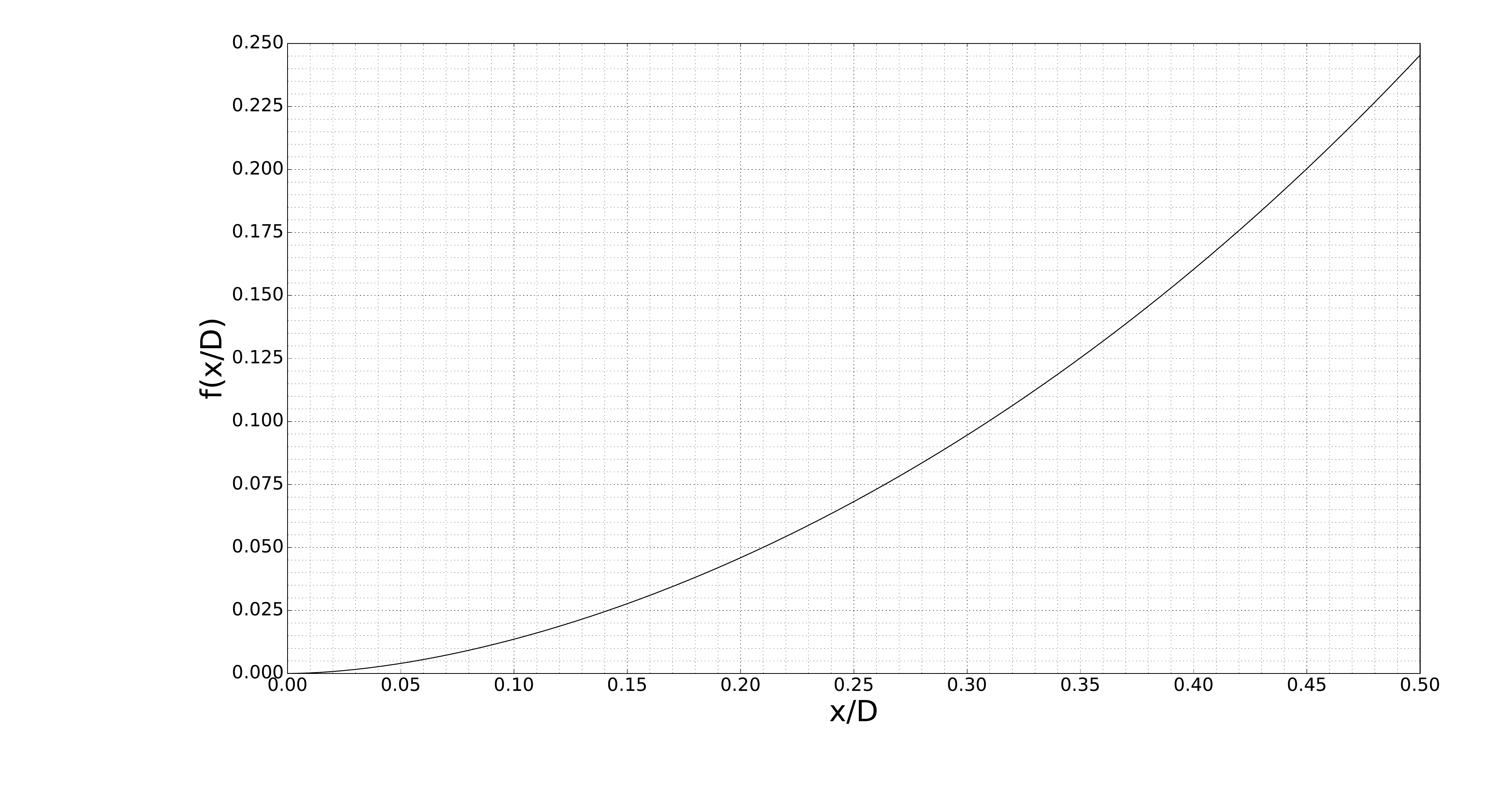}
\caption{Корекционната поправка $f(x_\mathrm{m}/D_\mathrm{m})$ от уравнение (\ref{BG:Correction})
като функция от безразмерното отношение на отместването от равновесното положение $x_\mathrm{m}$ и диаметъра на навивките $D_\mathrm{m}$.
}
\label{BG:f_delta}
\end{figure}

\item 
Използвайки точната формула \eqref{BG:Correction}
и графично представената корекционна функция от Фигура~\ref{BG:f_delta}
табулирана отгоре,
представете експерименталните данни графично:
по ордината 
$(1+f)x_\mathrm{m}^2$, 
по абсцисса $I^2$
и от наклона определете магнитната проницаемост $\mu_0$ на вакуума.
С колко процента се различават резултатите обработени чрез приближената и точната формула?
\end{enumerate}

\section{Определяне на скоростта на светлината}

Използвайте определените в предишните подусловия $\varepsilon_0$
и $\mu_0$ и пресметнете скоростта на светлината
\begin{equation}
c=\frac{1}{\sqrt{\varepsilon_0 \mu_0}}.
\label{cspeed}
\end{equation}
Не се смущавайте, ако резултатът се различава от известната Ви стойност, това е Вашето първо измерване на фундаментална константа.

\section{Теоретични задачи}

Използвайте формулите за силата на привличане между пластинките на безкраен плосък кондензатор и изведете формула (\ref{BG:eq:Epsilon_0}), която използвахте за определянето на $\varepsilon_0$.

Използвайте формулите за силите на привличане между безкрайни успоредни токове и изведете формула (\ref{BG:eq:Epsilon_0}), която използвахте за приближено определяне на $\mu_0.$

\section{Задача за домашна работа. Премия на Зоммерфелд 137 лева}

Потърсете в учебници по електродинамика, енциклопедии или в Интернет формулата за капацитет на кондензатор, при която се отчитат ефектите на края и изведете по-точната формула (\ref{BG:Effect_of_ends}) за определянето
на $\varepsilon_0$ с нашата постановка.

Аналогично потърсете в учебници формулите за взаимна индуктивност и сили на взаимодействие между токови пръстени и изведете точната формула (\ref{BG:Correction}).
Корекционната функция 
$f(x_\mathrm{m}/D_\mathrm{m})$
може да бъде пресметната числено или да се получи само първият член на
уравнение~(\ref{BG:function_2_rings}) на реда по степените на  $x_\mathrm{m}/D_\mathrm{m}$.

Решението пратете на epo@bgphysics.eu от адреса, на който сте се регистрирали до 07:00 на 24 април 2016 г.

Може да работите в колектив и да се консултирате с професори по теоретична физика, електродинамика или електротехника.
Премията се получава лично от участника и само в деня на обявяването на резултатите.


\newpage

\section{Решение на експерименталната задача}

\subsection{Електростатично определяне на $\varepsilon_0$}
\begin{enumerate}

\item Периодът $T$ на люлеене на висящата пластинка е $T= 1.5$ s.

\item Кръговата честота $\omega$ на люлеене на висящата пластинка е $\omega= 4.2$ rad/s.

\item Разстоянието $L_\mathrm{m}$ от точката на окачване до центъра на висящата пластинка е $L_\mathrm{m}=55.1$ cm.

\item Земното ускорение е $g=9.7$ m/s$^2$.

\item Отклонението от известната стойност $g=9.81$ m/s$^2$ е 1.1\%.


\section{Измерване на електродвижещите напрежения $\mathcal{E}_1$, $\mathcal{E}_2$, $\mathcal{E}_3$ и $\mathcal{E}_4$ на източниците за електричния експеримент с помощта само на амперметър и волтметър.}

\item Измерване на електродвижещите напрежения $\mathcal{E}_1$, $\mathcal{E}_2$, $\mathcal{E}_3$ и $\mathcal{E}_4$ на източника на напрежение.

Резултатите от измерването са дадени в Таблица~\ref{BG:tab:I123456}.
\begin{table}[h]
\caption{Таблица с експериментални резултати за токовете и напрежението измерени по схемите от Фигура~\ref{BG:Measure-battery} за различните дялове на източника на напрежение.}
\begin{tabular}{| r | r | r | r | r | r | r | }
\tableline
$i$& $I_\mathrm{A} \, [\mu \mathrm{A}]$ & $I_\mathrm{V} \,[\mu\mathrm{A}]$ & $U \,[\mathrm{V}]$ & $\mathcal{E} \,[\mathrm{V}]$ & $r \, [\Omega$] & $R_\mathrm{V} \, [M \Omega$]\\
\tableline
1 & 208 & 66.8 & 67.0 & 98.7 & 474 & 0.99 \\
2 & 209 & 67.1 & 66.9 & 98.5 & 471 & 1.00 \\
3 & 208 & 66.7 & 67.0 & 98.6 & 474 & 0.99 \\
4 & 208 & 66.9 & 66.8 & 98.5 & 473 & 1.00 \\
\tableline
\end{tabular}
\label{BG:tab:I123456}
\end{table}

\item Вътрешните съпротивления на различните дялове н източника на напрежение са изчислени в таблица~\ref{BG:tab:I123456}.

\item Вътрешните съпротивления на волтметъра е изчислено за различните измервания и е нанесено в таблица~\ref{BG:tab:I123456}.

\item Чрез заместване се проверява, че формулите ~\eqref{BG:eq:E}, \eqref{BG:eq:R}  и~\eqref{BG:eq:R} удовлетворяват уравненията за съответните електрични вериги от фигура~\ref{BG:Measure-battery}
\begin{equation}
 \mathcal{E}=r I_\mathrm{A}  =  (r+R_\mathrm{V}) I_\mathrm{V}
\end{equation}

\subsection{Измерване на електродвижещите напрежения $\mathcal{E}_1$, $\mathcal{E}_2$, $\mathcal{E}_3$ и $\mathcal{E}_4$ на източниците за електричния експеримент с помощта на амперметър и резистори}

\begin{figure}[ht]
\includegraphics[width=12cm]{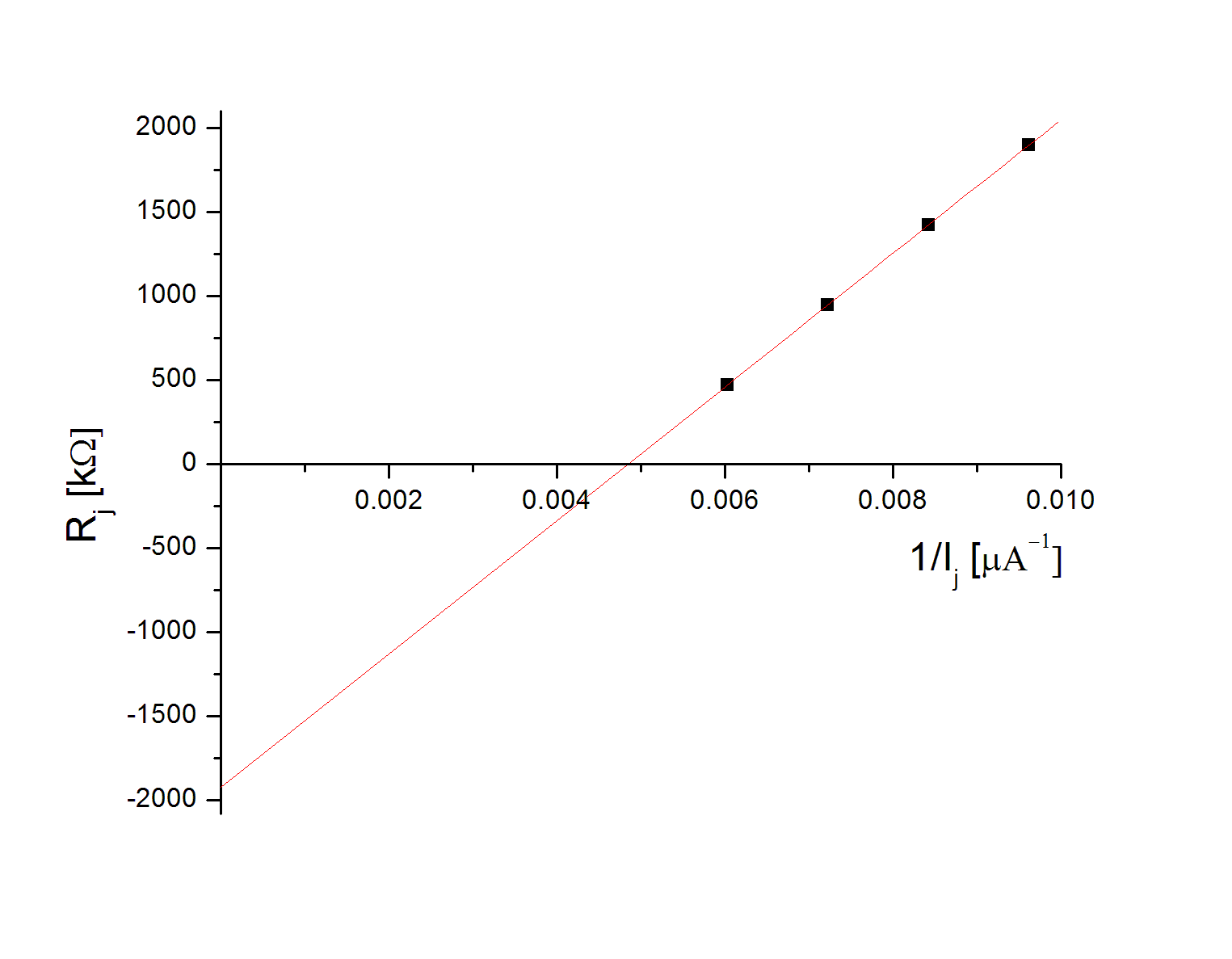}
\caption{Графично представяне на резултатите за измерването на електродвижещото напрежение $\mathcal{E}_\mathrm{tot}$ на батерията.}
\label{BG:fig:battery_measurement2}
\end{figure}

\item  виж таблица ~\ref{tab:E1234_2exp2}

\begin{table}[h]
\caption{Tаблица за експериментални резултати за съпротивленията $R^*_j$}
\begin{tabular}{| r | r | r | r | }
\tableline
j& $R^*_j \, [\mathrm{k}\Omega]$ \\
\tableline
1 & 465 \\
2 & 467 \\
3 & 465 \\
4 & 465 \\
\tableline
\end{tabular}
\label{tab:E1234_2exp2}
\end{table}

\item виж таблица ~\ref{tab:E1234_2exp}

\begin{table}[h]
\caption{Таблица за експериментални резултати за токовете и напрежението, измерени по схемите от Фигура~\ref{BG:Measure-battery2} за различните дялове на източниците на напрежение}
\begin{tabular}{| r | r | r | r | }
\tableline
j& $R_j \, [\mathrm{k}\Omega]$ & $I_j \,[\mu\mathrm{A}]$ &  $1/I_j \,[\mu\mathrm{A}^{-1}]$ \\
\tableline
1 & 465 & 167.5 & 0.00597\\
2 & 932 & 139.6 & 0.00716\\
3 & 1397 & 119.7 & 0.00835\\
4 & 1862 & 104.8 & 0.00954\\
\tableline
\end{tabular}
\label{tab:E1234_2exp}
\end{table}

\item без решение (виж фигура~\ref{BG:fig:battery_measurement2})

\item без решение (виж фигура~\ref{BG:fig:battery_measurement2})

\item без решение (виж фигура~\ref{BG:fig:battery_measurement2})

Резултатите от измерването са представени графично на фигура~\ref{BG:fig:battery_measurement2}.

\subsection{Електростатично определяне на $\varepsilon_0$}

\item Качествен експеримент

\item Измерванията на критичното разстояние $x_\mathrm{e}$ при различни напрежения $U$ между пластинките са дадена в Таблица~\ref{BG:tab:xc3_vs_U2}.

Таблично зависимостта на $x_\mathrm{e}^3$ от $U^2$ е изчислена и дадена в Таблица~\ref{BG:tab:xc3_vs_U2}

\begin{table}[h]
\caption{Таблица с експериментални резултати от електричния експеримент}
\begin{tabular}{| r | r | r | r | r | r | r |  }
\tableline
i& $U_\mathrm{e} \,[\mathrm{V}]$ & $x_\mathrm{e} \,[\mathrm{mm}]$ & $U_\mathrm{e}^2 \,[\mathrm{V}^2]$ & $x_\mathrm{e}^3 \,[\mathrm{mm}^3]$ & $x_\mathrm{e}^3 (1-\frac{4}{3 \pi} \frac{x_\mathrm{e}}{D_\mathrm{e}}) \,[\mathrm{mm}^3]$  \\
\tableline
1 & 98.7 & 3.0 & 9.74$\times10^3$ & 27.0 & 26.4 \\
2 & 197.2  & 5.0 & 38.9$\times10^3$ & 120 &  120 \\
3 & 295.9 & 7.0 & 87.5$\times10^3$ & 260 &  261 \\
4 & 394.3 & 8.5 & 155.5$\times10^3$ & 573 & 573 \\
5 & 493.0 & 9.5 & 243.1$\times10^3$ & 857 & 793 \\
6 & 591.6 & 11.0 & 349.9$\times10^3$ & 1331 & 1216 \\
7 & 690.2 & 12.0 & 476.3$\times10^3$ & 1728 & 1565 \\
8 & 788.7 & 13.5 & 621.9$\times10^3$ & 2460 & 2200 \\
\tableline
\end{tabular}
\label{BG:tab:xc3_vs_U2}
\end{table}

\item Зависимостта на $x_\mathrm{e}^3$ от $U^2$  е представена графично на Фигура~\ref{BG:fig:electric_experiment}.
Наклонът $k_\mathrm{e}$ на правата, която минава най-близо до експерименталните точки е $k_\mathrm{e}=0.00386\; \mathrm{mm}^3/\mathrm{V}^2$. 

\begin{figure}[h]
\includegraphics[width=12cm]{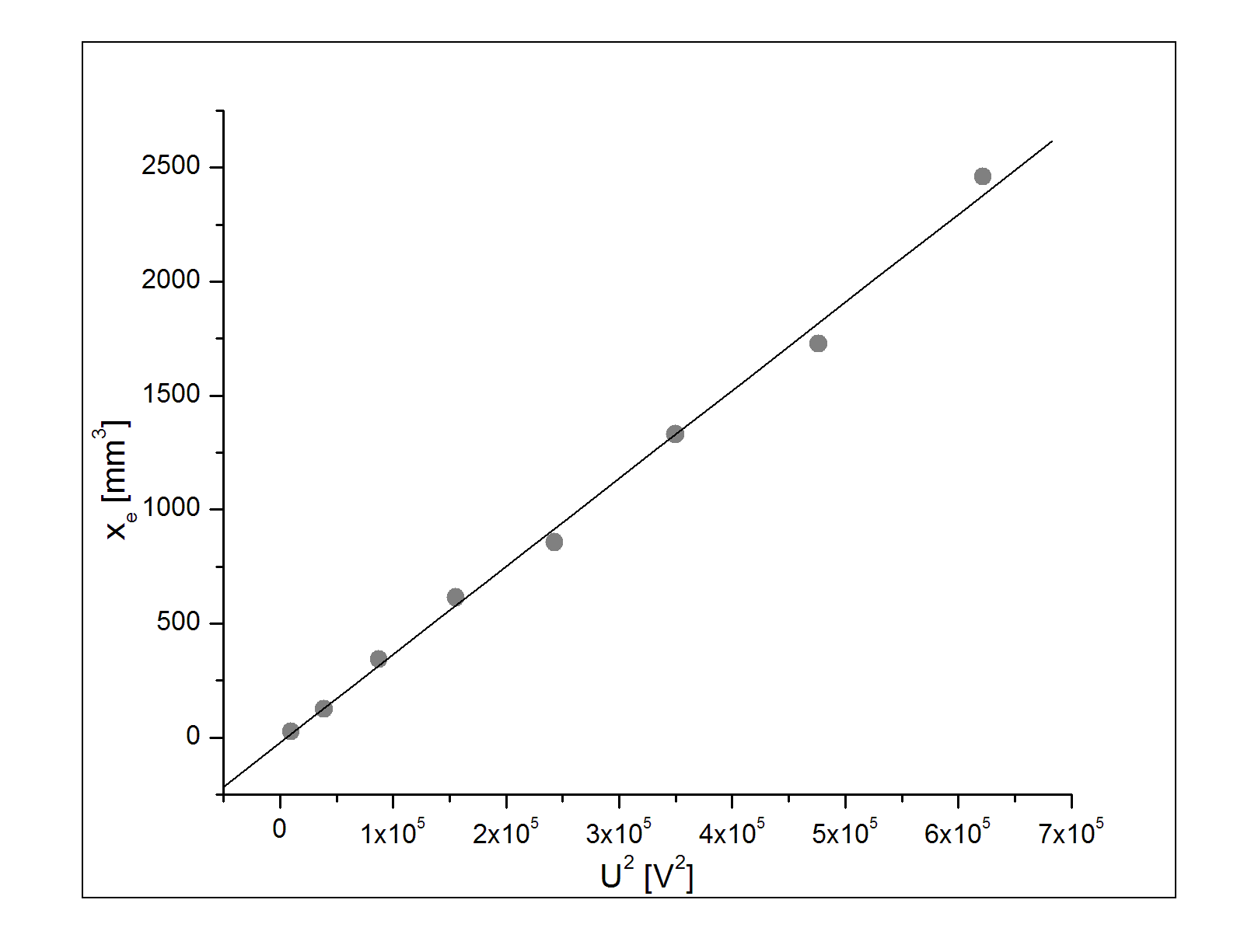}
\caption{Графично представяне на резултатите от електричния експеримент.}
\label{BG:fig:electric_experiment}
\end{figure}

\item Масата на висящата пластинки е $m_\mathrm{e}=1.14$ g.

\item Диаметърът на двете пластинки е $D_\mathrm{e}=5.4$ cm.

\item Дължината на махалото е $L_\mathrm{e}=57.4$ cm.

\item Диелектричната проницаемост на вакуума е $\varepsilon_0=9.83\times10^{-12}\; \mathrm{F/m}$.

\item Таблично зависимостта на $x_\mathrm{e}^3 (1-\frac{4}{3 \pi} \frac{x_\mathrm{e}}{D_\mathrm{e}})$ от $U^2$ е дадена в Таблица~\ref{BG:fig:electric_experiment}

Зависимостта на  $x_\mathrm{e}^3 (1-\frac{4}{3 \pi} \frac{x_\mathrm{e}}{D_\mathrm{e}})$ от $U^2$  е представена графично на Фигура~\ref{BG:fig:electric_experiment_Correction}.
Наклонът $k_\mathrm{e}$ на правата, която минава най-близо до експерименталните точки е $k_\mathrm{e}=0.00345\; \mathrm{mm}^3/\mathrm{V}^2$. 

\begin{figure}[h]
\includegraphics[width=12cm]{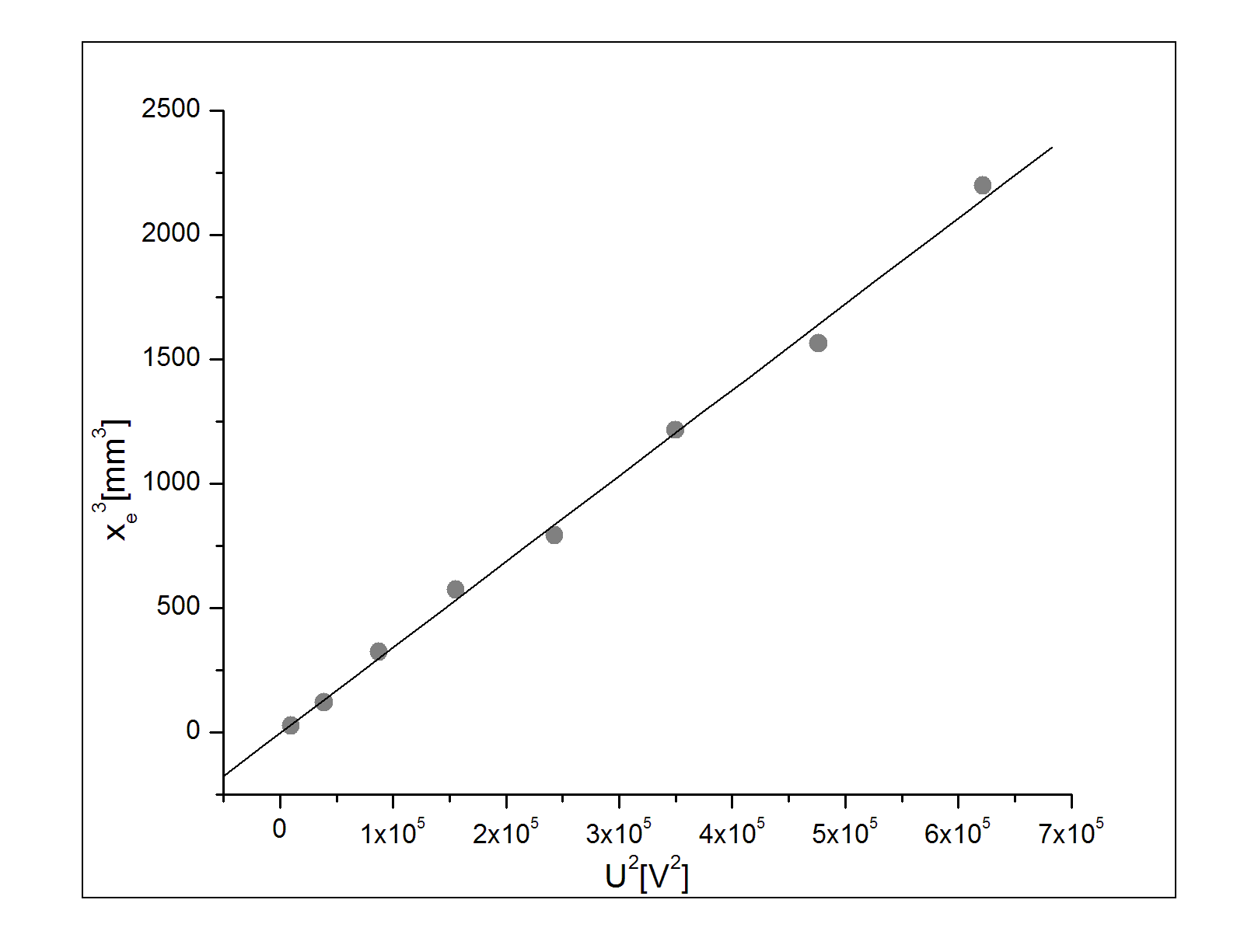}
\caption{Графично представяне на резултатите от електричния експеримент отчитайки краевите ефекти.}
\label{BG:fig:electric_experiment_Correction}
\end{figure}

Диелектричната проницаемост на вакуума е $\varepsilon_0=8.82\times10^{-12}\; \mathrm{F/m}$.

Разликата от двата метода е 11\% и благодарение на отчитането на поправката получаваме 0.3\% отклонение от истинската стойност на $\varepsilon_0$, която е $8.85\times 10^{-12}$~F/m.

\subsection{Магнитно определяне на $\mu_0$}

\item Качествен експеримент

\item Критичното разстояние $x_\mathrm{m}$ при различни токове $I$ през двете бобинки са дадени в Таблица~\ref{BG:m_measurements}.

\begin{table}[ht]
\caption{Експериментални резултати и корекционната функция за експеримента с токовите пръстени}
\begin{tabular}{| r | r | r | r | r | r | r | r |}
\tableline
$i$& $x_\mathrm{m}[\mathrm{mm}]$& $I [\mathrm{mA}]$  & $x_\mathrm{m}^2 [\mathrm{mm^2}]$ & $I^2 [\mathrm{mA^2}]$ &  $x_\mathrm{m}/D_\mathrm{m}$ & $1+f(x_\mathrm{m}/D_\mathrm{m})$ & $x_\mathrm{m}^2(1+f) [\mathrm{mm^2}] $ 
\\
\tableline
1 & 10 & 72.4 & 100 & 5241.76  & 0.15625 & 1.0296 & 102.96 \\
2 & 12 & 88.4 & 144 & 7814.56 & 0.1875 & 1.0405 & 149.832 \\
3 & 14 & 105.8 & 196 & 11193.64 & 0.21875	 & 1.053 & 206.388 \\
4 & 16 & 120 & 256 & 14400 & 0.25 & 1.0684 & 273.5104 \\
5 & 18 & 137.6 & 324 & 18933.76 & 0.28125 & 1.085 & 351.54 \\
6 & 20 & 153.5 & 400 & 23562.25 & 0.3125 & 1.103 & 441.2 \\
7 & 22 & 169  & 484 & 28561 & 0.34375 & 1.120 & 542.08 \\
8 & 24 & 185 & 576 & 34225 & 0.375 & 1.1423 & 657.9648 \\
\tableline
\end{tabular}
\label{BG:m_measurements}
\end{table}
\item Зависимостта на $x^2_\mathrm{m}$ от $I^2$ е дадена на Фигура~\ref{BG:m_measurements_graph}.
Наклонът на правата минаваща най-близо до експерименталните точки е $k_\mathrm{m}=0.0164\, \mathrm{A}^2/\mathrm{m}^2$.

\item Разстоянието от точката на окачване до центъра на висящата бобинка е $L_\mathrm{m}=55.8$ cm.

\item Масата на висящата бобинка е $m_\mathrm{m}=1.18$ g.

\item Диаметърът на двете бобинки е $D=65$ mm.

\begin{figure}[h]
\includegraphics[width=12cm]{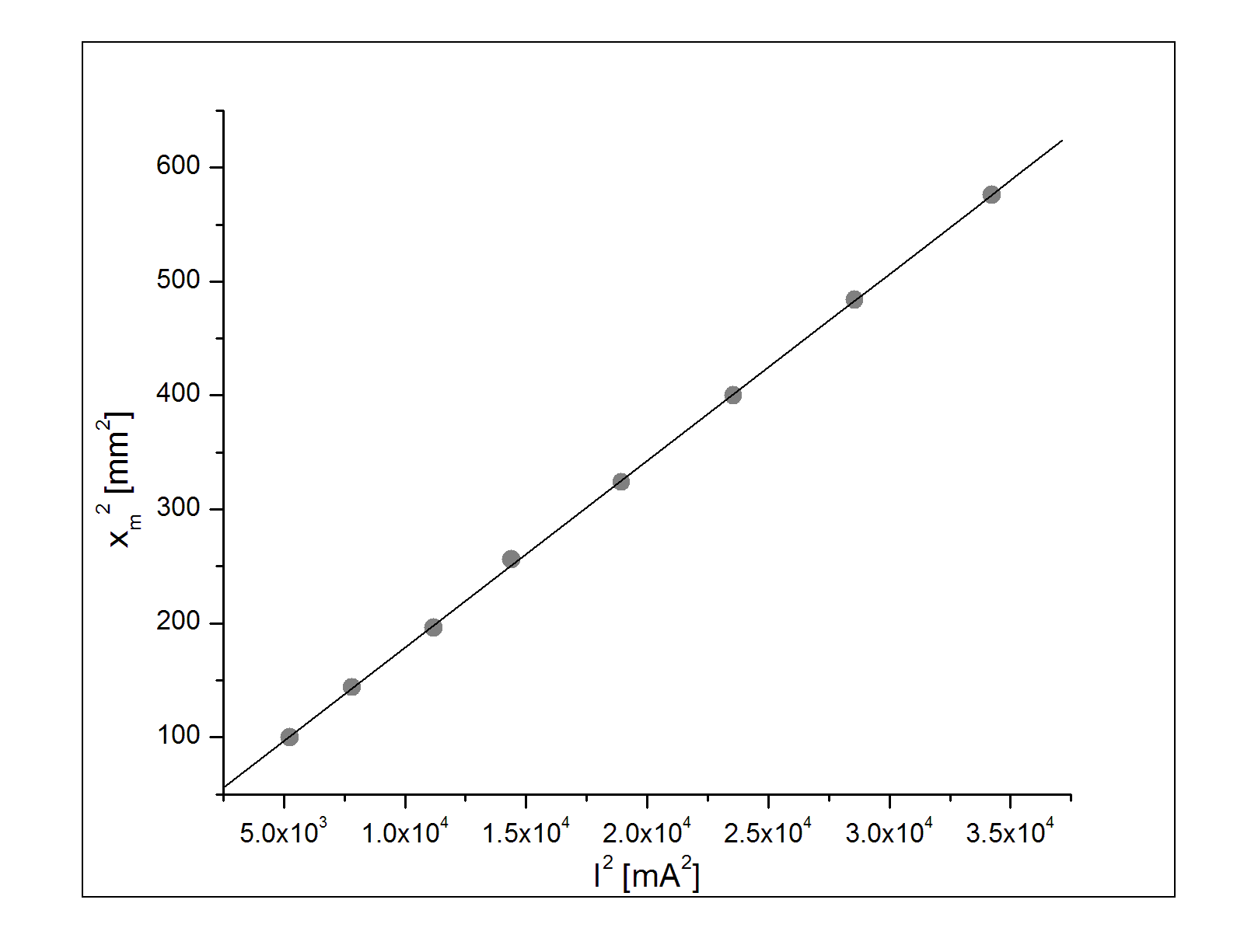}
\caption{Графично представяне на резултатите от магнитния експеримент.}
\label{BG:m_measurements_graph}
\end{figure}

\item Магнитната проницаемост във вакуум е $\mu_0=10.5\times 10^{-7}$ H/m.

\item Корекционният множител е определен от фигура~\ref{BG:f_delta} и са попълнени корекционните множители в съответните колони на таблица~\ref{BG:m_measurements}.

\item Наклонът на правата с корекционната поправка е $k_\mathrm{m}=0.0191\, \mathrm{A}^2/\mathrm{m}^2$ и зависимостта е дадена на Фигура~\ref{BG:m_measurements_graph_Correction}. 
Магнитната проницаемост във вакуум е $\mu_0=12.2\times 10^{-7}$~H/m. 

\begin{figure}[h]
\includegraphics[width=12cm]{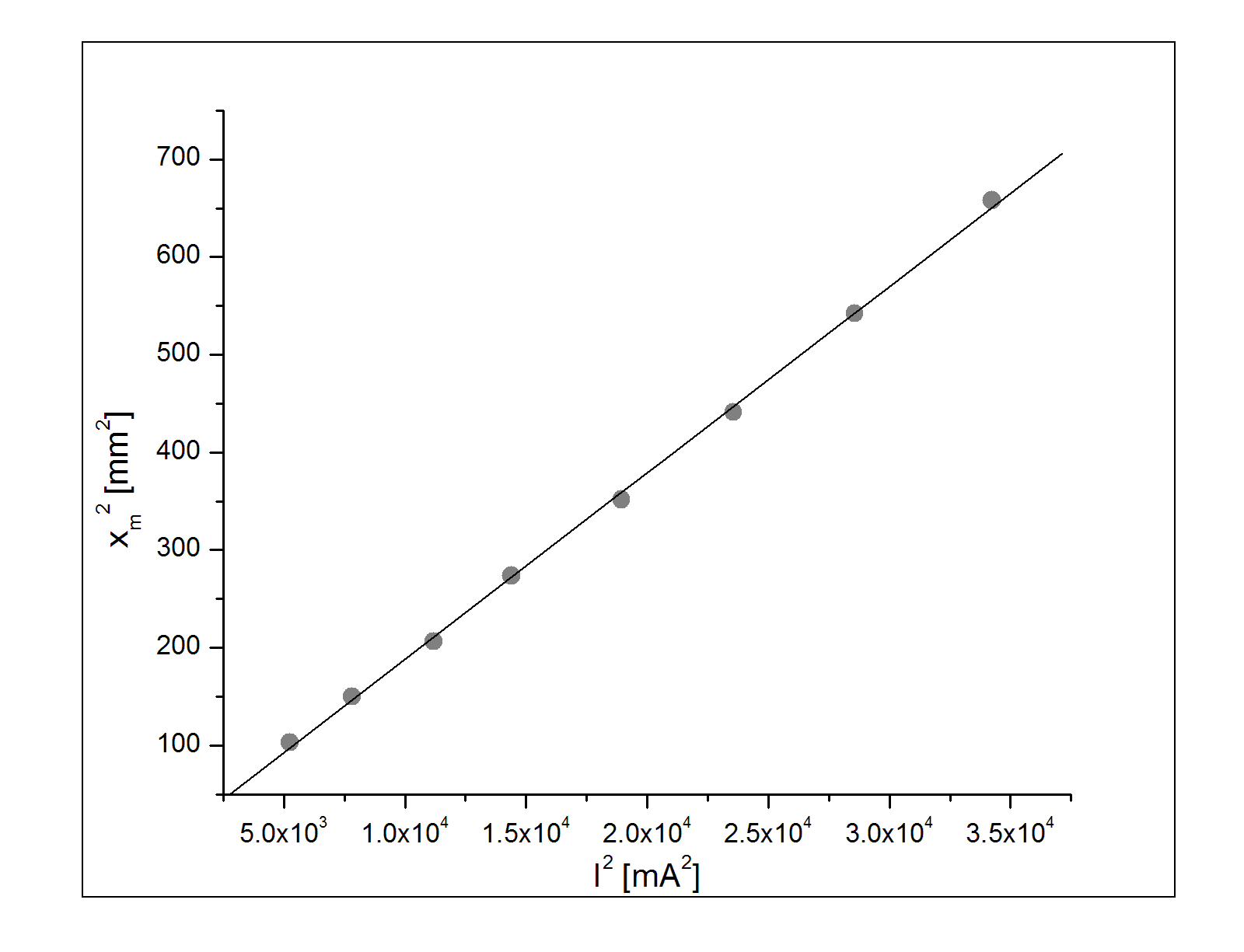}
\caption{Графично представяне на резултатите от магнитния експеримент с корекционна поправка.}
\label{BG:m_measurements_graph_Correction}
\end{figure}

Разликата от двата метода е 14\% и благодарение на корекцията постигаме точност 4\% до истинската стойност на $\mu_0$, която е $1.26\times 10^{-6}$~H/m.

\end{enumerate}

\section{Определяне на скоростта на светлината}

С получените стойности в предните две секции за $\varepsilon_0$ и $\mu_0$, пресмятаме скоростта на светлината по формула~\ref{cspeed} $c=3.048~\times10^{8}$~m/s. Отклонението от точната стойност за скоростта на светлината $c=2.998\times10^{8}$~m/s e 2\%.

\newpage

\section{Решение на теоретичните задачи}


\subsection{Извод на зависимостта на критичното разстояние $x_\mathrm{m}$, при което пластинките се залепват под действието на електричните сили, от напрежението $U$ върху пластинките}

Силата, която действат на висящата пластина е равна на разликата от електричната сила $F_\mathrm{e}$ на привличане на двете пластинки и хоризонталната проекцията на силата на тежестта $F_\mathrm{g}$
\begin{equation}
\label{BG:eq:F}
F=F_\mathrm{e}-F_\mathrm{g}.
\end{equation}

Потенциалната енергия на хоризонталната проекция на силата на тежестта е

\begin{equation}
\label{BG:eq:W_g}
W_\mathrm{g}= m g \frac{y^2}{2 L} = m g \frac{(x-z)^2}{2 L},
\end{equation}

а големината на силата е
\begin{equation}
\label{BG:eq:F_g}
F_\mathrm{g}=-W'(z)= - \frac{\mathrm{d} W_\mathrm{g}}{\mathrm{d} z} = m g \frac{x-z}{L},
\end{equation}

тук използваме че разстоянието $z$ между двете пластинки е равно на разликата от разстоянието $x$ на неподвижната пластинка спрямо равновесното положение на висящата пластинка, когато не е приложено напрежение върху тях,
и отклонението $y$ на висящата пластинка от равновесното и положение.

Потенциалната енергия на висящата пластинка намираща се в електричното поле на неподвижната пластинка е
\begin{equation}
\label{BG:eq:W_e}
W_\mathrm{e}=-\frac{C U^2}{2},
\end{equation}

където $C$ е капацитета на кондензатора образуван от двете кръгли пластинки с радиус $R$
\begin{equation}
C=\varepsilon_0 \frac{\pi R^2}{z},
\end{equation}

следователно
\begin{equation}
\label{BG:eq:W_e}
W_\mathrm{e}=-\varepsilon_0 \frac{\pi R^2 U^2}{2z}.
\end{equation}

Електричната сила, която действа на всяка от пластинките е
\begin{equation}
\label{BG:eq:F_e}
F_\mathrm{e}=-W'(z)=-\frac{\mathrm{d} W_\mathrm{e}}{\mathrm{d} z}=-\frac{\mathrm{d}}{\mathrm{d} z} \Big(-\varepsilon_0 \frac{\pi R^2 U^2}{2 z} \Big)=
- \varepsilon_0 \frac{\pi R^2  U^2}{2 z^2}
\end{equation}

Ако означим с $W$ потенциалната енергия на висящата пластина, то от~\eqref{BG:eq:W_e} и~\eqref{BG:eq:W_g} получаваме
 
\begin{equation}
W = W_\mathrm{e} + W_\mathrm{g} = -  \varepsilon_0 \frac{ \pi R^2 U^2}{2 z}  + m g \frac{(x-z)^2}{2 L}.
\end{equation}

Резултантната сила действаща на висящата пластина се получава като заместим~\eqref{BG:eq:F_e} и~\eqref{BG:eq:F_g} в~\eqref{BG:eq:F} 

\begin{equation}
F=-W'(z)=-\frac{\mathrm{d} W}{\mathrm{d} z}=-\varepsilon_0 \frac{\pi R^2 U^2}{2 z^2} + m g \frac{x - z}{L}
\end{equation}

Когато бавно приближаваме пластинката поставена на статив към висящата пластинка, 
при едно критично разстояние $x_\mathrm{e}$ и $z_\mathrm{e}$ висящата пластинка изведнъж се залепва за статичната пластинка.
Това разстояние може да определим от нулирането на първата и втората производна на потенциалната енергия, т.е. от условието максимума и минимума на потенциалната енергия да съвпадат.

От нулирането на първата производна на потенциалната енергия
\begin{equation}
W'(z)=\frac{\mathrm{d} W}{\mathrm{d} z} = 0
\end{equation}

получаваме
\begin{equation}
\varepsilon_0 \frac{\pi R^2 U^2}{2 z_\mathrm{e}^2} - m g \frac{x_\mathrm{e} - z_\mathrm{e}}{L} = 0,
\end{equation}

което още може да запишем по следния начин
\begin{equation}
\label{BG:eq:zero_of_first_deriv_W}
 z_\mathrm{e}^3 \Big(\frac{x_\mathrm{e}} {z_\mathrm{e}} - 1 \Big) = \varepsilon_0 \frac{L \pi R^2 U^2}{2 m g}.
\end{equation}

От нулирането на втората производна на потенциалната енергия
\begin{equation}
W''(z)=\frac{\mathrm{d^2} W}{\mathrm{d} z^2} = 0
\end{equation}

получаваме
\begin{equation}
2 \varepsilon_0 \frac{\pi R^2  U^2}{2 z_\mathrm{e}^3} -  \frac{m g}{L} = 0,
\end{equation}

от където намираме критичното разстояние между двете пластинки
\begin{equation}
\label{BG:eq:z_c}
z_\mathrm{e}^3 =  \varepsilon_0 \frac{L \pi R^2 U^2}{m g}.
\end{equation}

Ако заместим израза~\eqref{BG:eq:z_c} за $z_\mathrm{e}$ в~\eqref{BG:eq:zero_of_first_deriv_W}

\begin{equation}
\varepsilon_0 \frac{L \pi R^2 U^2}{m g} \Big(\frac{x_\mathrm{e}} {z_\mathrm{e}} - 1 \Big) = \varepsilon_0 \frac{L \pi R^2 U^2}{2 m g}
\end{equation}

и съкратим еднаквите членове
\begin{equation}
\frac{x_\mathrm{e}} {z_\mathrm{e}} - 1= \frac{1}{2},
\end{equation}

получаваме една проста връзка между двете критични разстояния
\begin{equation}
\label{BG:eq:x_c_vs_z_c}
x_\mathrm{e} = \frac{3}{2} z_\mathrm{e}.
\end{equation}

Като заместим равенство~\eqref{BG:eq:x_c_vs_z_c} в~\eqref{BG:eq:z_c} получаваме търсеното съотношение между $x_\mathrm{e}$ и $U$

\begin{equation}
x_\mathrm{e}^3 = \frac{27}{8}  \varepsilon_0 \frac{L \pi R^2 U^2}{m g}.
\end{equation}

\subsection{Извод на зависимостта на критичното разстояние $x_\mathrm{m}$, при което пластинките се залепват под действието на магнитните сили, от тока $I$ през бобинките}

Силата, която действат на висящата бобинка е равна на разликата от магнитната сила $F_\mathrm{m}$ на привличане на двете бобинки и хоризонталната проекцията на силата на тежестта $F_\mathrm{g}$
\begin{equation}
\label{BG:eq:F_magn}
F=F_\mathrm{m}-F_\mathrm{g}.
\end{equation}

Потенциалната енергия на хоризонталната проекцията на силата на тежестта е

\begin{equation}
\label{BG:eq:W_g_magn}
W_\mathrm{g}=  m g \frac{y^2}{2 L} =  m g \frac{(x-z)^2}{2 L},
\end{equation}

а големината на силата е
\begin{equation}
\label{BG:eq:F_g_magn}
F_\mathrm{g}=-m g \frac{x-z}{L},
\end{equation}

Потенциалната енергия на висящата бобинка намираща се в магнитното поле на неподвижната бобинка е
\begin{equation}
\label{BG:eq:W_m}
W_\mathrm{m}=-\mu_0 I^2 N^2 R \ln{z},
\end{equation}

където $I$ е токът през двете бобинки, $N$ е броят на навивки на всяка от бобинките, $R$ е радиусът на бобинките, а $z$ е разстоянието между тях.

Магнитната сила, която действа на всяка от бобинките е
\begin{equation}
\label{BG:eq:F_m}
F_\mathrm{m}=-W'(z)=-\frac{\mathrm{d} W_\mathrm{m}}{\mathrm{d} z}=-\frac{\mathrm{d}}{\mathrm{d} z}(-\mu_0 I^2 N^2 R \ln{z})=\mu_0 \frac{I^2 N^2 R}{z}.
\end{equation}

Ако означим с $W$ потенциалната енергия на висящата бобинка, то от~\eqref{BG:eq:W_m} и~\eqref{BG:eq:W_g_magn} получаваме
 
\begin{equation}
W = W_\mathrm{m} + W_\mathrm{g} = - \mu_0 I^2 N^2 R \ln{z} +  m g \frac{(x-z)^2}{2 L}.
\end{equation}

Резултанта сила действаща на висящата бобинките се получава като заместим~\eqref{BG:eq:F_m} и~\eqref{BG:eq:F_g_magn} в~\eqref{BG:eq:F_magn} 

\begin{equation}
F=-W'(z)=-\frac{\mathrm{d} W}{\mathrm{d} z}=\mu_0 \frac{I^2 N^2 R}{z} - m g \frac{x - z}{L}
\end{equation}

Когато бавно приближаваме бобинката поставена на статив към висящата бобинка, 
при едно критично разстояние $x_\mathrm{m}$ и $z_\mathrm{m}$ висящата бобинка изведнъж се залепва за статичната бобинка.
Това разстояние може да определим от нулирането на първата и втората производна на потенциалната енергия, т.е. от условието максимума и минимума на потенциалната енергия да съвпадат.

От нулирането на първата производна на потенциалната енергия
\begin{equation}
W'(z)=\frac{\mathrm{d} W}{\mathrm{d} z} = 0
\end{equation}

получаваме
\begin{equation}
\label{BG:eq:zeroing_mag_force_equation}
\mu_0 \frac{I^2 N^2 R}{z_\mathrm{m}} - m g \frac{x_\mathrm{m} - z_\mathrm{m}}{L} = 0.
\end{equation}

което още може да запишем по следния начин
\begin{equation}
\label{BG:eq:simplified_zeroing_mag_force_equation}
z_\mathrm{m}^2 \Big(\frac{x_\mathrm{m}}{z_\mathrm{m}} - 1 \Big) = \mu_0 \frac{L I^2 N^2 R}{mg} 
\end{equation}

От нулирането на втората производна на потенциалната енергия
\begin{equation}
W''(z) = \frac{\mathrm{d^2} W}{\mathrm{d} z^2} = 0
\end{equation}

получаваме
\begin{equation}
\label{BG:eq:zeroing_first_deriv_mag_force_equation}
-\mu_0 \frac{I^2 N^2 R}{z_\mathrm{m}^2} + \frac{m g}{L} = 0.
\end{equation}

от където намираме критичното разстояние между двете бобинки
\begin{equation}
\label{BG:eq:z_c_magn}
z_\mathrm{m}^2 = \mu_0 \frac{L I^2 N^2 R}{m g}
\end{equation}

Ако заместим израза~\eqref{BG:eq:z_c_magn} за $z_\mathrm{m}$ в~\eqref{BG:eq:simplified_zeroing_mag_force_equation}

\begin{equation}
\mu_0 \frac{L I^2 N^2 R}{m g} \Big(\frac{x_\mathrm{m}} {z_\mathrm{m}} - 1 \Big) = \mu_0 \frac{L I^2 N^2 R}{m g}
\end{equation}

и съкратим еднаквите членове получаваме една проста връзка между двете критични разстояния

\begin{equation}
\label{BG:eq:x_c_vs_z_c_magn}
x_\mathrm{m} = 2 z_\mathrm{m}.
\end{equation}

Като заместим равенство~\eqref{BG:eq:x_c_vs_z_c_magn} в~\eqref{BG:eq:z_c} получаваме търсеното съотношение между $x_\mathrm{m}$ и $I$
\begin{equation}
x_\mathrm{m}^2 = 4 \mu_0 \frac{L N^2 R}{m g}  I^2
\end{equation}

\newpage

\section{Translation into Macedonian: Услов на задачата}

\section{Задача}

Измерета ја брзината на светлината с користејќи  го дадениот експериментален  сет прикажан на Сл.1. Следете ги  подолу опишаните инструкции. После олимпијадата текстот на задачата и експерименталната постановка остануваат за училиштето.

\begin{figure}[h]
\includegraphics[width=9cm]{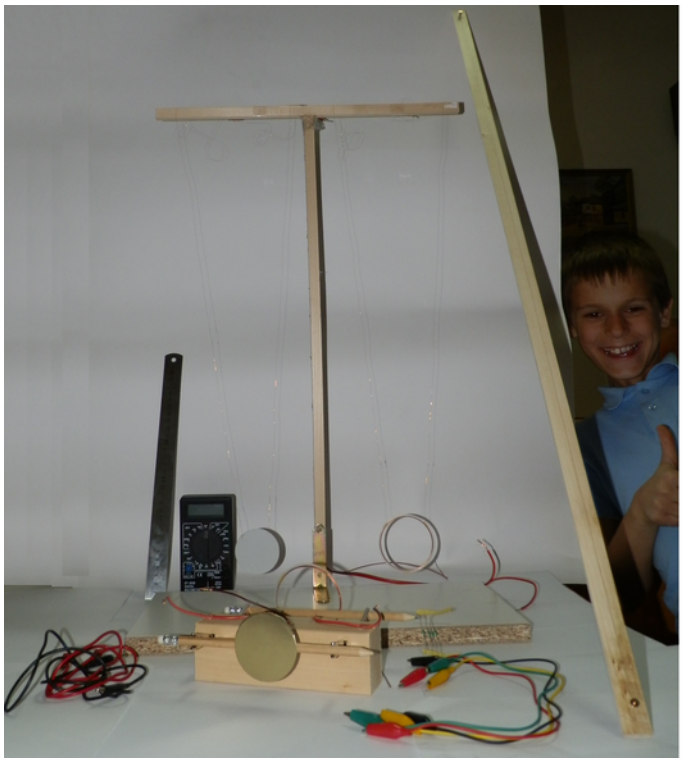}
\caption{Проверете дали експерименталниот  сет од  елементи прикажан на сликата е комплетен: статив  на кој се закачени кружна метална плоча и калем, прстен од бакарни намотки кои  висат на тенки бакарни жици; дрвено трупче на што е прицврстена друга метална плочка и друг прстен со бакарни намотки; две лежишта за четири  батерии со големина AA; 8 батерии од 1.5~V големина AA; два извори на напон поставени во цевки со три електроди-приклучоци; отпорник реостат, направен така што на тенка летва е оптегнат тенок проводник од кантал; алуминиумска плочка која се штипнува со крокодил спојка и може да се држи со рака се користи како лизгач; кабли  чии краеви се поврзани со крокодил спојки; 4 отпорници; метален ленир со поделоци од 0.5~mm и 1 мултиметар. Се претпоставува дека Вие носите уште еден мултиметар со неговите приклучни кабли.}
\label{MK:exp_setup_photo}
\end{figure}


\begin{enumerate}

\section{Експериментална задача. 
\large{Внимавајте да не ги прекинете жичките!}  
}
\subsection{Гравитација. Определување на земјиното забрзување $g$}

\begin{figure}[h]
\includegraphics[width=2.5cm]{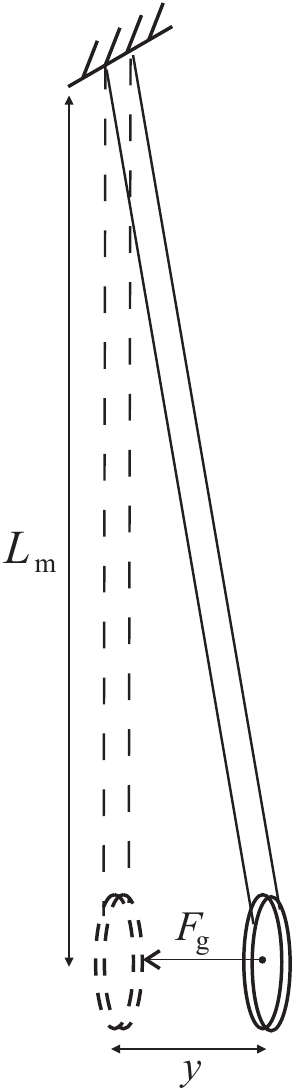}
\caption{
Калем во форма на прстен виси на две тенки жички. Растојанието од точката на прикачување до центарот на прстенот е $L_\mathrm{m}$. Ако нишалото се отклони од рамнотежната положба на мало растојание $y$, тогаш повратната сила е
$F_g=-m gy/L_\mathrm{m}$. 
Знакот минус ни покажува дека таа секогаш е спротивно насочена на поместувањето.}
\label{MK:setup_g}
\end{figure}

\item Измерете го периодот $T$ на нишање на висечкиот калем. 

Лесно поместете го прстенот во насока на оската на прстенот и избројте колкун осцилации $N_\mathrm{osc}$ може да наблудувате. 
Поместет го уште еднаш, ``залулете'' го прстенот и измерете го времето на тие осцилации $T=T_N/N_\mathrm{osc}$. 
Повторете го експериментот неколку пати и најдете ја средната вредност на периодот.

\item Определете ја кружната фреквенција $\omega=2\pi/T$. 

\item Измерете го растојанието $L_\mathrm{m}$ од точката на прикачување до центарот на висечкиот калем. (виж Сл.~\ref{MK:setup_g}). 

\item Пресметајте го земјиното забрзување $g=L_\mathrm{m}\omega^2$. 

Ве потсетуваме дека формулата за период на нишало e:
\begin{equation}
T=2 \pi \sqrt \frac{L_\mathrm{m}}{g}.
\end{equation}
\item За колку проценти вашиот резултат за земјино забрзување $g$ се разликува од познатата вредност за земјино забрзување? 

\subsection{Мерење на напоните на изворите на електромоторните сили $\mathcal{E}_1$, $\mathcal{E}_2$, $\mathcal{E}_3$ и $\mathcal{E}_4$ за електричниот експеримент само со амперметар и волтметар}
\label{MK:measure_varE}

Изворот на напон којшто ќе се користи во електричниот експеримент е составен од повеќе мали батерии со напон 12~V, димензија 23A, поставени во две пластични цевки. Секоја од цевките има по три приклучока  како електроди од кои 2 на краевите и еден на средината. По средината помеѓу две од електродите се залепени етикети со натписи $\mathcal{E}_1$, $\mathcal{E}_2$, $\mathcal{E}_3$ или $\mathcal{E}_4$.  Сериски т.е редно поврзано има високо-омски отпорник со отпор ri помеѓу секој од двата приклучока, како што е покажано на еквивалентната шема  на изворот за напојување  на Сл.~\ref{MK:HV-Battery}.
 Овој отпорник е поставен  заради безбедност при работата. Не го разглобувајте изворот за напојување и не отстранувајте го безбедносниот отпор!
Точното мерење на напонот  на електромоторната сила само со приложениот ви волтметар не е можно бидејќи внатрешниот  отпор $r_i$ на изворот е споредлив со внатрешниот отпор на волтметарот.

\begin{figure}[h]
\includegraphics[width=12cm]{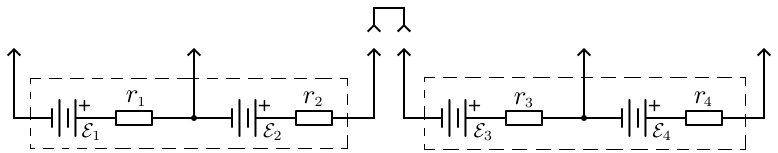}
\caption{
Еквивалентна шема на изворот на напонот за експериментот со електричното поле. Пластичните цевки се претставени со испрекинати линии. Приклучните електроди кои излегуваат од цевките во кои се сместени батериите ,на шемата се претставени со стрелки кон надвор,  а приклучните штипки ``крокодилки'' со  друга стрелка.
}
\label{MK:HV-Battery}
\end{figure}

\item Измерете го напонот на изворите на електромоторните сили натписи $\mathcal{E}_1$, $\mathcal{E}_2$, $\mathcal{E}_3$ и $\mathcal{E}_4$ само со помош на амперметар и волтметар.

\begin{figure}[h]
\includegraphics[width=12cm]{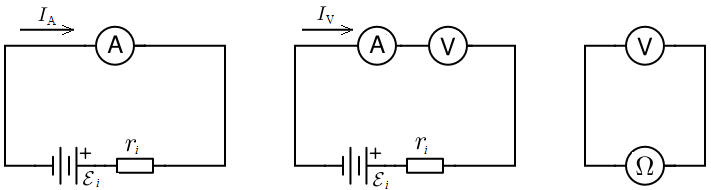}
\caption{
Мерење нана напонот на еден дел од изворот на напојување $\mathcal{E}_i$ за потребниот напон на електричното поле помеѓу металните плочки (електричниот експеримент). Внатре во цевките има скриено отпорник со голем отпор $r_i$ ($i=1,\,2,\,3,\,4$, ограничувачки отпорници. На сликата лево, помеѓу две соседни електроди вклучувате амперметар и струјата која протекува е $I_\mathrm{A}$.Средната слика: Кога амперметарот е вклучен сериски со волтметарот струјата $I_\mathrm{V}$ е помала , а волтметарот покажува напон $U.$ На сликата десно имаме директно мерење на внатрешниот отпор $R_\mathrm{V}$ на волтметарот, искористувајќи го другиот мултиметар како ом-метар.}
\label{MK:Measure-battery}
\end{figure}

Поврзете го амперметарот со еден дел од изворот за напојување како што е прикажано на сликата~\ref{MK:Measure-battery} и измерете ја јачината на струја $I_\mathrm{V}$ која тече низ него и напонот $U$, кој го покажува волтметарот. Повторете го мерењето за сите четири дела од изворите за напојување и нанесете ги резултатите во табела, како што е покажано во дадената табела како пример, Табела~\ref{MK:tab:I12345}.
 
\begin{table}[h]
\caption{
Табела пример за запис на експерименталните резултати за струите и напонот измерени според шемите од сликата~\ref{MK:Measure-battery}, како и за пресметаната вредност за напонот  електромоторните сили за различните делови на изворот на напојување. }
\begin{tabular}{| r | r | r | r | r | r | r | }
\tableline
$i$& $I_\mathrm{A} \, [\mathrm{mA}]$ & $I_\mathrm{V} \,[\mathrm{mA}]$ & $U \,[\mathrm{V}]$ & $\mathcal{E} \,[\mathrm{V}]$ & $r \, [\Omega$] & $R_\mathrm{V} \, [\Omega$]\\
\tableline
1 & & & & & & \\
2 & & & & & & \\
3 & & & & & & \\
4 & & & & & & \\
\tableline
\end{tabular}
\label{MK:tab:I12345}
\end{table}

За секој еден дел од изворот за напојување запишете ги пресметките  во примерната Табела~\ref{MK:tab:I12345} направени според формулата: 
\begin{equation}
\label{MK:eq:E}
\mathcal{E}=\frac{U}{1-I_\mathrm{V}/I_\mathrm{A}}.
\end{equation}

\item Колку изнесува внатрешниот отпор на секој дел изворот за напојување?

Искористете ги податоците од мерењата внесени во Табелата~\ref{MK:tab:I12345} определувајќи го внатрешниот отпор на секој дел од изворот според формулата:
\begin{equation}
\label{MK:eq:r}
r=\frac{U}{I_\mathrm{A}-I_\mathrm{V}},
\end{equation}
за секое од четирите мерења.

\item Измерете го внатрешниот отпор на волтметарот.

Искористувајќи ги податоците од Табелата~\ref{MK:tab:I12345} определете го внатрешниот отпор на волтметарот
\begin{equation}
\label{MK:eq:r}
R_\mathrm{V}=\frac{U}{I_\mathrm{V}},
\end{equation}
за секое од четирите мерења и и пресметајте средна вредност од сите мерења.
Внатрешниот отпор на волтметарот може да го измерете директно според шемата десно од сликата~\ref{MK:Measure-battery}.

\item Изведете ги формулите~\eqref{MK:eq:E}, \eqref{MK:eq:r}  и~\eqref{MK:eq:r}.

\subsection{
Определување на електромоторните сили $\mathcal{E}_1$, $\mathcal{E}_2$, $\mathcal{E}_3$ и $\mathcal{E}_4$ на изворите на напојување за електричниот експеримент со помошта на амперметар и отпорници}
\textit{Оваа серија од задачи е за учениците со помала возраст, за учениците од Ѕ категорија. Повозрасните ученици може да ги решаваат откако ќе ги решат останатите задачи.}

\item{Измерете ги отпорот на дадените четири отпорници $R_1^*,$ $R_2^*,$ $R_3^*$ и $R_4^*$, и  и претставете ги резултатите графички.}

\item 11. Составете ја шемата како на Слика~\ref{MK:Measure-battery2} и измерете ја струјата Ij, кога ќе се постави отпорникот $R_1=R_1^*,$ $R_2=R_1^*+R_2^*,$ $R_3=R_1^*+R_2^*+R_3^*$ и $R_4=R_1^*+R_2^*+R_3^*+R_4^*$.

\begin{figure}[h]
\includegraphics[width=10cm]{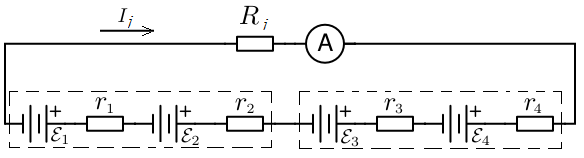}
\caption{
Мерење на напонот на еден дел од изворот за напон, за електростатскиот експеримент. Јачината на струјата $I_j$ се мери при различни надворешни отпори $R_j$ и така се определува внатрешниот отпор на изворот сместен во цевката.}
\label{MK:Measure-battery2}
\end{figure}

Нанесете ги измерените вредности во табела, како што е покажано во табелата~\ref{MK:tab:E1234_2}. Во последната колона на табелата запишете ја реципрочната вредност на јачината на струјата.

\begin{table}[h]
\caption{Табела за внесување на експерименталните резултати од мерењата за струите измерени според шемата од слика~\ref{MK:Measure-battery2} за различните делови од изворот за напојување.}
\begin{tabular}{| r | r | r | r | }
\tableline
i& $R_j \, [\mathrm{k}\Omega]$ & $I_j \,[\mu\mathrm{A}]$ &  $1/I_j \,[\mu\mathrm{A}^{-1}]$ \\
\tableline
1 &  &  &\\
2 & &  &\\
3 & & &\\
4 & &  &\\
\tableline
\end{tabular}
\label{MK:tab:E1234_2}
\end{table}

\item Искористувајќи ги резултатите од Табелата~\ref{MK:tab:E1234_2} претставете ја графички зависноста на отпорот $R_j$ од реципрочната вредност на струјата $1/I_j$.

Отпорот $R_j$ и реципрочната вредност на струјата $I_j$  се поврзани со со следната зависност
\begin{equation}
\label{MK:eq:r_j}
R_j=\mathcal{E}_\mathrm{tot} \frac{1}{I_j} - r_\mathrm{tot},
\end{equation}
каде $\mathcal{E}_\mathrm{tot}$ е вкупната електромоторна сила  на сите сериски поврзани извори на струја, а $r_\mathrm{tot}$  е нивниот вкупен внатрешен отпор.

\item Определете ја елктромоторната сила $\mathcal{E}_\mathrm{tot}=\mathcal{E}_1 + \mathcal{E}_2 + \mathcal{E}_3 + \mathcal{E}_4$ и внатрешниот отпор $r_\mathrm{tot}=r_1+r_2+r_3+r_4$ на двата сериски поврзани извори на струја.

Нацртајте права  од експерименталните точки, која најблиску минува низ нив. Од коефициентот на правец (аголниот коефициент) на правата може да се определи електромоторната сила $\mathcal{E}_\mathrm{tot}$, а од пресекот помеѓу правата и ординатната оска може да го определете внатрешниот отпор на батеријата $r_\mathrm{tot}$ .

\item Изведете ја формулата ~\eqref{MK:eq:r_j}.

\subsection{Електростатичко определување на $\varepsilon_0$}

\item Квалитативен електричен експеримент

Со помош на кабли “крокодилки“ поврзете ги сериски двата извори на напојување без да сврзете кон нив волтметри. Внимавајте на поларитетот на изворите. Сврзете (+) од едниот извор со (-) од другиот извор. Вкупниот напон $U_4=\mathcal{E}_1+\mathcal{E}_2+\mathcal{E}_3+\mathcal{E}_4$ се пресметува како сума од електромоторните сили дефинирани во поглавјето ~\ref{MK:measure_varE}. Поврзете ги краевите на таквиот извор со помош на кабли крокодилки со краевите на проводниците кои излегуваат од кружните плочи. Нека на почетокот плочките се оддалечени една од друга на растојание поголемо од 1 центиметар, притоа се паралелни поставени. Погледнете ја шемата на слика~\ref{MK:setup_e}. Постепено придвижувајте го трупчето, додека висечката плочка целосно не се прилепи кон неподвижната плочка прицврстена за трупче. Внимателно повторете го експериментот исчекувајќи осцилациите да престанат. Придвижувајте го трупчето така што секогаш плочките да се паралелно поставени. Откако нишалото ќе загуби рамнотежа и плочките се залепат прекинете го напонот.Тогаш плочките се одлепуваат и кога ќе престанат да се нишат одредете го растојанието $x_\mathrm{e}$, на кое плочките на тој плочест кондензатор почнуваат да се привлекуваат.

\item Повторете го внимателно експериментот мерејќи го растојанието $x_\mathrm{e}$, помеѓу плочките со точност до 0,5 mm за различни напони:
$U_4=\mathcal{E}_1+\mathcal{E}_2+\mathcal{E}_3+\mathcal{E}_4,$
$U_3=\mathcal{E}_1+\mathcal{E}_2+\mathcal{E}_3,$
$U_2=\mathcal{E}_1+\mathcal{E}_2,$ и
$U_1=\mathcal{E}_1.$

Запишете ги резултатите од мерењата во табелата  како табелата за пример~\ref{MK:tab:xe3_vs_U2}. Со помош на калкулатор пополнете уште две дополнителни колони
$x_\mathrm{e}^3$ и $U^2$.

\begin{table}[h]
\caption{Табела со резултати од електричниот експеримент.}
\begin{tabular}{| r | r | r | r | r | r |}
\tableline
i& $U \,[\mathrm{V}]$ & $x_\mathrm{e} \,[\mathrm{mm}]$ & $U^2 \,[\mathrm{V}^2]$ & $x_\mathrm{e}^3 \,[\mathrm{mm}^3]$ \\
\tableline
1 &  &  &  & \\
2 &   & &  &  \\
3 &   &  &  &  \\
4 &  & &  &  \\
\tableline
\end{tabular}
\label{MK:tab:xe3_vs_U2}
\end{table}

\item Претставете ја графички зависноста $x_\mathrm{e}^3$ од $U^2$ и определете го наклон на правата $k_\mathrm{e}$ која минува најблиску низ експерименталните точки.

\begin{figure}[h]
\includegraphics[width=7cm]{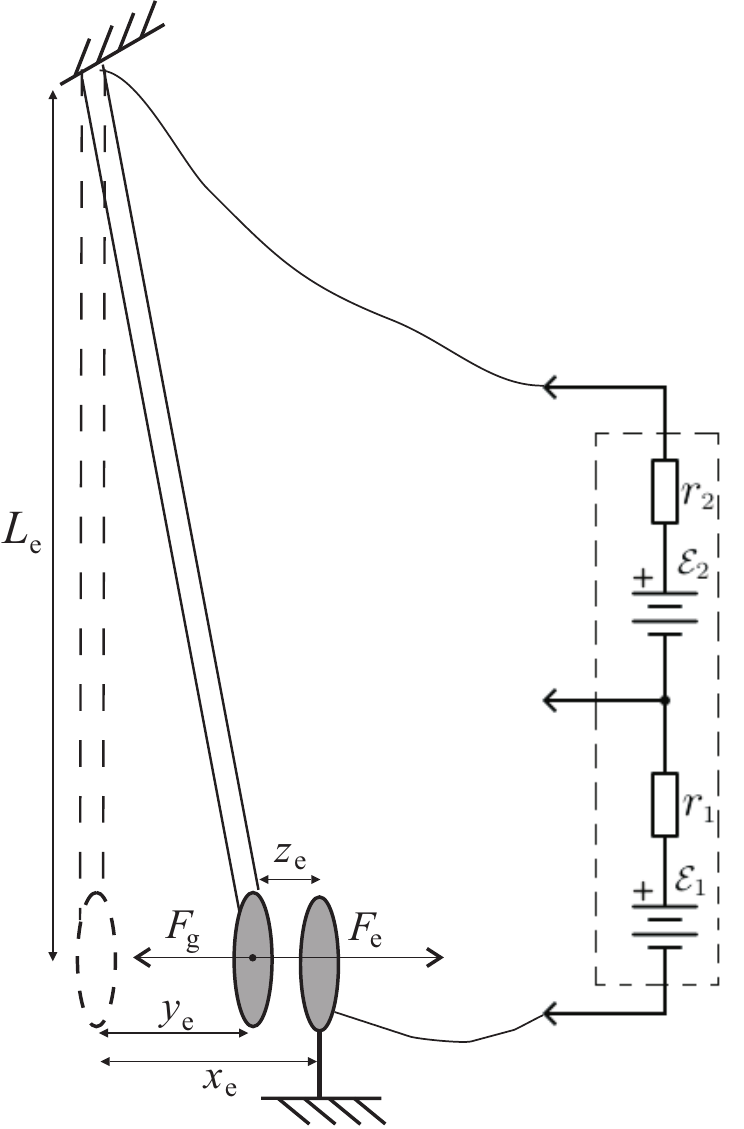}
\caption{
Електрично нишало за мерење на диелектричната константа на вакуумот $\varepsilon_0$. Должината на нишалото е $L_\mathrm{e}$. Со $x_\mathrm{e}$ го означуваме растојанието на нишалото до другата плочка поставена на трупчето.Поместувањето на нишалото од рамнотежната положба пред да се залепи од неподвижната плочка е $y_\mathrm{e}=x_\mathrm{e}-z_\mathrm{e}.$. 
Со $z_\mathrm{e}$ го означуваме  растојанието помеѓу плочката непосредно пред нишалото да се прилепи кон плочката со трупче. 
Eлектичното поле помеѓу плочките се создава од сериско поврзаните батерии како извори на електромоторни сили $\mathcal{E}_1$ и $\mathcal{E}_2$. 
Помегу батереиите  има и отпорници со големи отпори кои ја ограничуваат струјата  и се означени  со $r_1$ и $r_2$. Отпорниците како и батериите се сместени во пластична цевка означена симболично со испрекината линија. Во случај н а рамнотежа електричната привлечна  сила помеѓу плочките $F_\mathrm{e}$ се урамнотежува со гравитационото сила $F_\mathrm{g}$.}
\label{MK:setup_e}
\end{figure}

\item Измерете ја масата $m_\mathrm{e}$ на висечката плочка.

За таа цел искористете ја електронската вага , која се наоѓа кај тестаторот. При мерењето не ја откачувајте висечката плочка од стативот. Подигнете ја вагата на повисоко ниво. Извршете го мерењето на масата додека плочката е закачена на тенката бакарна жичка, внивајќи да не даде дополнителен товар за време на мерењето.

\item Измерете го дијаметарот $D_\mathrm{e}$ на двете плочки.  

\item Измерете го растојанието од центарот на плочката до стативот за кај што е прикачено $L_\mathrm{e}$.

\item Определете ја диелектричната константа на вакуумот $\varepsilon_0$ преку приближната формула за поврзаност помеѓу $x_\mathrm{e}$ и $U$, 
за случај кога е исполнет условот $x_\mathrm{e}\ll D_\mathrm{e}$.

\begin{equation}
x_\mathrm{e}^3 =k_\mathrm{e}  U^2,
\end{equation}
каде $k_\mathrm{e}$ е линеарен коефициент 
\begin{equation}
\label{MK:eq:Epsilon_0}
k_\mathrm{e} 
=\frac{27}{32} \pi  \varepsilon_0 \frac{L_\mathrm{e} D_\mathrm{e}^2}{m_\mathrm{e} g}.
\end{equation}

Преку  формулата за kе  може да се пресмета диелектрината константа $k_\mathrm{e}$ на воздухот.
\begin{equation}
\label{MK:eq:k_e}
\varepsilon_0=
\frac{32}{27\pi}\frac{m_\mathrm{e} g}{L_\mathrm{e} D_\mathrm{e}^2}k_\mathrm{e}. 
\end{equation}

\item Кон таблицата додадете уште еден столб $x_\mathrm{e}^3 (1-\frac{4}{3 \pi} \frac{x_\mathrm{e}}{D_\mathrm{e}})$, пресметувајќи ја малата корекција
$\frac{4}{3 \pi} \frac{x_\mathrm{e}}{D_\mathrm{e}}$.
Повторете го погоре опишаниот метод  за поточно определување на диелектричната константа на вакуумот преку формулата
\begin{equation}
\label{MK:Effect_of_ends}
x_\mathrm{e}^3 \left(1-\frac{4}{3 \pi} \frac{x_\mathrm{e}}{D_\mathrm{e}} \right) =
\frac{27}{32} \pi  \varepsilon_0 \frac{L_\mathrm{e} D_\mathrm{e}^2}{m_\mathrm{e} g}  U^2.
\end{equation}
Оваа формула овозможува постигнување на поголема процентна точност при користење на дадената постановка.

Каква е разликата во определувањето на $\varepsilon_0$ според двата метода?

\subsection{Магнетостатичко определување на $\mu_0$}

\item Квалитативен магнетен експеримент

Поставете ги двата кружни калеми во форма на прстен паралелно на растојание $x_\mathrm{m}= \mathrm{20\;mm}$, сметано од центарот на прстените. 
Искористувајќи кабли со крокодилки составете шемата покажана како на сликата~\ref{MK:setup_m}. 
Проследете го патот на струјата кој започнува од (+) на батеријата. Првин поминува низ канталовата отпорна жица , оптегната врз летвата. Таквиот променлив отпор уште се нарекува реохорд. Веригата се соединува со алуминиумов лизгач прицврстен со крокодилка штипка која се држи со рака. Потоа струјата поминува низ амперметар, низ висечкиот прстен, неподвижниот прстен кој има ист број навивки $N=50$,  како и висечкиот прстен и со (-) полот се затвора колото.
Кога ќе го допрете лизгачот до отпорната жица  намотките затреперуваат. Ако прстените се одбиваат сменете го поларитетот.При вклучување на струјата прстените потребно е да се привлекуваат. Започнете со слаби струи, поместувајќи го лизгачот, гледајќи истовремено и во амперметарот  и запомнете го растојанието при кое нишалото ја губи рамнотежата и висечкиот прстен се привлекува. Повторете го експериментот , чекајќи осцилациите да се смират. Запишете ја најмалата струја $I$ (критичната) при која висечкиот прстен се прилепува за неподвижниот.

\begin{figure}[h]
\includegraphics[width=7cm]{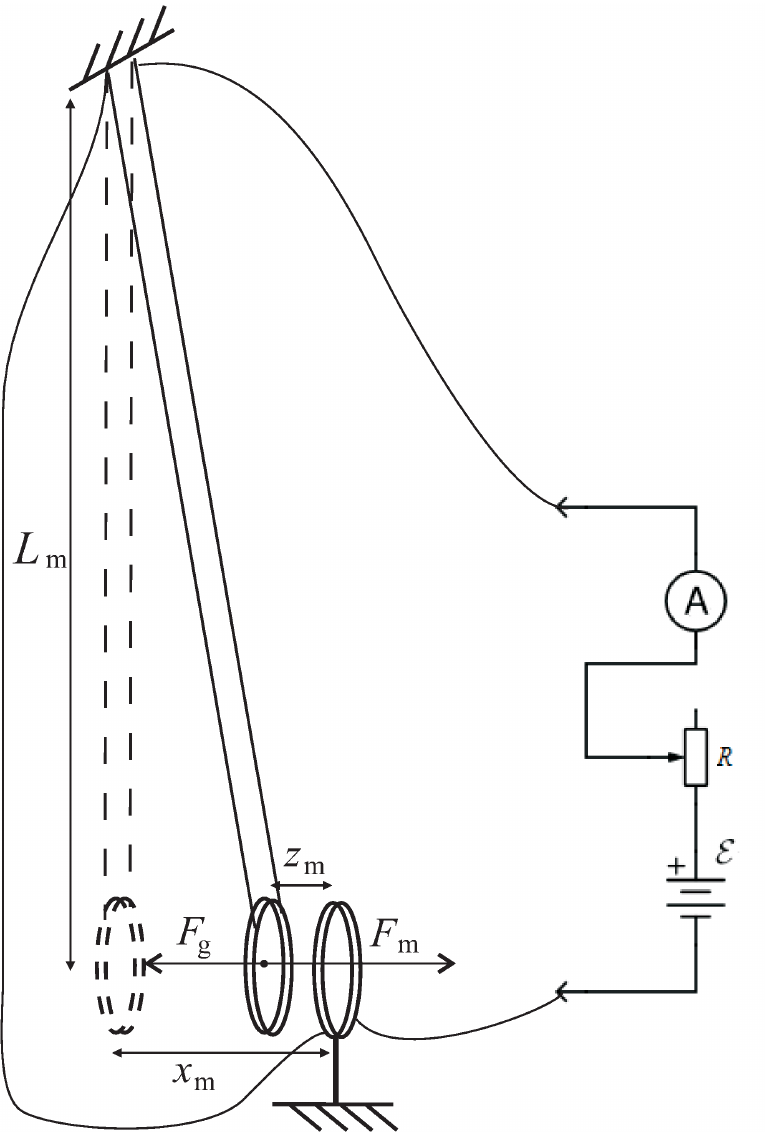}
\caption{
Магнетно нишало за мерење на $\mu_0$, магнетната пермеабилност (пропустливост) на вакуумот. Должината на нишалото е $L_\mathrm{m}$. Со $x_\mathrm{m}$ ја означуваме оддалеченоста на нишалото од другата бакарна намотка. Растојанието помеѓу бакарните намотки - струјните прстени во моментов пред да се привлечат е $z_\mathrm{m}$. 
Магнетното поле помеѓу намотките се создава од струјата на 4 или 8 сериско поврзани батерии од 1.5 V поставени во нивните лежишта. 
Струјата се мери од амперметар и се регулира со променлив отпор $R$, поместувајќи го лизгачот во различни точки на канталовата растегната жица на летвата. Во рамнотежа состојба магнетна сила на привлекување помеѓу прстените со паралелни струи $F_\mathrm{m}$ се урамнотежува  со гравитационото сила $F_\mathrm{g}$.}
\label{MK:setup_m}
\end{figure}

\item Повторете го експериментот и измерете ја критичната струја $I$ за различни вредности на $x_\mathrm{m}$, например 25, 20, 15, 10, 5~mm. 

Резултатите за растојанието $x_\mathrm{m}$, и струјата $I$ претставете ги табеларно во првите две колони на табелата~\ref{MK:tab:mu0}. 
Со помош на калкулатор направете уште две  дополнителни колони за $x_\mathrm{m}^2$ и $I^2$ или со други зборови направете ја зависноста $x_\mathrm{m}^2$ од $I^2$ 
За малите растојанија искористете само еден носач на батерии.

\item Претставете ја графички зависноста $x_\mathrm{m}^2$ од $I^2$ и определете го наклонот $k_\mathrm{m}$ (коефициентот на правец) на правата исцртана од експерименталните точки

\item Измерете ја должината на нишалото т.е растојанието од местото на прикачување до центарот на висечкиот калем $L_\mathrm{m}$ .

\item Измерете ја масата $m_\mathrm{m}$ на висечкиот прстен.

\item Измерете го дијаметарот $D_\mathrm{m}$ на кружните калеми.

\item  Определете ја магнетната пермеабилност на вакуумот $\mu_0$ искористувајќи ја приближната формула,
като използвате приближената формула
\begin{equation}
x_\mathrm{m}^2 =k_\mathrm{m}  I^2,
\end{equation}
каде $k_\mathrm{m}$ е коефициентот на правец и изнесува
\begin{equation}
\label{MK:eq:k_m}
k_\mathrm{m} =2 \mu_0 \frac{L_\mathrm{m} N^2 D_\mathrm{m}}{m g}.
\end{equation}
Таа формула важи за случај кога $x_\mathrm{m}\ll D_\mathrm{м}$
и проводниците се прави.

Со помош на формулата за $k_\mathrm{m}$ може да се пресмета магнетната пермеабилност на вакуумот $\mu_0$
\begin{equation}
\label{MK:eq:k_m}
\mu_0=\frac{m gk_\mathrm{m}}{2 L_\mathrm{m} N^2 D_\mathrm{m}}.
\end{equation}

\item Силата на привлекување помеѓу два струјни прстена е позната магнетостатичка задача  која има повеќе решенија. Точната формула за определување на $\mu_0$ преку нашата постановка се дава со
\begin{equation}
\label{MK:Correction}
x_\mathrm{m}^2 \left[1+f \left(\frac{x_\mathrm{m}}{D_\mathrm{m}}\right) \right] =k_\mathrm{m}  I^2,
\end{equation}
каде функцијата $f(x_\mathrm{m}/D_\mathrm{m})$ 
е претставена и графички на Слика ~\ref{MK:f_delta}.

При точноста со која работиме корекциона поправка може да се направи и преку приближната формула
\begin{equation}
\label{MK:function_2_rings}
f \left(\frac{x_\mathrm{m}}{D_\mathrm{m}}\right)\approx
\frac{1}{16}\left(-5+6 \log{ \frac{8 D_\mathrm{m}}{x_\mathrm{m}} } \right)
 \frac{x_\mathrm{m}^2}{D_\mathrm{m}^2}.
\end{equation}

Пример за подредување на експерименталните податоци и корекциониот множител се дадени во табелата~\ref{MK:tab:mu0}.  Пополнете ја вашата табела. 

\begin{table}[h]
\caption{Табела како пример  за подредување на експерименталните резултати и корекционата функција со струјните прстени. Помеѓу $I^2$ и $x_\mathrm{m}^2(1+f)$ има линеарна зависност и коефициентот на пропорционалност ја определува $\mu_0.$}
\begin{tabular}{| r | r | r | r | r | r | r | r| }
\tableline
$i$& $x_\mathrm{m}[\mathrm{mm}]$& $I [\mathrm{mA}]$  & $x_\mathrm{m}^2 [\mathrm{mm^2}]$ & $I^2 [\mathrm{mA^2}]$ &  $x_\mathrm{m}/D_\mathrm{m}$ & $1+f(x_\mathrm{m}/D_\mathrm{m})$ & $x_\mathrm{m}^2(1+f) [\mathrm{mm^2}] $ 
\\
\tableline
1 &  5 & && & & & \\
2 & 10& && & & & \\
3 & 15& && & & & \\
4 & 20& & && & & \\
5 & 25& && & & & \\
\tableline
\end{tabular}
\label{MK:tab:mu0}
\end{table}

\begin{center}
\begin{tabular}{| c | c || c | c || c | c || c |  c || c | c |}
		\hline
			\multicolumn{1}{| c |}{ } & \multicolumn{1}{ c ||}{ } & \multicolumn{1}{ c |}{ } & \multicolumn{1}{ c ||}{ } & \multicolumn{1}{ c |}{ } & \multicolumn{1}{ c ||}{ } 
			& \multicolumn{1}{ c |}{ } & \multicolumn{1}{ c ||}{ } & \multicolumn{1}{ c |}{ } & \multicolumn{1}{ c |}{ } \\
			\boldmath$x/D$ & \boldmath$f(x/D)$ & \boldmath$x/D$ &\boldmath$f(x/D)$ & \boldmath$x/D$ & \boldmath$f(x/D)$ & \boldmath$x/D$ & \boldmath$f(x/D)$  & \boldmath$x/D$ & \boldmath$f(x/D)$ \\[15pt] \hline 
			0.005	&	0.0001	&	0.105	&	0.0148	&	0.205	&	0.0480	&	0.305	&	0.0976	&	0.405	&	0.1644	\\	\hline
0.010	&	0.0002	&	0.110	&	0.0162	&	0.210	&	0.0502	&	0.310	&	0.1005	&	0.410	&	0.1682	\\	\hline
0.015	&	0.0005	&	0.115	&	0.0175	&	0.215	&	0.0522	&	0.315	&	0.1034	&	0.415	&	0.1721	\\	\hline
0.020	&	0.0008	&	0.120	&	0.0188	&	0.220	&	0.0545	&	0.320	&	0.1066	&	0.420	&	0.1760	\\	\hline
0.025	&	0.0012	&	0.125	&	0.0202	&	0.225	&	0.0566	&	0.325	&	0.1096	&	0.425	&	0.1800	\\	\hline
0.030	&	0.0016	&	0.130	&	0.0216	&	0.230	&	0.0589	&	0.330	&	0.1126	&	0.430	&	0.1840	\\	\hline
0.035	&	0.0021	&	0.135	&	0.0231	&	0.235	&	0.0613	&	0.335	&	0.1156	&	0.435	&	0.1880	\\	\hline
0.040	&	0.0027	&	0.140	&	0.0245	&	0.240	&	0.0634	&	0.340	&	0.1190	&	0.440	&	0.1920	\\	\hline
0.045	&	0.0034	&	0.145	&	0.0262	&	0.245	&	0.0659	&	0.345	&	0.1221	&	0.445	&	0.1965	\\	\hline
0.050	&	0.0040	&	0.150	&	0.0278	&	0.250	&	0.0684	&	0.350	&	0.1252	&	0.450	&	0.2006	\\	\hline
0.055	&	0.0048	&	0.155	&	0.0294	&	0.255	&	0.0709	&	0.355	&	0.1287	&	0.455	&	0.2047	\\	\hline
0.060	&	0.0056	&	0.160	&	0.0312	&	0.260	&	0.0732	&	0.360	&	0.1319	&	0.460	&	0.2092	\\	\hline
0.065	&	0.0064	&	0.165	&	0.0328	&	0.265	&	0.0757	&	0.365	&	0.1355	&	0.465	&	0.2134	\\	\hline
0.070	&	0.0073	&	0.170	&	0.0345	&	0.270	&	0.0783	&	0.370	&	0.1390	&	0.470	&	0.2181	\\	\hline
0.075	&	0.0083	&	0.175	&	0.0365	&	0.275	&	0.0810	&	0.375	&	0.1423	&	0.475	&	0.2223	\\	\hline
0.080	&	0.0092	&	0.180	&	0.0382	&	0.280	&	0.0837	&	0.380	&	0.1460	&	0.480	&	0.2270	\\	\hline
0.085	&	0.0103	&	0.185	&	0.0401	&	0.285	&	0.0864	&	0.385	&	0.1497	&	0.485	&	0.2317	\\	\hline
0.090	&	0.0114	&	0.190	&	0.0421	&	0.290	&	0.0891	&	0.390	&	0.1530	&	0.490	&	0.2361	\\	\hline
0.095	&	0.0125	&	0.195	&	0.0440	&	0.295	&	0.0919	&	0.395	&	0.1568	&	0.495	&	0.2408	\\	\hline
0.100	&	0.0136	&	0.200	&	0.0461	&	0.300	&	0.0947	&	0.400	&	0.1606	&	0.500	&	0.2456	\\	\hline

		\hline
\end{tabular}
\end{center}

\begin{figure}[h]
\includegraphics[width=18cm]{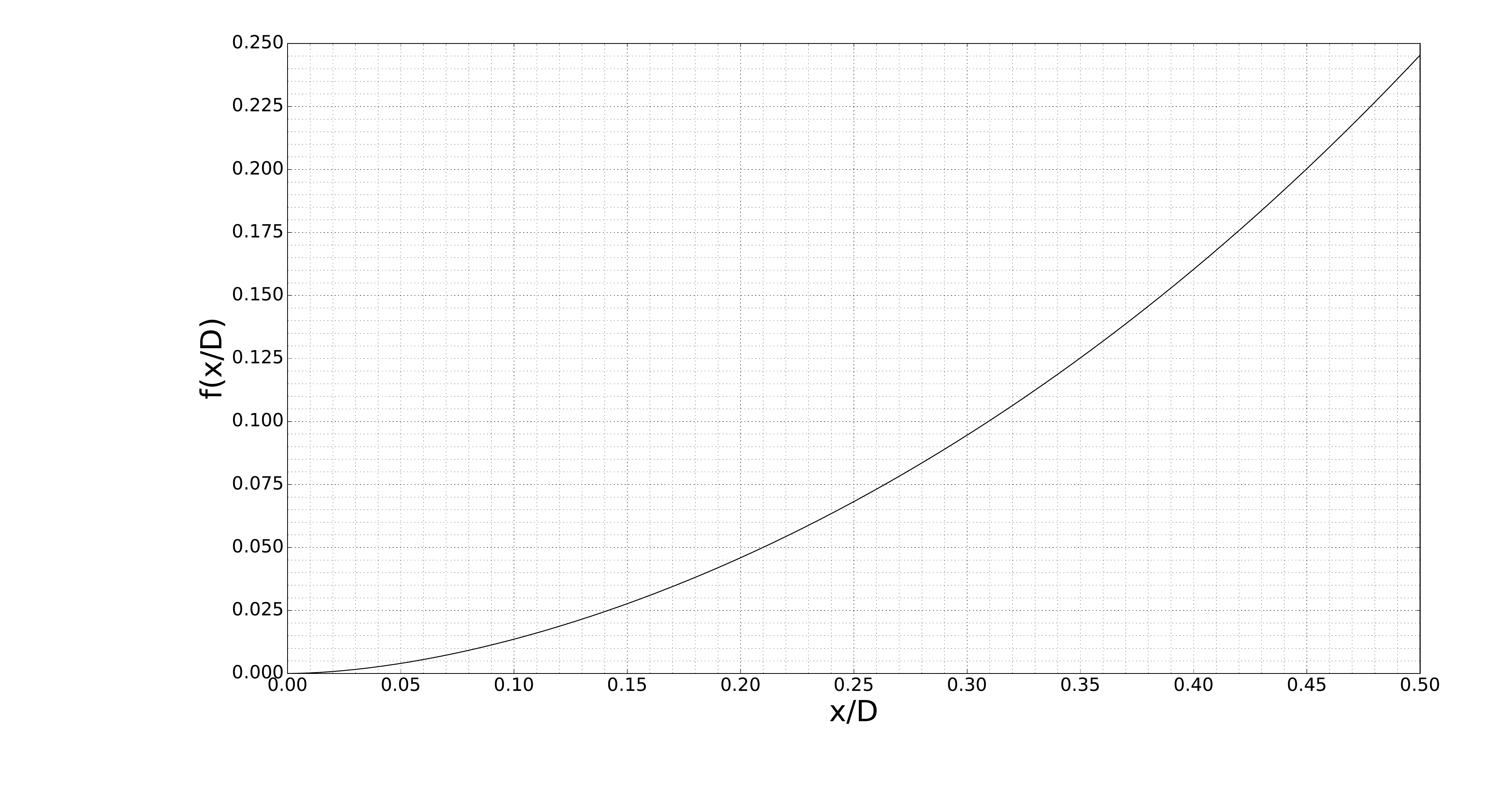}
\caption{
 Корекциониот фактор $f(x_\mathrm{m}/D_\mathrm{m})$ од равенката (\ref{MK:Correction}) како функција од бездимензионалниот однос на поместувањето од рамнотежната положба $x_\mathrm{m}$ и дијаметарот на висечкиот кружен калем $D_\mathrm{m}$.
}
\label{MK:f_delta}
\end{figure}

\item Искористувајќи ја точната формула \eqref{MK:Correction} и табеларно претставените резултати од табелата~\ref{MK:f_delta}, претставете ги податоците на график. на ордината $(1+f)x_\mathrm{m}^2$, а на апциса $I^2$ и од наклонот определете ја магнетната пермеабилност на вакуумот $\mu_0$. Колку  е процентната разлика од резултатите што се добиваат со точната и приближната формула?
\end{enumerate}

\section{Определување на брзината на светлината}

Искористувајќи ги определените вредности за $\varepsilon_0$
и $\mu_0$ пресметајте ја брзината на светлината
\begin{equation}
c=\frac{1}{\sqrt{\varepsilon_0 \mu_0}}.
\end{equation}
Не разочарувајте се  ако резултатот ви се разликува од позната вредност за с; тоа е вашето прво определување на фундаментална константа.

\section{Теориска задача}

Искористувајќи ги формулите за привлечна сила помеѓу плочи на бескраен плочест кондензатор  изведете ја формулата (\ref{MK:eq:Epsilon_0}), за определување на $\varepsilon_0$.

Искористувајќи ги формулите за привлечна сила помеѓу бескрајни паралелни струи изведете ја формулата (\ref{MK:eq:Epsilon_0}) за приближно определување на $\mu_0.$

\section{Домашна задача. Награда на Зомерфелд од 137 лева}

Побарајте во уебници по електродинамика, енциклопедии или на интернет формула за капацитет на кондензатор при кои се земаат во предвид ефектите на краевите  и изведете ја точната формула (\ref{MK:Effect_of_ends}) за определување на $\varepsilon_0$ со нашата постановка.
Аналогно побарајте и формули за взаемна индуктивност и сили на взаемно дејство на на струјни прстени и изведете ја точната формула (\ref{MK:Correction}). Корекционата функција $f(x_\mathrm{m}/D_\mathrm{m})$ може да биде пресметана или да се добие само првиот член на равенката ~(\ref{MK:function_2_rings}) на редот по степени на $x_\mathrm{m}/D_\mathrm{m}$.
Решението на домашната задача пратете го на epo@bgphysics.eu  од е-маил адреса со која сте регистрирани на олимпијадата до 7:00 на 24 април 2016.
Може да работите и заеднички, да консултирате професори по теориска физика, електродинамика или електротехника. Премијата ќе му биде доделена лично на учесникот и само на денот на објавувањето на резултатите.

\newpage

\section{Russian translation by Ana Bozhankova. Задача}
\begin{figure}[h]
		\includegraphics[width=9cm]{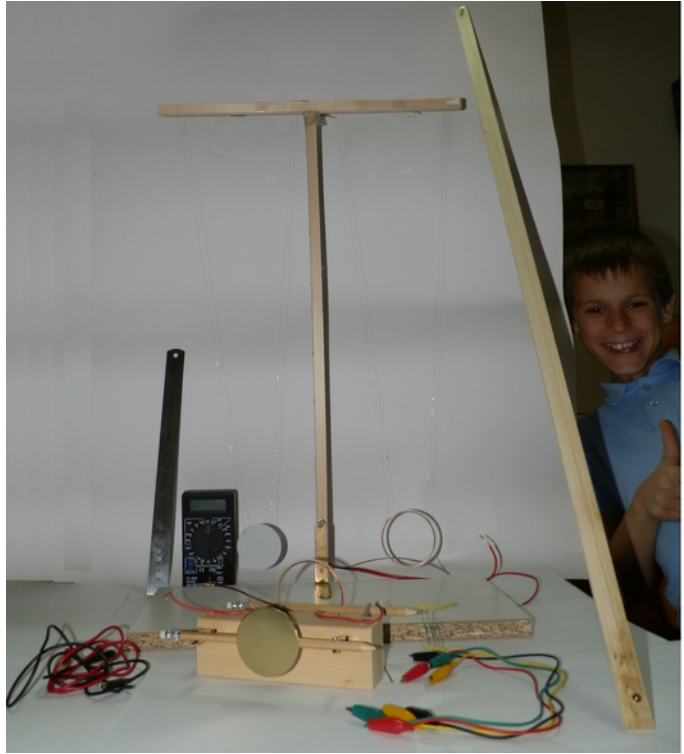}
		\caption{Проверьте если есть у вас полный набор элементов, показанных на фотографии:
			стенд на котором повешены металлические пластины и кольцо медных обмоток, висячих на тонких медных проволок;
			деревянный блок, на котором прикреплены другая металлическая пластинка и другое кольцо с медными обмотками;
			два вчетвером держателя батарейки размера АА; 
			8 батареи 1.5 V размера AA; 
			два источника напряжения, 
			поставленные в трубах, каждая с тремя металлическими выходами (электродами); реостат, изготовленный из тонкой планки, на которой натянута тонкая проволока сопротивления; 
			алюминиевая пластина захваченная крокодилом 
			которая держится рукой и используется как слайда;
			кабели с разъемами крокодил-крокодил; 
			4 сопротивления;
			металлическая линейка для черчения с делениями 0,5~мм и 1~мм. Предполагается, что вы несете другой мультиметр 
			вместе с присоединительными кабелями.
		}
		\label{RS:exp_setup_photo}
	\end{figure}

	Измерить скорости света 
	$c$, используя данного экспериментального набора, который показан на Фигуре~\ref{RS:exp_setup_photo}.
	Вы можете следовать данные ниже инструкции.
	После олимпиады, настоящий текст и экспериментальная установка остаются у участников.
	
	\begin{enumerate}
		
		\section{Экспериментальная задача.
			\large{Будьте осторожны, чтобы не сорват тонкие провода!}  
		}
		\subsection{Гравитационная часть. Надо измерить земное ускорение $g$}
		
		\begin{figure}[h]
			\includegraphics[width=2.5cm]{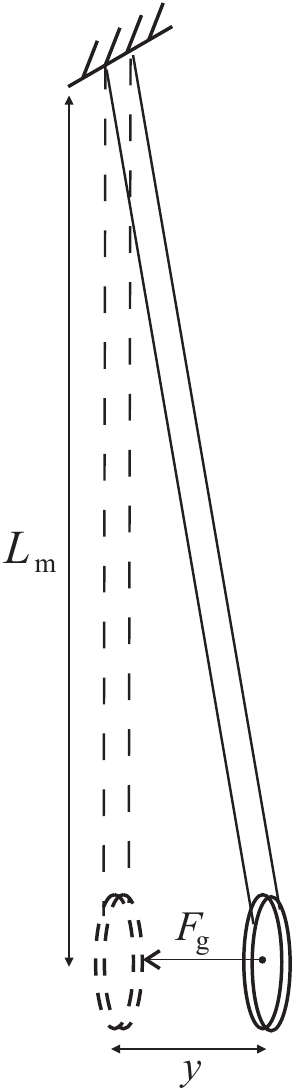}
			\caption{
				Кольцо с витым проводом висит на двух тонких проволок.
				Расстояние от точки опоры до центра кольца
				$L_\mathrm{m}$. 
				Когда маятник отклонится от положения равновесия на небольшом расстоянии $y$, возвращающая сила 
				$F_g=-m gy/L_\mathrm{m}$. 
				Знак минус означает, что сила направлена против смещения.
			}
			\label{RS:setup_g}
		\end{figure}
		
		\item 
		Измерьте период колебания $T$ (в секундах) висячего кольца.
		Осторожно встряхните кольцо в направлении своей оси
		и подсчитайте, сколько колебаний $N_\mathrm{osc}$  можно наблюдать.
		Снова встряхните и измерьте время этих колебаний $T_N$.
		Определите период как отношение полного времени и числа колебаний 
		$T=T_N/N_\mathrm{osc}$. 
		Повторите опыт несколько раз и найдите среднее значение измеряемого периода.
		
		\item 
		Определите угловую частоту $\omega=2\pi/T$.
		
		\item 
		Измерьте расстояние $L_\mathrm{m}$ (в метрах) от точки подвеса до центра висячего кольца (см рисунок~\ref{RS:setup_g}).
		
		\item Определите земное ускорение $g=L_\mathrm{m}\omega^2$. 
		
		Напоминаем известную формулу для периода маятника
		\begin{equation}
			T=2 \pi \sqrt \frac{L_\mathrm{m}}{g}.
		\end{equation}
		\item Насколько процентов ваш результат земного ускорение $g$ се различается от известной вам стоимости?
		
		\subsection{Измерение электродвижущих напряжении $\mathcal{E}_1$, $\mathcal{E}_2$, $\mathcal{E}_3$ и $\mathcal{E}_4$ источников напряжения для электрического эксперимента с помощью только амперметра и вольтметра}
		\label{RS:measure_varE}
		
		Источник напряжения, которого будете использовать для электрического эксперимента, составлен от множества маленьких батареек 12~V, с размером 23A, поставленных в двух пластмассовых трубах.
		У каждого из трубок есть три металлические вывода (электрода), два в конце и один в середине.
		В середине между двумя электродами приклеен этикет с надписями
		$\mathcal{E}_1$, $\mathcal{E}_2$, $\mathcal{E}_3$ или $\mathcal{E}_4$.
		В соответствующих разделах последовательно батареям есть поставлено высокоомное сопротивление $r_i$ между каждыми двумя металлическими выводами, как показано на эквивалентной схеме источника напряжения на Фигуре~\ref{RS:HV-Battery}.
		Это сопротивление поставлено с предохранительной цели. 
		Не разбирайте на части источника напряжения и не убирайте предохранительного сопротивления!
		
		Точное измерение электродвижущего напряжения только с предоставленным вам вольтметром невозможно, так как внутреннее сопротивление $r_i$ источника сравнимо по стоимости внутренним сопротивлением вольтметра.
		
		\begin{figure}[h]
			\includegraphics[width=12cm]{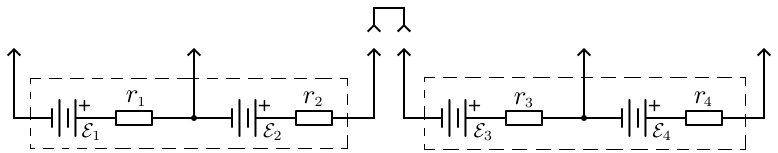}
			\caption{
				Эквивалентная схема источника напряжения для электрического эксперимента. Пластмассовые трубы, в которых поставлены батарейки и сопротивление, обозначены с пунктирными линиями.
				Винты (электроды), которые выходят из труб обозначены схематично со стрелками, а захватывающие их крокодилы - с дополнительным знаком.
			}
			\label{RS:HV-Battery}
		\end{figure}
		
		\item Измерьте электродвижущего напряжения $\mathcal{E}_1$, $\mathcal{E}_2$, $\mathcal{E}_3$ и $\mathcal{E}_4$
		только с помощью амперметра и вольтметра.
		
		\begin{figure}[h]
			\includegraphics[width=12cm]{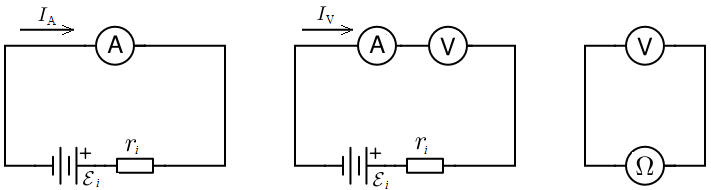}
			\caption{Измерение напряжения одной части источника напряжения $\mathcal{E}_i$ для электростатического эксперимента. 
				Внутри трубы между каждыми двумя болтами скрыто большое сопротивление $r_i$ ($i=1,\,2,\,3,\,4$), ограничивающее тока.
				Левая фигура: Между двумя соседними болтами (электродами) 
				включите амперметр и ток, которы протекает е $I_\mathrm{A}$.
				Средняя фигура: Когда к амперметру подключен последовательно и вольтметр, 
				ток $I_\mathrm{V}$ меньше, a вольтметр показывает напряжение $U.$
				Правая фигура: Схема прямого измерения внутреннего сопротивления вольтметра $R_\mathrm{V}$, используя друг мультиметр, переключенный как омметр.
			}
			\label{RS:Measure-battery}
		\end{figure}
		
		Свяжите амперметра с одним из частях источника напряжение, как это показано в левой схеме Фигуры~\ref{RS:Measure-battery} и измерьте тока $I_\mathrm{A}$, который протекает через него.
		В источнике напряжения есть высокоомное сопротивление, которое ограничивает тока и нет опасности от протекания большого тока через амперметра.
		Повторите измерение всех четырех частей источников напряжения и нанесите результаты в таблице, как это показано в примерной Таблице~\ref{RS:tab:I12345}.
		
		Свяжите амперметра и вольтметра последовательно к источнику напряжения, как это показано в середине схемы Фигуры~\ref{RS:Measure-battery}, и измерьте тока $I_\mathrm{V}$, который протекает через цепью и напряжение $U$, которое показывает вольтметр.
		Повторите измерение всех четырех частей источников напряжения и нанесите результат в таблице, как это показано в примерной Таблице~\ref{RS:tab:I12345}.
		
		\begin{table}[h]
			\caption{Примерная таблица экспериментальных результатов токов и напряжений измерены по схемом Фигуры~\ref{RS:Measure-battery}, как и для вычисленных значении электродвижущего напряжения различных частей источников напряжения.}
			\begin{tabular}{| r | r | r | r | r | r | r | }
				\tableline
				$i$& $I_\mathrm{A} \, [\mathrm{mA}]$ & $I_\mathrm{V} \,[\mathrm{mA}]$ & $U \,[\mathrm{V}]$ & $\mathcal{E} \,[\mathrm{V}]$ & $r \, [\Omega$] & $R_\mathrm{V} \, [\Omega$]\\
				\tableline
				1 & & & & & & \\
				2 & & & & & & \\
				3 & & & & & & \\
				4 & & & & & & \\
				\tableline
			\end{tabular}
			\label{RS:tab:I12345}
		\end{table}
		
		Для каждых частей источника напряжения запишите в примерной Таблице~\ref{RS:tab:I12345} электродвижущих напряжении, вычисленных по формуле
		\begin{equation}
			\label{RS:eq:E}
			\mathcal{E}=\frac{U}{1-I_\mathrm{V}/I_\mathrm{A}}.
		\end{equation}
		
		\item Какое внутреннее сопротивление каждой части источников напряжения? 
		
		Используя данных Таблицы~\ref{RS:tab:I12345} определите внутреннее сопротивление каждой части источника напряжения
		\begin{equation}
			\label{RS:eq:r}
			r=\frac{U}{I_\mathrm{A}-I_\mathrm{V}},
		\end{equation}
		для каждого из четырех измерений.
		
		\item Определите внутреннее сопротивление вольтметра.
		
		Используя данных Таблицы~\ref{RS:tab:I12345} определите внутреннее сопротивление вольтметра
		\begin{equation}
			\label{RS:eq:R}
			R_\mathrm{V}=\frac{U}{I_\mathrm{V}},
		\end{equation}
		для каждого из четырех измерений и подсчитайте средние значение всех измерений.
		
		Внутреннее сопротивление омметра можете измерить прямо по правой схеме Фигуры~\ref{RS:Measure-battery}. 
		Сравните подсчитанное среднее значение с прямо измеренным. Сколько процентов есть разница?
		
		\item Выведите формулы~\eqref{RS:eq:E}, \eqref{RS:eq:r}  и~\ref{RS:eq:R}.
		
		\subsection{Измерение электродвижущих напряжении $\mathcal{E}_1$, $\mathcal{E}_2$, $\mathcal{E}_3$ и $\mathcal{E}_4$ источников напряжения для электрического эксперимента с помощью амперметра и резисторов}
		\textit{Это серия задач (секция В) предназначена для маленьких школьников, состязающихся в категории S. Старшие школьники могут вернутся к ней, после окончания остальных задач.}
		
		\item{Измерьте сопротивлении $R_1^*,$ $R_2^*,$ $R_3^*$ и $R_4^*$ представленных вам четырех резисторов и представьте результаты в таблице.}
		
		\item Свяжите схему, данную на Фигуре~\ref{RS:Measure-battery2} и измерьте тока $I$, когда вы поставили резистор $R_1=R_1^*,$ $R_2=R_1^*+R_2^*,$ $R_3=R_1^*+R_2^*+R_3^*$ и $R_4=R_1^*+R_2^*+R_3^*+R_4^*$.
		
		\begin{figure}[h]
			\includegraphics[width=10cm]{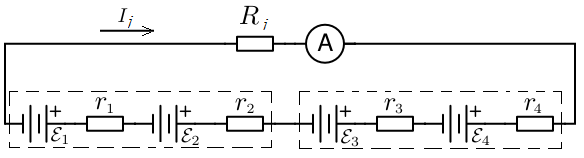}
			\caption{Измерение напряжения одной части источника напряжения для электростатического эксперимента.
				Ток через цепь $I_j$ измеряется 
				при разных внешних сопротивлениях $R_j$
				и так определяется внутреннее сопротивление 
				источника скрытого в трубе.  
			}
			\label{RS:Measure-battery2}
		\end{figure}
		
		Нанесите измеренные значения в таблице, как это показано в примерной Таблице~\ref{RS:tab:E1234_2}.
		В последней колонке таблицы запишите обратное значение тока.
		
		\begin{table}[h]
			\caption{Примерная таблица для экспериментальных результатов токов и напряжении, измерены по схеме Фигуры~\ref{RS:Measure-battery2} для разных частей источников напряжения.}
			\begin{tabular}{| r | r | r | r | }
				\tableline
				i& $R_j \, [\mathrm{k}\Omega]$ & $I_j \,[\mu\mathrm{A}]$ &  $1/I_j \,[\mu\mathrm{A}^{-1}]$ \\
				\tableline
				1 &  &  &\\
				2 & &  &\\
				3 & & &\\
				4 & &  &\\
				\tableline
			\end{tabular}
			\label{RS:tab:E1234_2}
		\end{table}
		
		\item Используя результатов Таблицы~\ref{RS:tab:E1234_2} представьте чески зависимость сопротивления $R_j$ от обратной стоимости тока $1/I_j$.
		
		Сопротивление $R_j$ и обратное значение тока $I_j$ связаны со следующей зависимостью
		\begin{equation}
			\label{RS:eq:R_j}
			R_j=\mathcal{E}_\mathrm{tot} \frac{1}{I_j} - r_\mathrm{tot},
		\end{equation}
		где $\mathcal{E}_\mathrm{tot}$ есть суммарное электродвижущее напряжение всех последовательно связанных источников напряжения, а $r_\mathrm{tot}$ есть их суммарное внутреннее сопротивление.
		
		\item Определите электродвижущее напряжениe $\mathcal{E}_\mathrm{tot}=\mathcal{E}_1 + \mathcal{E}_2 + \mathcal{E}_3 + \mathcal{E}_4$ и внутреннее сопротивление $r_\mathrm{tot}=r_1+r_2+r_3+r_4$ двух последовательно связанных источников напряжения.
		
		Начертите прямую линию, которая проходит ближайшей до экспериментальных точек.
		От углового коэффициента можно определить электродвижущее напряжение $\mathcal{E}_\mathrm{tot}$ батареи, а от пресечения прямой линии и ординаты можно определить внутреннее сопротивление батареи $r_\mathrm{tot}$.
		
		\item Выведите формулу~\eqref{RS:eq:R_j}.
		
		\subsection{Электростатическое определение на $\varepsilon_0$}
		
		\item Качественный электрический эксперимент
		
		С помощью кабели с крокодильями свяжите последовательно двух источников напряжения, не связывая к ним вольтметр.
		Осторожно с полярностью батареи. Свяжите (+) с (-) двух источников.
		Суммарное напряжение
		$U_4=\mathcal{E}_1+\mathcal{E}_2+\mathcal{E}_3+\mathcal{E}_4$ подсчитывается как сумма от определенных раньше в секции ~\ref{RS:measure_varE} электродвижущих напряжении.
		Используя сново связанные кабели с крокодильями, дайте это напряжение к круглым пластинкам через выходящих из них проводники.
		Пусть сначало пластинки будут далеко одна от другой (дальше 1~cm), но они параллельно расположены на одной оси, которая проходит через их центр.
		Посмотрите еще и схему показанную на Фигуре~\ref{RS:setup_e}.
		Постепенно двигайте блока, пока висячая пластина не приклеится внезапно к неподвижной пластинки, закреплена к блоку.
		Осторожно повторите эксперимент медленнее так, что дожидаясь затихания колебании. 
		Передвигайте блока так, что пластины были параллельными.
		После как маятник потерял равновесие и пластины приклеились, прекратите их связи к источнику напряжение.
		Используйте кабель с крокодильями и свяжите на короткое замыкание проволок выходящих из двух пластинок.
		Тогда пластины отделятся одна от другой и после как колебания затихнут, измерьте расстояние $x_\mathrm{e}$ между пластинами этого плоского конденсатора.
		
		\item 
		Повторите осторожно этот эксперимент измеряя расстояние  $x_\mathrm{e}$ между пластинами, с точностью 0.5~mm, для разных напряжении:
		$U_4=\mathcal{E}_1+\mathcal{E}_2+\mathcal{E}_3+\mathcal{E}_4,$
		$U_3=\mathcal{E}_1+\mathcal{E}_2+\mathcal{E}_3,$
		$U_2=\mathcal{E}_1+\mathcal{E}_2,$ и
		$U_1=\mathcal{E}_1.$
		
		Запишите результаты измерении в примерной таблице~\ref{RS:tab:xe3_vs_U2}.
		С помощью калькулятора добавьте в таблице две дополнительные колонки
		$x_\mathrm{e}^3$ и $U^2$.
		
		\begin{table}[h]
			\caption{Таблица с экспериментальными результатами электрического эксперимента.}
			\begin{tabular}{| r | r | r | r | r | r |}
				\tableline
				i& $U \,[\mathrm{V}]$ & $x_\mathrm{e} \,[\mathrm{mm}]$ & $U^2 \,[\mathrm{V}^2]$ & $x_\mathrm{e}^3 \,[\mathrm{mm}^3]$ \\
				\tableline
				1 &  &  &  & \\
				2 &   & &  &  \\
				3 &   &  &  &  \\
				4 &  & &  &  \\
				\tableline
			\end{tabular}
			\label{RS:tab:xe3_vs_U2}
		\end{table}

		\item Представьте графически на миллиметровую бумагу зависимость $x_\mathrm{e}^3$ от $U^2$, и определите наклон $k_\mathrm{e}$ прямой линии, которая проходит ближайшей к экспериментальным точкам.

		\begin{figure}[h]
			\includegraphics[width=7cm]{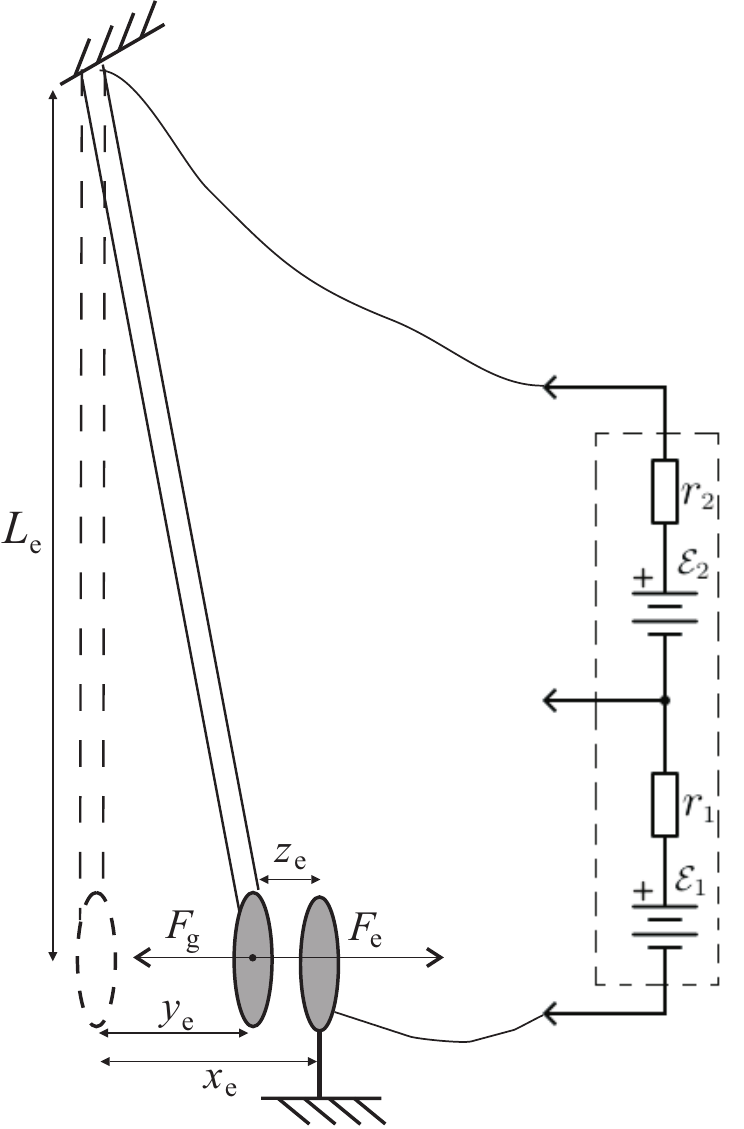}
			\caption{
				Электрический маятник для измерения диэлектрической проницаемости  $\varepsilon_0$ вакуума.
				Длина маятника есть $L_\mathrm{e}$. 
				Сдвиг от положения равновесия маятника
				немного перед тем как приклеится к притягивающей его неподвижной пластинки есть $y_\mathrm{e}=x_\mathrm{e}-z_\mathrm{e}.$ Здесь 
				$z_\mathrm{e}$ есть расстояние между пластинками немного перед тем как маятник устремился к блоку. А $x_\mathrm{e}$ есть расстояние между положением равновесия маятника и неподвижной пластинки.
				Электрическое поле между пластинками создается от последовательно связанных батарей с электродвижущими напряжениями
				$\mathcal{E}_1$ и $\mathcal{E}_2$. 
				Большие сопротивления, которые ограничивают тока обозначены как
				$r_1$ и $r_2$. 
				Сопротивления, как и батареи, спрятаны в пластмассовой трубе,
				обозначенной символично с пунктирными линиями.
				В положение равновесии электрическая сила притяжении между пластинами $F_\mathrm{e}$ уравновешивается гравитационной силой $F_\mathrm{g}$.
			}
			\label{RS:setup_e}
		\end{figure}
		
		\item Измерьте массу $m_\mathrm{e}$ висячей пластинки. 
		
		Для этой цели используйте электронных весов, которые находятся у квестора.
		Для измерения не передвигайте висячую пластинку от стенда. 
		Поставьте весы на высокую подушечку, которая вместе с ней. 
		Сделайте измерение пока пластинка укреплена к тонкой медной нити и будьте осторожны чтобы нить не оказывала дополнительных сил на пластинку во время измерении.
		
		\item Измерьте диаметр $D_\mathrm{e}$ двух пластинок. 
		
		\item Измерьте расстояние от центра пластинки до планки подвески $L_\mathrm{e}$.
		
		\item 
		Определите диэлектрическую проницаемость вакуума $\varepsilon_0$ через приближенную зависимость между $x_\mathrm{e}$ и $U$, 
		которая выведена при предположении, что $x_\mathrm{e}\ll D_\mathrm{e}$
		\begin{equation}
			x_\mathrm{e}^3 =k_\mathrm{e}  U^2,
		\end{equation}
		где $k_\mathrm{e}$ есть линейный коэффициент
		\begin{equation}
			\label{RS:eq:epsilon_0}
			k_\mathrm{e} 
			=\frac{27}{32} \pi  \varepsilon_0 \frac{L_\mathrm{e} D_\mathrm{e}^2}{m_\mathrm{e} g}.
		\end{equation}
		
		Используя эту формулу и экспериментально измеренное значение 
		$k_\mathrm{e} $, определите диэлектрической проницаемости вакуума
		\begin{equation}
			\label{RS:eq:k_e}
			\varepsilon_0=
			\frac{32}{27\pi}\frac{m_\mathrm{e} g}{L_\mathrm{e} D_\mathrm{e}^2}k_\mathrm{e}.
		\end{equation}
		
		\item 
		К таблице добавьте еще один столб $x_\mathrm{e}^3 (1-\frac{4}{3 \pi} \frac{x_\mathrm{e}}{D_\mathrm{e}})$, пересчитывая маленькую коррекцию
		$\frac{4}{3 \pi} \frac{x_\mathrm{е}}{D_\mathrm{e}}$.
		Повторите описанный выше метод для одного боле точного определения диэлектрической проницаемости вакуума через формулу
		\begin{equation}
			\label{RS:Effect_of_ends}
			x_\mathrm{e}^3 \left(1-\frac{4}{3 \pi} \frac{x_\mathrm{e}}{D_\mathrm{e}} \right) =
			\frac{27}{32} \pi  \varepsilon_0 \frac{L_\mathrm{e} D_\mathrm{e}^2}{m_\mathrm{e} g}  U^2.
		\end{equation}
		Эта формула позволяет достичь процентной точности при использовании данной постановки.
		
		Какая разница в определении $\varepsilon_0$ по двум методам?
		
		\subsection{Магнитостатическое определение $\mu_0$}
		
		\item Качественный магнитный эксперимент
		
		Поставьте двух колец параллельно расстоянии 
		$x_\mathrm{m}= \mathrm{20\;mm}$, считано между серединами обмоток.
		Используйте кабели с крокодильями и соберите схему показанную на Фигуре~\ref{RS:setup_m}.
		Давайте проследим путь тока, который идет от (+) батареи.
		Сначало проходит через сопротивительную канталовую проволоку, натянутую на планке.
		Такое переменное сопротивление называется еще реохорд.
		Цепь соединяется алюминиевым слайдом прикреплен с крокодилом, который надо держать рукой.
		Потом ток проходит через амперметр, висячее токовое кольцо,
		неподвижное токовое кольцо с тоже самым числом обмоток $N=50$
		и через (-) батареи возвращается в источник питания.
		
		Когда прикоснете слайд к сопротивительной проволоки
		обмоток начинают вибрировать. 
		Если бобины отталкиваются, смените полярность связи одной из них.
		Токи должны быть параллельными и при включении тока катушек привлекаться.
		Двигайте слайд по проволоке, смотря в тоже самое время амперметр и висячую обмотку.
		Начните с маленькими стоимостями токов и двигая слайда запоминайте положение, при котором маятник теряет устойчивость и висячее кольцо устремляется к неподвижному.
		Повторите эксперимента медленно ожидая затихания колебаний.
		Запишите самый маленький (критичный) ток $I$, при котором спокойно висячее кольцо внезапно приклеивается к неподвижному.
		
		\begin{figure}[h]
			\includegraphics[width=7cm]{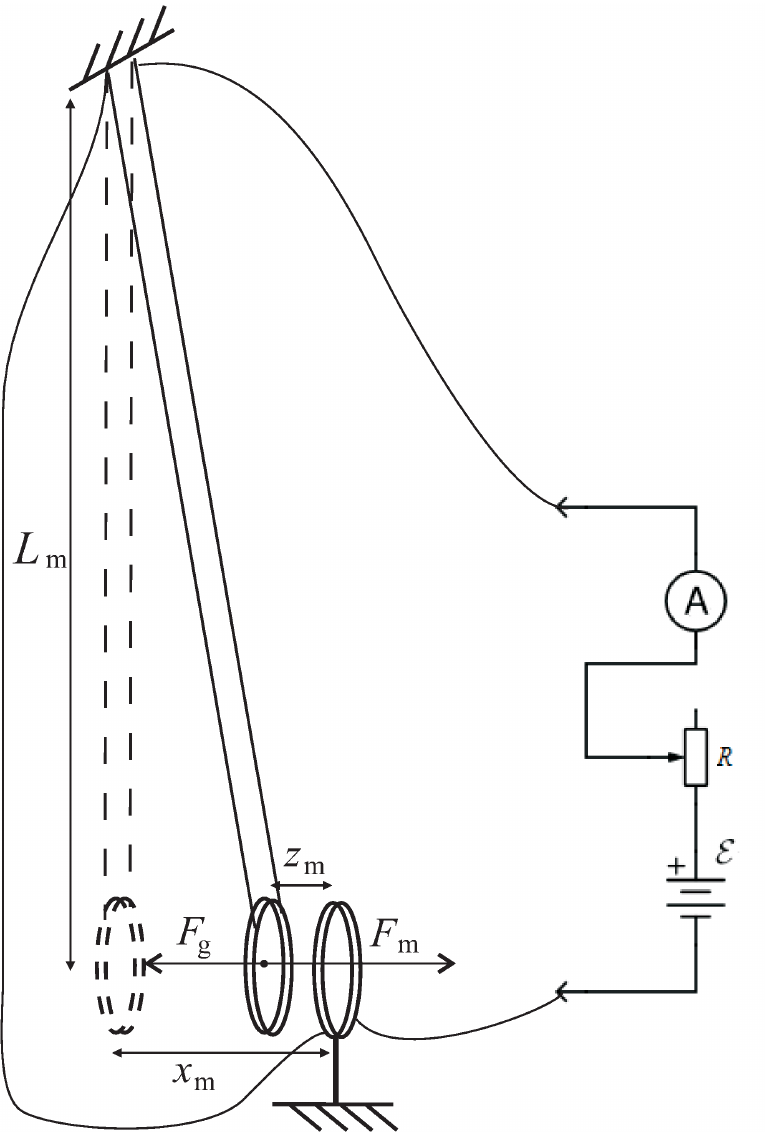}
			\caption{
				Магнитный маятник для измерения $\mu_0$, магнитную проницаемость вакуума.
				Длина маятника есть $L_\mathrm{m}$. 
				Расстояние от неподвижной обмотки до равновесного положения маятника есть $x_\mathrm{m}$.
				А расстояние между токовыми кольцами в момент перед тем как висячее устремляется к неподвижному есть $z_\mathrm{m}$.
				Магнитное поле между обмотками создают ток 4 или 8 последовательно связанных батареи 1.5~V, поставлены в держателе.
				Ток измеряется амперметром
				и регулируется переменном сопротивлением $R$
				с помощью слайда, скользящий по длине 
				канталовы проволоки, натянутой на планке.
				В равновесии магнитная сила притяжения между кольцами с параллельными токами $F_\mathrm{m}$ уравновешивается гравитационной силой $F_\mathrm{g}$.
			}
			\label{RS:setup_m}
		\end{figure}

		\item Повторите эксперимент и измерьте критичный ток $I$ для разных стоимостях $x_\mathrm{m}$, например 25, 20, 15, 10, 5~mm. 
		
		Результаты для расстояния $x_\mathrm{m}$ и тока $I$ представьте таблично в первых двух колонах таблицы~\ref{RS:tab:mu0}.
		С помощью калькулятора посчитайте дополнительные колоны для 
		$x_\mathrm{m}^2$ и $I^2$. 
		С другими словами, представьте таблично зависимость  $x_\mathrm{m}^2$ от $I^2$.
		
		Для маленьких расстоянии используйте только один держатель с четырьмя батареями.

		\item Представьте графически зависимость $x_\mathrm{m}^2$ от $I^2$ и определите наклона $k_\mathrm{m}$ прямой, которая проходит ближайшей к экспериментальным точкам.
		Это есть метод для экспериментального определения наклона $k_\mathrm{m}$.

		\item Измерьте длину маятника, т.е. расстояние $L_\mathrm{m}$ от планки до центра висячей бобины.
		
		\item Измерьте массу $m_\mathrm{m}$ висячей бобины.
		
		\item Измерьте диаметр $D_\mathrm{m}$ двух бобин.
		
		\item Определите магнитную проницаемость вакуума $\mu_0$,
		используя приблизительную формулу
		\begin{equation}
			x_\mathrm{m}^2 =k_\mathrm{m}  I^2,
		\end{equation}
		где $k_\mathrm{m}$ линейный коэффициент
		\begin{equation}
			\label{RS:eq:k_m}
			k_\mathrm{m} =2 \mu_0 \frac{L_\mathrm{m} N^2 D_\mathrm{m}}{m g}.
		\end{equation}
		Эта формула можно приложить когда $x_\mathrm{m}\ll D_\mathrm{м}$
		и можно принять, что проводники есть прямы.
		
		Эта формула дает возможность для выражения 
		магнитной проницаемости $\mu_0$ вакуума
		через экспериментально определенный наклон
		\begin{equation}
			\label{RS:eq:k_m}
			\mu_0=\frac{m gk_\mathrm{m}}{2 L_\mathrm{m} N^2 D_\mathrm{m}}.
		\end{equation}
		
		\item Сила притяжения между двумя токовыми кольцами есть известная магнитостатическая задача, которая имеет много приложении.
		Точную формулу для определения $\mu_0$ через нашу установку дана формулой
		\begin{equation}
			\label{RS:Correction}
			x_\mathrm{m}^2 \left[1+f \left(\frac{x_\mathrm{m}}{D_\mathrm{m}}\right) \right] =k_\mathrm{m}  I^2,
		\end{equation}
		где функция $f(x_\mathrm{m}/D_\mathrm{m})$ 
		представлена таблично ниже и графически на Фигуре~\ref{RS:f_delta}.
		
		При точности, с которы работаем, коррекцию можно посчитать и через приблизительную формулу
		\begin{equation}
			\label{RS:function_2_rings}
			f \left(\frac{x_\mathrm{m}}{D_\mathrm{m}}\right)\approx
			\frac{1}{16}\left(-5+6 \log{ \frac{8 D_\mathrm{m}}{x_\mathrm{m}} } \right)
			\frac{x_\mathrm{m}^2}{D_\mathrm{m}^2}.
		\end{equation}

		Пример для приведения в порядок экспериментальных данных и коррекционного множителя даны в Таблице~\ref{RS:tab:mu0}.
		Заполните вашу таблице.
		
		\begin{table}[h]
			\caption{Примерная таблица для приведения в порядок экспериментальных результатов и коррекционной функции для эксперимента с токовыми кольцами. Между $x_\mathrm{m}^2(1+f)$ и $I^2$ есть линейная зависимость и коэффициент пропорциональности $k_\mathrm{m}$ определяет $\mu_0.$}
			\begin{tabular}{| r | r | r | r | r | r | r | r| }
				\tableline
				$i$& $x_\mathrm{m}[\mathrm{mm}]$& $I [\mathrm{mA}]$  & $x_\mathrm{m}^2 [\mathrm{mm^2}]$ & $I^2 [\mathrm{mA^2}]$ &  $x_\mathrm{m}/D_\mathrm{m}$ & $1+f(x_\mathrm{m}/D_\mathrm{m})$ & $x_\mathrm{m}^2(1+f) [\mathrm{mm^2}] $ 
				\\
				\tableline
				1 &  5 & && & & & \\
				2 & 10& && & & & \\
				3 & 15& && & & & \\
				4 & 20& & && & & \\
				5 & 25& && & & & \\
				\tableline
			\end{tabular}
			\label{RS:tab:mu0}
		\end{table}
		
\begin{center}	
		\begin{tabular}{| c | c || c | c || c | c || c |  c || c | c |}
			\hline
			\multicolumn{1}{| c |}{ } & \multicolumn{1}{ c ||}{ } & \multicolumn{1}{ c |}{ } & \multicolumn{1}{ c ||}{ } & \multicolumn{1}{ c |}{ } & \multicolumn{1}{ c ||}{ } 
			& \multicolumn{1}{ c |}{ } & \multicolumn{1}{ c ||}{ } & \multicolumn{1}{ c |}{ } & \multicolumn{1}{ c |}{ } \\
			\boldmath$x/D$ & \boldmath$f(x/D)$ & \boldmath$x/D$ &\boldmath$f(x/D)$ & \boldmath$x/D$ & \boldmath$f(x/D)$ & \boldmath$x/D$ & \boldmath$f(x/D)$  & \boldmath$x/D$ & \boldmath$f(x/D)$ \\[15pt] \hline 
			0.005	&	0.0001	&	0.105	&	0.0148	&	0.205	&	0.0480	&	0.305	&	0.0976	&	0.405	&	0.1644	\\	\hline
0.010	&	0.0002	&	0.110	&	0.0162	&	0.210	&	0.0502	&	0.310	&	0.1005	&	0.410	&	0.1682	\\	\hline
0.015	&	0.0005	&	0.115	&	0.0175	&	0.215	&	0.0522	&	0.315	&	0.1034	&	0.415	&	0.1721	\\	\hline
0.020	&	0.0008	&	0.120	&	0.0188	&	0.220	&	0.0545	&	0.320	&	0.1066	&	0.420	&	0.1760	\\	\hline
0.025	&	0.0012	&	0.125	&	0.0202	&	0.225	&	0.0566	&	0.325	&	0.1096	&	0.425	&	0.1800	\\	\hline
0.030	&	0.0016	&	0.130	&	0.0216	&	0.230	&	0.0589	&	0.330	&	0.1126	&	0.430	&	0.1840	\\	\hline
0.035	&	0.0021	&	0.135	&	0.0231	&	0.235	&	0.0613	&	0.335	&	0.1156	&	0.435	&	0.1880	\\	\hline
0.040	&	0.0027	&	0.140	&	0.0245	&	0.240	&	0.0634	&	0.340	&	0.1190	&	0.440	&	0.1920	\\	\hline
0.045	&	0.0034	&	0.145	&	0.0262	&	0.245	&	0.0659	&	0.345	&	0.1221	&	0.445	&	0.1965	\\	\hline
0.050	&	0.0040	&	0.150	&	0.0278	&	0.250	&	0.0684	&	0.350	&	0.1252	&	0.450	&	0.2006	\\	\hline
0.055	&	0.0048	&	0.155	&	0.0294	&	0.255	&	0.0709	&	0.355	&	0.1287	&	0.455	&	0.2047	\\	\hline
0.060	&	0.0056	&	0.160	&	0.0312	&	0.260	&	0.0732	&	0.360	&	0.1319	&	0.460	&	0.2092	\\	\hline
0.065	&	0.0064	&	0.165	&	0.0328	&	0.265	&	0.0757	&	0.365	&	0.1355	&	0.465	&	0.2134	\\	\hline
0.070	&	0.0073	&	0.170	&	0.0345	&	0.270	&	0.0783	&	0.370	&	0.1390	&	0.470	&	0.2181	\\	\hline
0.075	&	0.0083	&	0.175	&	0.0365	&	0.275	&	0.0810	&	0.375	&	0.1423	&	0.475	&	0.2223	\\	\hline
0.080	&	0.0092	&	0.180	&	0.0382	&	0.280	&	0.0837	&	0.380	&	0.1460	&	0.480	&	0.2270	\\	\hline
0.085	&	0.0103	&	0.185	&	0.0401	&	0.285	&	0.0864	&	0.385	&	0.1497	&	0.485	&	0.2317	\\	\hline
0.090	&	0.0114	&	0.190	&	0.0421	&	0.290	&	0.0891	&	0.390	&	0.1530	&	0.490	&	0.2361	\\	\hline
0.095	&	0.0125	&	0.195	&	0.0440	&	0.295	&	0.0919	&	0.395	&	0.1568	&	0.495	&	0.2408	\\	\hline
0.100	&	0.0136	&	0.200	&	0.0461	&	0.300	&	0.0947	&	0.400	&	0.1606	&	0.500	&	0.2456	\\	\hline

			\hline
		\end{tabular}
\end{center}		
		\begin{figure}[h]
			\includegraphics[width=18cm]{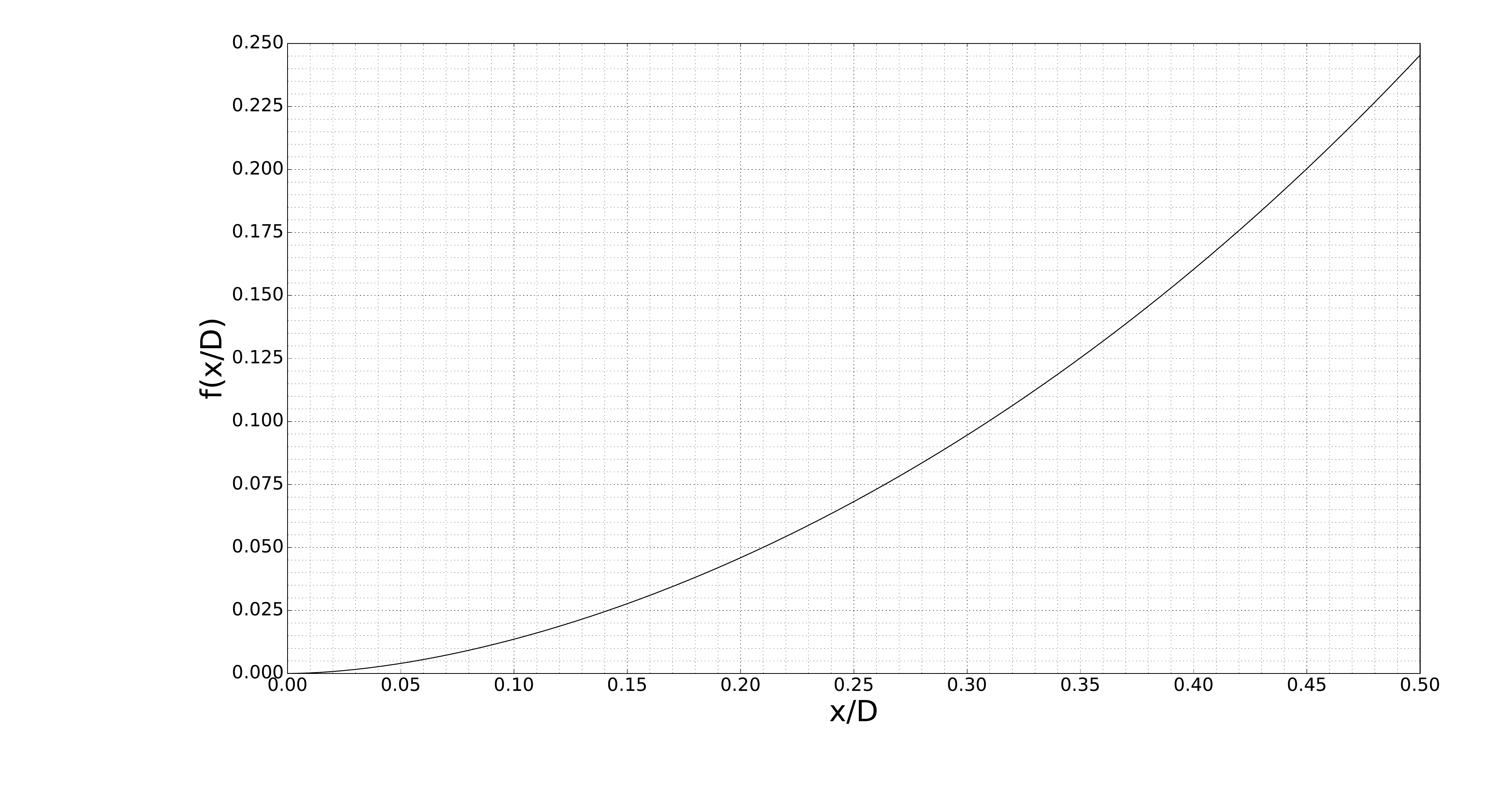}]
			\caption{Коррекция $f(x_\mathrm{m}/D_\mathrm{m})$ из уравнения (\ref{RS:Correction})
				как функция безразмерного отношения смещения от равновесного положения $x_\mathrm{m}$ и диаметра обмоток $D_\mathrm{m}$.
			}
			\label{RS:f_delta}
		\end{figure}
		
		\item 
		Используя точную формулу \eqref{RS:Correction}
		и таблично представленные результаты Фигуры~\ref{RS:f_delta},
		представьте данные графически:
		по ординате 
		$(1+f)x_\mathrm{m}^2$, 
		по абсциссе $I^2$
		и от наклона $k_\mathrm{e}$ определите магнитную проницаемость $\mu_0$ вакуума.
		С колких процентов различаются результаты от приблизительной и точной формул?
	\end{enumerate}
	
	\section{Определение скорости света}
	
	Используйте определенных в предыдущих условиях $\varepsilon_0$
	и $\mu_0$ и посчитайте скорость света
	\begin{equation}
		c=\frac{1}{\sqrt{\varepsilon_0 \mu_0}}.
	\end{equation}
	Не смущайтесь, если ваш результат различается от известны вам стоимости, это есть ваше первое измерение фундаментальной константы.

	\section{Теоретическая задача}
	
	Используйте формул силы притяжении между пластинками бесконечного плоского конденсатора и выведите формулу (\ref{RS:eq:epsilon_0}), которою вы использовали при определении $\varepsilon_0$.
	
	Используйте формул сил притяжении между бесконечными параллельными токами и выведите формулу (\ref{RS:eq:epsilon_0}), которою вы использовали для приблизительного определения $\mu_0.$
	
	\section{Задача надом. Премия Зоммерфельда 137 лев}
	
	Поищите в учебниках электродинамики, энциклопедии или в Интернете формулу емкости конденсатора, при которой отсчитываются эффекты конца и выведите точную формулу (\ref{RS:Effect_of_ends}) для определения $\varepsilon_0$с нашей установки.
	
	Аналогично поищите в учебниках формул взаимной индукции и сил взаимодействия между токовыми кольцами и выведите точную формулу (\ref{RS:Correction}).
	Коррекционная функция 
	$f(x_\mathrm{m}/D_\mathrm{m})$
	может быть подсчитана численно или можно получить только первый член
	уравнения~(\ref{RS:function_2_rings}) реда по степеней $x_\mathrm{m}/D_\mathrm{m}$.
	
	Решение пошлите epo@bgphysics.eu по адресу, где вы регистрировались до 07:00 24-ого апреля 2016 г.
	
	Можете работает колективно и консультируются с профессорами теоретической физики, электродинамики или электротехники.
	Участник получает премию лично и только в день объявления результатов.

\newpage

\section{Translation into Serbian: Условие задачи}

\section{Eksperimentalni zadatak}

\begin{figure}[h]
\includegraphics[width=9cm]{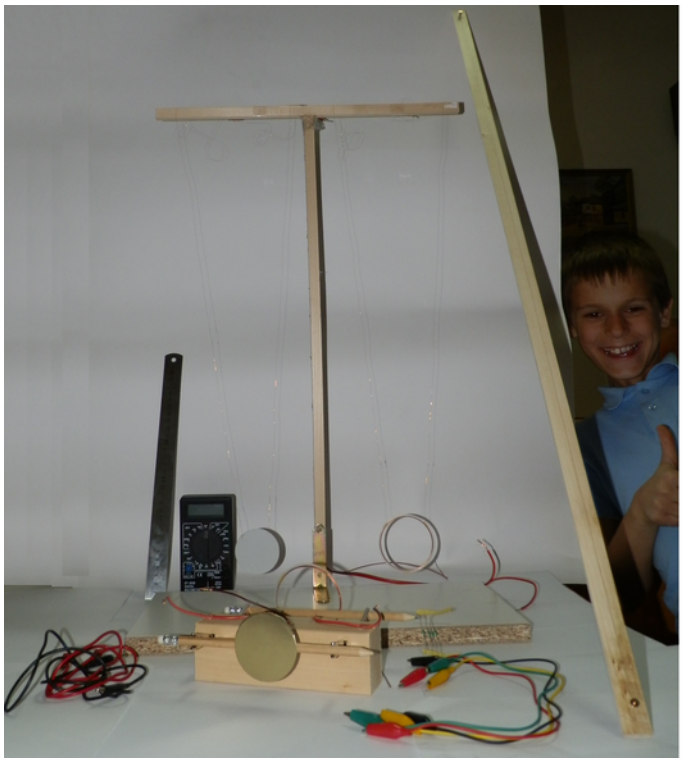}
\caption{Provjerite dali je kompletan eksperimentalni skup elemenata prikazanih na slici: stativ na kojem su okačene kružne metalne ploče i zavojnice, prsten od bakrene zavojnice što vise na tankim bakarnim žicama; drveni  na kome je pričvršćena druga metalna pločica i drugi prsten sa bakarnim zavojnicama, dva ležišta za četiri AA baterije; 8 AA baterije veličine 1.5 V; dva izvora napona postavljenih u cevima u kojih ima 3 elektrode; otpornik reostat napravljen tako što na tankoj letvici je rastegnut tanak kantalov provodnik; aluminijumska pločica koja se uštine krokodil-spojkom i može se držati rukom; kablovi čiji su krajevi povezani sa krokodil-spojkama; 4 otpornika; metalni lenjir sa podeljcima od 0,5 mm i 1 multimetar. Predpostavlja se da Vi nosite još jedan multimetar sa njegovim priključnim kablovima.}
\label{SR:exp_setup_photo}
\end{figure}

Pomoću datog kompleta eksperimentalnih uređaja (set uređaja) prikazanog na slici  izmjerite brzinu svjetlosti  $c$. Pratite upute opisane u nastavku Фигура~\ref{SR:exp_setup_photo}. Nakon Olimpijade tekst zadatka i eksperimentalnih uređaja (set uređaja) eksperimentalni su tvoje tj vašoj školi.

\begin{enumerate}

\section{Eksperimentalan zadatak.
\emph{Pazite da ne prekinete žice!}  
}
\subsection{Gravitacija. Određivanje Zemljinog ubrzanja $g$}

\begin{figure}[h]
\includegraphics[width=2.5cm]{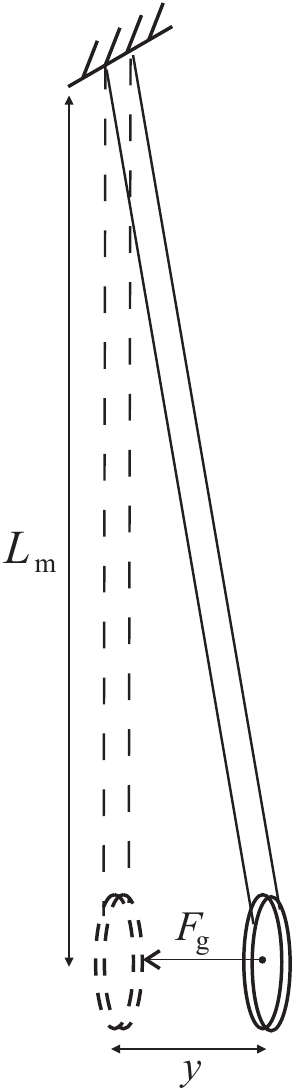}
\caption{
Kalem u obliku prstena visi na dve tanke žice. Rastojanje od tačke veyivanja do centra prstena je
$L_\mathrm{m}$. 
Pomeraj $y$ i povratne sile $F_g=-m gy/L_\mathrm{m}$ su uvijek suprotno usmjerena.
}
\label{SR:setup_g}
\end{figure}

\item Izmerite perio nihanja visećeg kalema. 

Jednostavno pomaknite prsten u smjeru osi prstena i brojati koliko oscilacije $N_\mathrm{osc}$.
Pomestete ga opet, "zalulete" prsten i mjeri vrijeme ovih oscilacija $T_N$.
Ponovite eksperiment par puta i nađite srednju vrednost perioda  $T=T_N/N_\mathrm{osc}$.
Povtorite nekoliko puta, eksperimentirati i pronaći prosječnu vrijednost perioda oscilovanja.

\item Presmetajte kružnu frekvenciju. $\omega=2\pi/T$.  

\item Izmerite udaljenost $L_\mathrm{m}$ od tačke prikačivanja do centra visećeg kalema (Slika~\ref{SR:setup_g}).

\item Izračunajte Zemljino ubrzanje $g=L_\mathrm{m}\omega^2$.
$T$, se može izračunati po formuli za period njihala:
\begin{equation}
T=2 \pi \sqrt \frac{L_\mathrm{m}}{g}.
\end{equation}
\item Za koji postotak rezultat zemljinog ubrzanja $g$ se razlikuje od poznate vrijednosti za Zemljinog ubrzanje?

\subsection{Merenje napona izvora elektromotornih sila $\mathcal{E}_1$, $\mathcal{E}_2$, $\mathcal{E}_3$ i $\mathcal{E}_4$ za električni eksperiment samo sa ampermetrom i voltmetrom}
\label{SR:measure_varE}

Izvor napona koji će biti upotrebljen u električnom eksperimentu je stvoren od više malih 23~A baterija koje imaju napon od 12~V, postavljene u dve plastične cevi. Svaka od tih cevi ima tri priključka poput elektroda od kojih 2 na krajevima i jedan na sredini. Visoko-omski otpornik sa otporom r je serijski povezan između svakog.
Od oba priključka, kao što se to vidi na ekvivalentnoj šemi za napajanje na Sl.~\ref{SR:HV-Battery}. Ovaj otpornik je postavljen zbog bezbednosti dok se radi. Nemojte razglobiti izvor napajanja i ne odstranjivajte sigurnosni otpor. 
Tačno merenje napona elektromotorne sile samo sa priloženim voltmetrom je nemoguće zato što unutrašnji otpor $r_i$  uporedljiv samo sa unutrašnjim otporom voltmetra.

\begin{figure}[h]
\includegraphics[width=12cm]{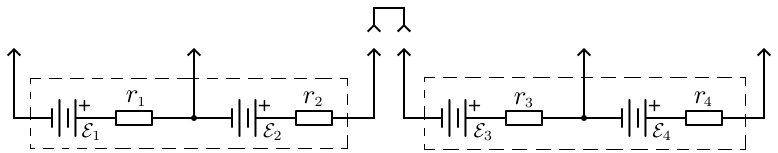}
\caption{
Ekvivalentna šema izvora za napajanje eksperimenta s električnim poljem. Prikljčene elektrode koje izlaze iz cevi u kojima su smeštene baterije na šemi su prikazani sa strelicama koje idu ka spoljašnosti, a krokodil-štpalice su prikazane drugom strelicom.
}
\label{SR:HV-Battery}
\end{figure}

\item Izmerite napon izvora elektromotornih sila $\mathcal{E}_1$, $\mathcal{E}_2$, $\mathcal{E}_3$ i $\mathcal{E}_4$ samo us pomoću ampermetra i voltmetra.

\begin{figure}[h]
\includegraphics[width=12cm]{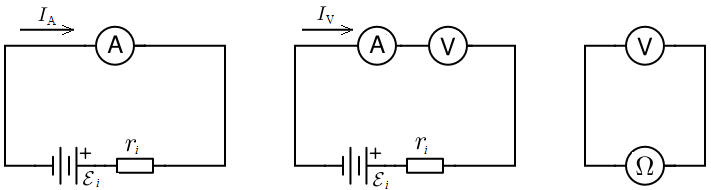}
\caption{Merenje jednog dela izvora za napajanje za potrebnog napona električnog polja između metalnih pločica (električnog eksperimenta). Ista se šema može upotrebiti za merenje napona elektromotorne sile $\mathcal{E}_i$ serijsko povezanih delova izvora za napajanje. Slika sa desne strane imamo direktno mjerenje unutrašnjeg otpora voltmetar $R_\mathrm{V}$, uzimajući drugi multimetar kao ohm metar.
}
\label{SR:Measure-battery}
\end{figure}

Povežite ampermetar s jednim delom izvora za napajanje kao što je to prikazano na slici~\ref{SR:Measure-battery} i izmerite jačinu struje $I_\mathrm{A}$ koja teče kroz njega. U izvoru za napajanje ima otpornik s velikim otporom koji ograničava struju i nema opasnosti da proteče jaka struja kroz ampermetar. Proverite merenje za sva četiri dela izvora za napajanje. 
Nanesite rezultate u tabelu kao što je prikazano na Tabeli~\ref{SR:tab:I12345}.

\begin{table}[h]
\caption{Primer tabela za upisivanje eksperimentalnih rezultata o strujama i napona mereni prema šemi iz slike~\ref{SR:Measure-battery}, kao i izračunate vrednosti za napon.}
\begin{tabular}{| r | r | r | r | r | r | r | }
\tableline
$i$& $I_\mathrm{A} \, [\mathrm{mA}]$ & $I_\mathrm{V} \,[\mathrm{mA}]$ & $U \,[\mathrm{V}]$ & $\mathcal{E} \,[\mathrm{V}]$ & $r \, [\Omega$] & $R_\mathrm{V} \, [\Omega$]\\
\tableline
1 & & & & & & \\
2 & & & & & & \\
3 & & & & & & \\
4 & & & & & & \\
\tableline
\end{tabular}
\label{SR:tab:I12345}
\end{table}

Za svaki deo izvora za napajanje upišite računanje u primernoj Tabeli~\ref{SR:tab:I12345} prema formuli
\begin{equation}
\label{SR:eq:E}
\mathcal{E}=\frac{U}{1-I_\mathrm{V}/I_\mathrm{A}}.
\end{equation}

\item Koliko iznosi unutrašnji otpor svakog dela izvora za napajanje?

Iskoristite podatke iz merenja unesenih u Tabeli~\ref{SR:tab:I12345} determinirajući unutrašnji otpor svakog dela izvora prema formuli
\begin{equation}
\label{SR:eq:r}
r=\frac{U}{I_\mathrm{A}-I_\mathrm{V}}.
\end{equation}

\item Izmerite unutrašnji otpor voltmetra.

Iskoristite podatke iz merenja unesenih u Tabeli~\ref{SR:tab:I12345}, 1 determinirajući unutrašnji otpor svakog dela izvora prema formuli

\begin{equation}
\label{SR:eq:r}
R_\mathrm{V}=\frac{U}{I_\mathrm{V}}.
\end{equation}


\item Izvedite formule~\eqref{SR:eq:E}, \eqref{SR:eq:r}  i~\eqref{SR:eq:r}.

\subsection{C. Determiniranje elektromotornih sila $\mathcal{E}_1$, $\mathcal{E}_2$, $\mathcal{E}_3$ i $\mathcal{E}_4$ na izvorima za napajanje električnog eksperimenta us pomoć ampermetra i otpornika}
\textit{Ova serija zadatka je namenuta za đake manjeg uzrasta i učenike S kategorije. Odrasli učenici mogu rešiti ove zadatke kad budu rešili ostatak.}

Us pomoću 4 otpornika u eksperimentalnom setu možete iskombinovati 4 otpornika sa suodvetnim otporima $R_1^*,$ $R_2^*,$ $R_3^*$ i $R_4^*$.

\item{Izmerite vrednosti datih otpornika i pretstavite tabelarno.}

\item Sklopite šemu kao na slici~\ref{SR:Measure-battery2} i izmerite struju $I$ kad se postavi otpornik $R_1=R_1^*,$ $R_2=R_1^*+R_2^*,$ $R_3=R_1^*+R_2^*+R_3^*$ i $R_4=R_1^*+R_2^*+R_3^*+R_4^*$.

\begin{figure}[h]
\includegraphics[width=10cm]{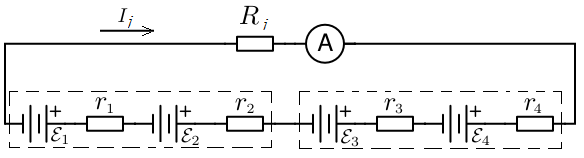}
\caption{Određivanje napona jednog dela izvora napajanja za elektrostatičkog eksperimenta. Jačina struje $I_j$ se meri preko različitih otpornika $R_j$  i tako se odredi unutrašnji otpor izvora struje sakriven u cevi.
}
\label{SR:Measure-battery2}
\end{figure}

Unesite izmereno u tabelu kao što je prikazano na tabeli~\ref{SR:tab:E1234_2}. U zadnjoj koloni tabele upišite recipročnu vrednost jačine struje.

\begin{table}[h]
\caption{Tabela namenjena za unos eksperimetalnih rezultata od merenja struja merenih prema šemi od slike~\ref{SR:Measure-battery2} za različite delove izvora napajanja.}
\begin{tabular}{| r | r | r | r | }
\tableline
i& $R_j \, [\mathrm{k}\Omega]$ & $I_j \,[\mu\mathrm{A}]$ &  $1/I_j \,[\mu\mathrm{A}^{-1}]$ \\
\tableline
1 &  &  &\\
2 & &  &\\
3 & & &\\
4 & &  &\\
\tableline
\end{tabular}
\label{SR:tab:E1234_2}
\end{table}

\item Koristeći rezultate iz tabele~\ref{SR:tab:E1234_2} predstavite grafički zavisnost otpora od recipročne vrednosti struje $1/I_j$. Otpor $R_j$  i recipročna vrednost struje $I_j$ su povezane sa zavisnošću: 

\begin{equation}
\label{SR:eq:r_j}
R_j=\mathcal{E}_\mathrm{tot} \frac{1}{I_j} - r_\mathrm{tot}.
\end{equation}

\item Odredite elektromotornu silu  $\mathcal{E}_\mathrm{tot}=\mathcal{E}_1 + \mathcal{E}_2 + \mathcal{E}_3 + \mathcal{E}_4$ i unutrašnji otpor $r_\mathrm{tot}=r_1+r_2+r_3+r_4$ na oba serijski povezana izvora struje. Nacrtajte pravu od tih eksperimentalnih tačaka koja im se najviše približava. Od kojeficijenta pravca (ugaonog k.) prave može se odrediti elektromotorna sila $\mathcal{E}_\mathrm{tot}$ a od preseka između prave i ordinatne ose možete odrediti unutrašnji otpor baterije  $r_\mathrm{tot}$.

\item Izvedite formulu~\eqref{SR:eq:r_j}.

\subsection{Elektrostatichko utvrđivanje $\varepsilon_0$}

\item Kvalitativni električni eksperiment

Korištenje kabela "aligator" spojiti serijski dva izvora energije, bez potrebe za spajanje voltmetra na njih. Obratite pozornost na polaritet izvora. Povezivanje (+) iz jednog izvora za (-) iz drugog izvora.
Ukupni napon
$U_4=\mathcal{E}_1+\mathcal{E}_2+\mathcal{E}_3+\mathcal{E}_4$ izračunava se kao zbroj elektromotornih sila definirano u Odjeljku~\ref{SR:measure_varE}.
Spojite krajeve takvog izvora pomoću kabela aligator sa krajevima žice koje dolaze iz kružne ploče. Neka pločice su udaljene jedna od druge na udaljenosti većoj od 1 cm, dok su paralelno postavljena.
Pogledajte dijagram na slici~\ref{SR:setup_e}.
Postupno premjestiti dok visecha ploča u potpunosti prilepiti, dobliziti na fiksnoj pločici pričvršćenoj na blok. Pažljivo ponoviti eksperiment čekajuci oscilacije da prestanu. Pazte da  ploče budu uvijek paralelno postavljene.
Odrediti udaljenost $x_\mathrm{e}$, na koje ploča kondenzatora počinju da se privukuju.

\item Ponovite eksperiment, pažljivo mjerenjem udaljenosti  $x_\mathrm{e}$ 16. tačnosču 0,5 mm,  za različite napone:
$U_4=\mathcal{E}_1+\mathcal{E}_2+\mathcal{E}_3+\mathcal{E}_4,$
$U_3=\mathcal{E}_1+\mathcal{E}_2+\mathcal{E}_3,$
$U_2=\mathcal{E}_1+\mathcal{E}_2,$ и
$U_1=\mathcal{E}_1.$

Evidentirati rezultate mjerenja u tablici koja sluzi za primjer ,tab.~\ref{SR:tab:xe3_vs_U2}. Koristeći kalkulator ispuniti još dva stupca $x_\mathrm{e}^3$ i $U^2$.

\begin{table}[h]
\caption{}
\begin{tabular}{| r | r | r | r | r | r |}
\tableline
i& $U \,[\mathrm{V}]$ & $x_\mathrm{e} \,[\mathrm{mm}]$ & $U^2 \,[\mathrm{V}^2]$ & $x_\mathrm{e}^3 \,[\mathrm{mm}^3]$ \\
\tableline
1 &  &  &  & \\
2 &   & &  &  \\
3 &   &  &  &  \\
4 &  & &  &  \\
\tableline
\end{tabular}
\label{SR:tab:xe3_vs_U2}
\end{table}

\item Uvesti grafički ovisnost $x_\mathrm{e}^3$ $U^2$ i odrediti nagib linije $k_\mathrm{e}$ koja prolaze, dobita iz eksperimentalnih točaka.

\begin{figure}[h]
\includegraphics[width=7cm]{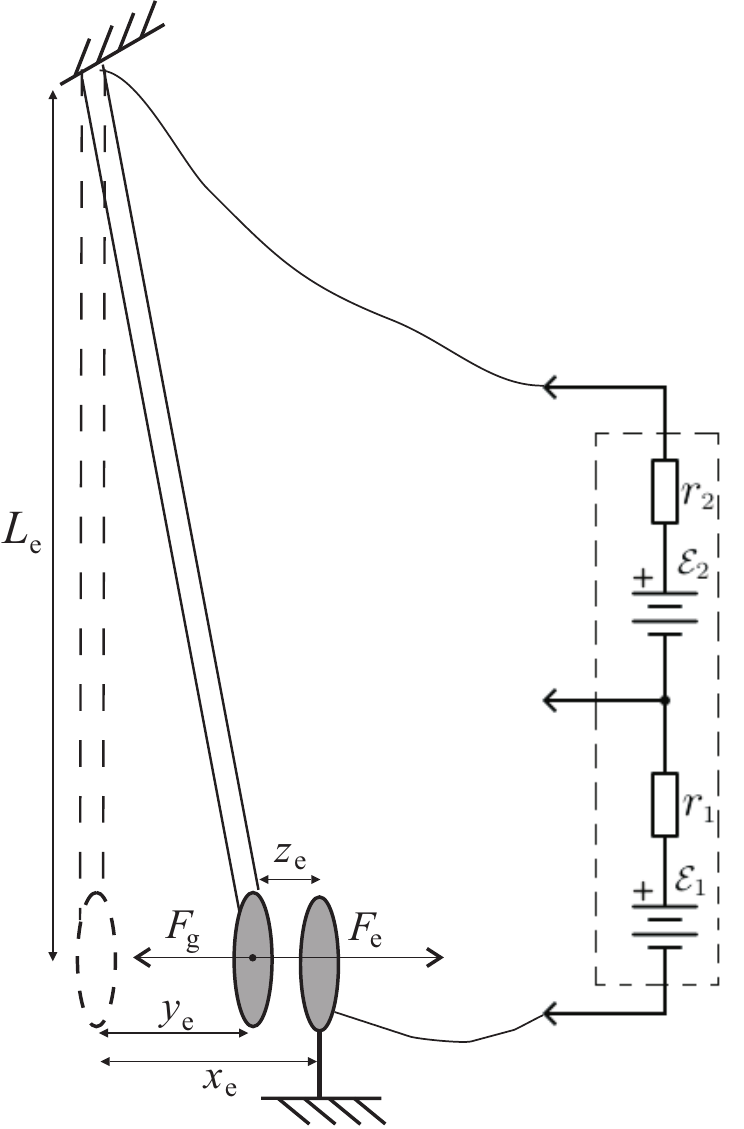}
\caption{
Električno njihalo za merenje dielektrične konstante vakuuma $\varepsilon_0$.
Dužina njihala je $L_\mathrm{e}$. 
$z_\mathrm{e}$ označava udaljenost između pločice neposredno ispred njihala da se priključi ka pločici sa testije.
$x_\mathrm{e}$ označava udaljenost između njihala do druge pločice koja je postavljena na testije.
Električno polje između pločica stvara se pomoću serijsko povezanih baterija kao izvori elektromotornih snaga
$\mathcal{E}_1$ i $\mathcal{E}_2$. 
Između baterija ima i otpornika sa velikim otporima koji ograničavaju struju i označavaju se sa
$r_1$ i $r_2$. 
Otpornici i baterije su smešteni u plastičnoj cevi simbolično predstavljena isprekidanom linijom. 
U slučaju ravnoteže električna privlačna sila između pločica $F_\mathrm{e}$ uravnoteži se sa gravitacionom silom $F_\mathrm{g}$.
}
\label{SR:setup_e}
\end{figure}

\item Izmjerite masu $m_\mathrm{e}$ viseče ploče.

Za taj cilj upotrebite elektronsku vagu koja se nalazi kod testatoru u prostoriju. Da bi izmerili masu ne otklanjajte obesenu pločku od stativa. Podignite elektronsku vagu na višem nivou. Napravite merenje masom, dok je pločka prikačena na tanji bakarni nit, pazite da ne dadate dodatni teret za vreme merenja.

\item Izmerite dijamtar $D_\mathrm{e}$ 19. obome kružne pločke.

\item Izmerite rastojanje $L_\mathrm{e}$ od mesta za koje je prikačena do centru viseče kružne pločke.

\item 
Opredelite dielektričnu konstantu vazduha $\varepsilon_0$ preko približne formule o vezanost između $x_\mathrm{e}$ i $U$,
slučaju kad je ispunjen uvet $x_\mathrm{e}\ll D_\mathrm{e}$. Teoriski se moye dokazat da pri t $x_\mathrm{e}\ll D_\mathrm{e}$ imamo linearnu ovisnost između  $x_\mathrm{e}^3$ i $U^2$.
\begin{equation}
x_\mathrm{e}^3 =k_\mathrm{e}  U^2,
\end{equation}
gde $k_\mathrm{e}$ je linearni kojeficijent
\begin{equation}
\label{SR:eq:Epsilon_0}
k_\mathrm{e} 
=\frac{27}{32} \pi  \varepsilon_0 \frac{L_\mathrm{e} D_\mathrm{e}^2}{m_\mathrm{e} g}.
\end{equation}

Preko formule za $k_\mathrm{e}$, može se presmetati dielektričnu konstantu $\varepsilon_0$ vakuuma
\begin{equation}
\label{SR:eq:k_e}
\varepsilon_0=
\frac{32}{27\pi}\frac{m_\mathrm{e} g}{L_\mathrm{e} D_\mathrm{e}^2}k_\mathrm{e}. 
\end{equation}

\item 
Prema tabeli dodaj još jedan stupac, računajući male korekcije $x_\mathrm{e}^3 (1-\frac{4}{3 \pi} \frac{x_\mathrm{e}}{D_\mathrm{e}})$.
Ponovite gore opisani način za precizno određivanje dielektrična konstanta vakuuma formulom 
\begin{equation}
\label{SR:Effect_of_ends}
x_\mathrm{e}^3 \left(1-\frac{4}{3 \pi} \frac{x_\mathrm{e}}{D_\mathrm{e}} \right) =
\frac{27}{32} \pi  \varepsilon_0 \frac{L_\mathrm{e} D_\mathrm{e}^2}{m_\mathrm{e} g}  U^2.
\end{equation}
Ova formula omogućava postizanje veći postotak preciznost. 

Koja je razlika u određivanju dielektričnu konstantu $\varepsilon_0$ vakuuma preku ove dvije metode?

\subsection{Magnetostatichko određivanje $\mu_0$}

\item Kvalitativni magnetski eksperiment

Postavite dva kružna namotaja u obliku prstena paralelno sa udaljenosti 
$x_\mathrm{m}= \mathrm{20\;mm}$, počevši od centra prstenova.
Iskoristite kablove sa aligator shtipalke, i sastavite shema prikazana kao slika~\ref{SR:setup_m}.
Pratite put struje, počevši od (+) baterije, zatim žicu otpornik od kantala, optegnuta preko drvene letve.
Takav promenliv otpor se zove reohord.
Struja prolazi kroz ampermetar kroz viseći prsten, stacionarne prsten (ima isti broj navoja,  $N=50$), pa viseći prsten i (-) pol zatvara krug.

Kada dodirnete klizač otpor žice namotaje počinju da se trese. Ako se prstenovi odbijaju promjene polaritetot. Pri uključivanje struje, prstena mora biti privučeni.
Počnite sa slabe struje, pomicanjem klizača, gledajuči istovremeno i ampermetar i zapamtite udaljenost na kojoj se klatno gubi ravnotežu i viseći prsten privlači.
Ponovite eksperiment, čekajući oscilacija da se smiri.
Zapišite najmanji struja $I$ (kritična) u kojoj je viseći prsten privlači.

\begin{figure}[h]
\includegraphics[width=7cm]{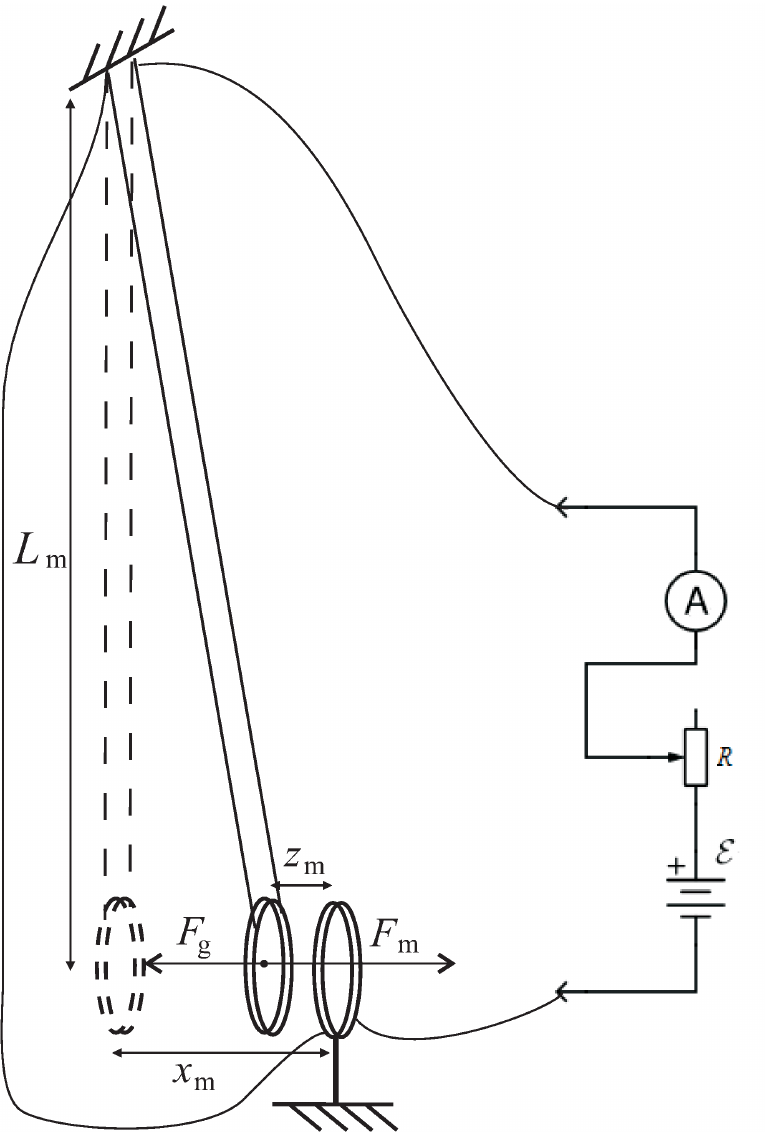}
\caption{
Magnetno njihalo za merenje magnetne permeabilnosti vakuuma $\mu_0$.
Dužina njihala je $L_\mathrm{m}$. 
$x_\mathrm{m}$ označava udaljenost njihala od druge bakrene zavojnice..
Udaljenost između bakarnih zavojnica-strujnih prstena u trenutku pre nego što se privuku je $z_\mathrm{m}$.
Magnetno polje između zavojnica se stvara na struju od 4 ili 8 serijsko povezanih baterija od 1,5~V postavljene u njihovim ležištima.
Struja se meri s ampermetrom i reguliše se sa promenljivim otporom $R$, mrdajući klizač u različitim tačkama kantalove rastegnute žice na letvi. U stanju ravnoteže magnetna sila privlačenja između prstena sa paralelnim strujama $F_\mathrm{m}$ se uravnotežava sa gravitacionom silom $F_\mathrm{g}$. 
}
\label{SR:setup_m}
\end{figure}

\item Ponovite eksperiment i mjerenje kritične struje za različite vrijednosti $x_\mathrm{m}$, npr 25, 20, 15, 10, 5~mm.

Rezultati za  $x_\mathrm{m}$ daljinu, i struja $I$ postaviti i u prve dvije kolone Tabele~\ref{SR:tab:mu0}.
Koristeći kalkulator napraviti dve dodatne kolone t.j  pretstavite tabelarno ovisnost za $x_\mathrm{m}^2$ i $I^2$.

\item Uvesti grafički ovisnost $x_\mathrm{m}^2$ i $I^2$ i odrediti nagib $k_\mathrm{m}$ prave iscrtana iz eksperimentalnih tačaka.

\item Izmerite rastojanje $L_\mathrm{m}$ od mesta za koje je prikačen do centru viseči kružni kalem.

\item Izmerite masu $m_\mathrm{m}$ visečem kružni kalem.

\item Izmerite dijametar $D_\mathrm{m}$ kružnim kalemima.

\item Opredelite magnetnu permabilnost vazduha $\mu_0$. Teorijski se može dokazati da ako  $x_\mathrm{m}\ll D_\mathrm{m}$ ima linearnu ovisnost na 
\begin{equation}
x_\mathrm{m}^2 =k_\mathrm{m}  I^2,
\end{equation}
gde $k_\mathrm{m}$ je linearni kojeficijent
\begin{equation}
\label{SR:eq:k_m}
k_\mathrm{m} =2 \mu_0 \frac{L_\mathrm{m} N^2 D_\mathrm{m}}{m g}.
\end{equation}

\begin{equation}
\label{SR:eq:k_m}
\mu_0=\frac{m gk_\mathrm{m}}{2 L_\mathrm{m} N^2 D_\mathrm{m}}.
\end{equation}

\item Snaga privlačnosti između dva prstena je poznati magnetostatichki zadatak koji ima više rješenja. Tačna formula za utvrđivanje $\mu_0$ kroz naše postanovke daje 
\begin{equation}
\label{SR:Correction}
x_\mathrm{m}^2 \left[1+f \left(\frac{x_\mathrm{m}}{D_\mathrm{m}}\right) \right] =k_\mathrm{m}  I^2.
\end{equation}

\begin{equation}
\label{SR:function_2_rings}
f \left(\frac{x_\mathrm{m}}{D_\mathrm{m}}\right)\approx
\frac{1}{16}\left(-5+6 \log{ \frac{8 D_\mathrm{m}}{x_\mathrm{m}} } \right)
 \frac{x_\mathrm{m}^2}{D_\mathrm{m}^2}.
\end{equation}

Primjer tabela za  eksperimentalnih podataka i faktor korekcije su dati u tabeli~\ref{SR:tab:mu0}. 
Popunite tabelu sa svojim proračunima.

\begin{table}[h]
\caption{Primjer tabela za  eksperimentalnih podataka i faktor korekcije su dati. Među $I^2$ i $x_\mathrm{m}^2(1+f)$, postoi linearni koeficijent proporcionalnosti  i gdje ga se može utvrditi $\mu_0.$}
\begin{tabular}{| r | r | r | r | r | r | r | r| }
\tableline
$i$& $x_\mathrm{m}[\mathrm{mm}]$& $I [\mathrm{mA}]$  & $x_\mathrm{m}^2 [\mathrm{mm^2}]$ & $I^2 [\mathrm{mA^2}]$ &  $x_\mathrm{m}/D_\mathrm{m}$ & $1+f(x_\mathrm{m}/D_\mathrm{m})$ & $x_\mathrm{m}^2(1+f) [\mathrm{mm^2}] $ 
\\
\tableline
1 &  5 & && & & & \\
2 & 10& && & & & \\
3 & 15& && & & & \\
4 & 20& & && & & \\
5 & 25& && & & & \\
\tableline
\end{tabular}
\label{SR:tab:mu0}
\end{table}

\begin{center}
\begin{tabular}{| c | c || c | c || c | c || c |  c || c | c |}
		\hline
			\multicolumn{1}{| c |}{ } & \multicolumn{1}{ c ||}{ } & \multicolumn{1}{ c |}{ } & \multicolumn{1}{ c ||}{ } & \multicolumn{1}{ c |}{ } & \multicolumn{1}{ c ||}{ } 
			& \multicolumn{1}{ c |}{ } & \multicolumn{1}{ c ||}{ } & \multicolumn{1}{ c |}{ } & \multicolumn{1}{ c |}{ } \\
			\boldmath$x/D$ & \boldmath$f(x/D)$ & \boldmath$x/D$ &\boldmath$f(x/D)$ & \boldmath$x/D$ & \boldmath$f(x/D)$ & \boldmath$x/D$ & \boldmath$f(x/D)$  & \boldmath$x/D$ & \boldmath$f(x/D)$ \\[15pt] \hline 
			0.005	&	0.0001	&	0.105	&	0.0148	&	0.205	&	0.0480	&	0.305	&	0.0976	&	0.405	&	0.1644	\\	\hline
0.010	&	0.0002	&	0.110	&	0.0162	&	0.210	&	0.0502	&	0.310	&	0.1005	&	0.410	&	0.1682	\\	\hline
0.015	&	0.0005	&	0.115	&	0.0175	&	0.215	&	0.0522	&	0.315	&	0.1034	&	0.415	&	0.1721	\\	\hline
0.020	&	0.0008	&	0.120	&	0.0188	&	0.220	&	0.0545	&	0.320	&	0.1066	&	0.420	&	0.1760	\\	\hline
0.025	&	0.0012	&	0.125	&	0.0202	&	0.225	&	0.0566	&	0.325	&	0.1096	&	0.425	&	0.1800	\\	\hline
0.030	&	0.0016	&	0.130	&	0.0216	&	0.230	&	0.0589	&	0.330	&	0.1126	&	0.430	&	0.1840	\\	\hline
0.035	&	0.0021	&	0.135	&	0.0231	&	0.235	&	0.0613	&	0.335	&	0.1156	&	0.435	&	0.1880	\\	\hline
0.040	&	0.0027	&	0.140	&	0.0245	&	0.240	&	0.0634	&	0.340	&	0.1190	&	0.440	&	0.1920	\\	\hline
0.045	&	0.0034	&	0.145	&	0.0262	&	0.245	&	0.0659	&	0.345	&	0.1221	&	0.445	&	0.1965	\\	\hline
0.050	&	0.0040	&	0.150	&	0.0278	&	0.250	&	0.0684	&	0.350	&	0.1252	&	0.450	&	0.2006	\\	\hline
0.055	&	0.0048	&	0.155	&	0.0294	&	0.255	&	0.0709	&	0.355	&	0.1287	&	0.455	&	0.2047	\\	\hline
0.060	&	0.0056	&	0.160	&	0.0312	&	0.260	&	0.0732	&	0.360	&	0.1319	&	0.460	&	0.2092	\\	\hline
0.065	&	0.0064	&	0.165	&	0.0328	&	0.265	&	0.0757	&	0.365	&	0.1355	&	0.465	&	0.2134	\\	\hline
0.070	&	0.0073	&	0.170	&	0.0345	&	0.270	&	0.0783	&	0.370	&	0.1390	&	0.470	&	0.2181	\\	\hline
0.075	&	0.0083	&	0.175	&	0.0365	&	0.275	&	0.0810	&	0.375	&	0.1423	&	0.475	&	0.2223	\\	\hline
0.080	&	0.0092	&	0.180	&	0.0382	&	0.280	&	0.0837	&	0.380	&	0.1460	&	0.480	&	0.2270	\\	\hline
0.085	&	0.0103	&	0.185	&	0.0401	&	0.285	&	0.0864	&	0.385	&	0.1497	&	0.485	&	0.2317	\\	\hline
0.090	&	0.0114	&	0.190	&	0.0421	&	0.290	&	0.0891	&	0.390	&	0.1530	&	0.490	&	0.2361	\\	\hline
0.095	&	0.0125	&	0.195	&	0.0440	&	0.295	&	0.0919	&	0.395	&	0.1568	&	0.495	&	0.2408	\\	\hline
0.100	&	0.0136	&	0.200	&	0.0461	&	0.300	&	0.0947	&	0.400	&	0.1606	&	0.500	&	0.2456	\\	\hline

		\hline
\end{tabular}
\end{center}

\begin{figure}[h]
\includegraphics[width=18cm]{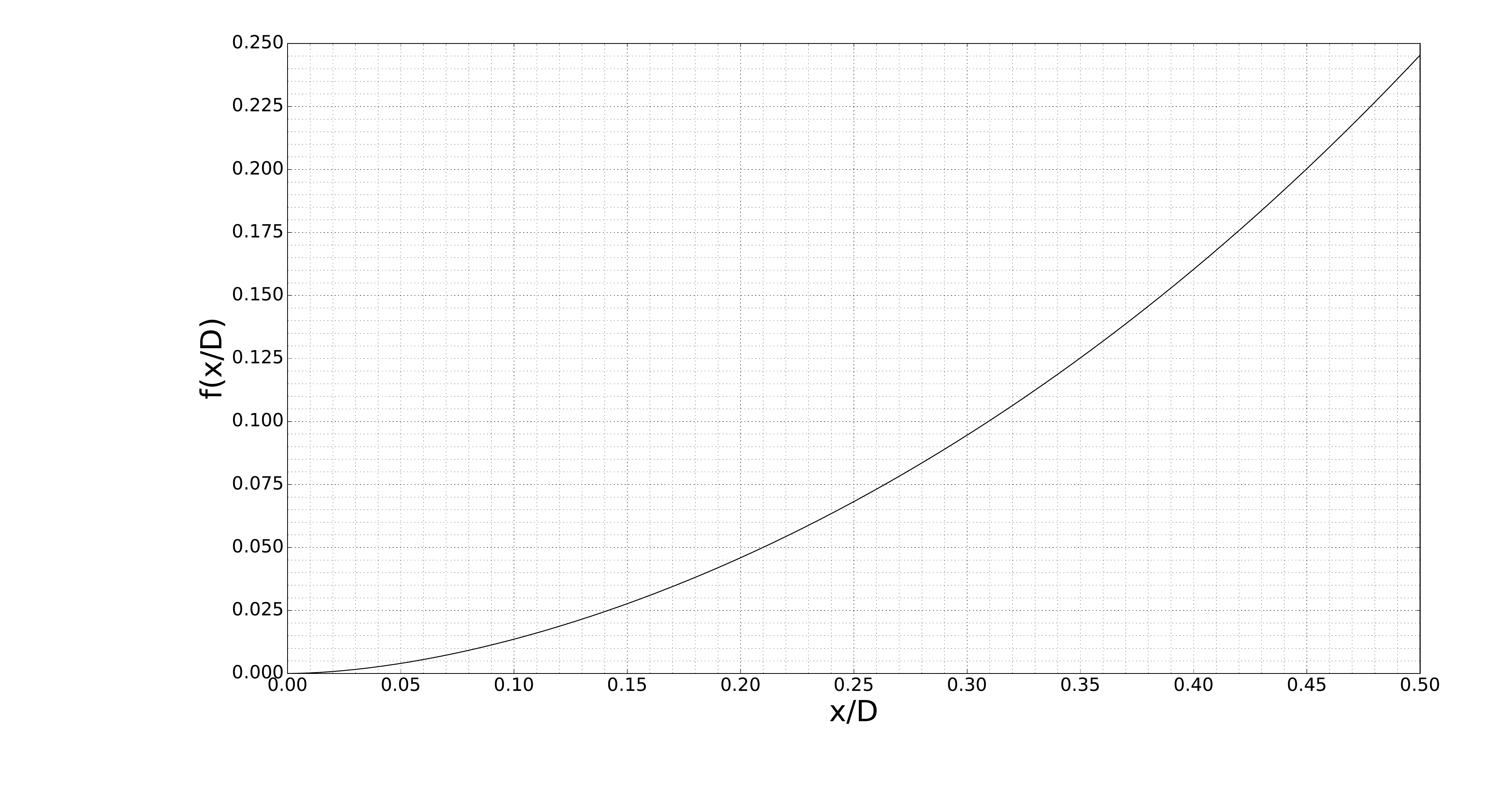}
\caption{Korekcijski faktor $f(x_\mathrm{m}/D_\mathrm{m})$ jednadžbe (\ref{SR:Correction}) kao bezdimenzionalna funkcija.
}
\label{SR:f_delta}
\end{figure}

\item 
Uzimajući točnu formulu \eqref{SR:Correction}
i u tablicama prikazani rezultati tablici~\ref{SR:f_delta},
pretstavite na grafik po ordinati: $(1+f)x_\mathrm{m}^2$, a po apcisu $I^2$. Iz nagiba odredite magnetsku permeabilnost vakuuma $\mu_0$. 

\end{enumerate}

\section{Određivanje brzine svjetlosti}

Iskorištavanjem navedenih vrijednosti $\varepsilon_0$
i $\mu_0$ i izračunati brzinu svjetlosti
\begin{equation}
c=\frac{1}{\sqrt{\varepsilon_0 \mu_0}}.
\end{equation}
Da se ne razočarate od rezultat,  ako se razlikuje od poznate vrijednosti za $c$; to je vaše prvo određivanje fundamentalne konstante.

\section{Teoriski zadatak}

Potražite u  udžbenicima iz elektrodinamike, enciklopedije ili na internetu formula za kapacitet kondenzatora u kojima se uzima u obzir učinke na krajevima i izvedite točnu ispravnu formulu~(\ref{SR:Effect_of_ends}) za  određivanje $\varepsilon_0$.

Iskorištavanjem formulu za privlačnu silu između paralelnih beskrajne struje  izvedite formulu~(\ref{SR:eq:Epsilon_0} za približnu procjenu i određivanje $\mu_0$.

\section{Nagradni domaći zadatak. Zomerfeldova premija 137 leva}

Potražite u  udžbenicima iz elektrodinamike, enciklopedije ili na internetu formula za kapacitet kondenzatora u kojima se uzima u obzir učinke na krajevima i izvedite točnu ispravnu formulu~(\ref{SR:Effect_of_ends}) za određivanje  $\varepsilon_0$.

Analogno prethodnom potražite  formule o obostranu induktivnost i zajedničko djelovanje snaga struje prstena i izvedite točnu formulu~(\ref{SR:Correction}).
Korekcionu funkciju $f(x_\mathrm{m}/D_\mathrm{m})$ može se izračunati ili da se uzme samo prvi član jednadžbe~(\ref{SR:function_2_rings}) u redoslijedu stupnjeva $x_\mathrm{m}/D_\mathrm{m}$.

Odgovor, rješenje, poslati na e-mail:  epo@bgphysics.eu  adresu od e-pošte s kojom ste se registrirali za Olimpijadu do 7:00 dana 24. travnja 2016. godine.
Možete raditi zajedno, da se savjetuje sa profesore teorijskoj fizici, elektrodinamike ili elektrotehnici. Premija će se predati osobno sudioniku, i samo na dan objave rezultata.


\begin{thebibliography}{4}
%
\bibitem{EPO_IYL2015}
 \textit{Second Experimental Physics Olympiad: The Day of the Photon in the International Year of Light, Sofia, 25 April (2015)}\\
 \url{http://www.light2015.org/Home/Event-Programme/2015/Competition/Bulgaria-Second-Experimental-Physics-Olympiad--25-April-2015-in-Sofia.-The-Day-of-the-Photon-in-the-International-Year-of-the-Light.html}

\bibitem{EPO1}
 V.~G.~Yordanov, P.~V.~Peshev, S.~G.~Manolev, and T.~M.~Mishonov,
 \textit{Charging of capacitors with double switch. The principle of operation of auto-zero and chopper-stabilized DC amplifiers},
 arXiv:1511.04328 [physics.ed-ph], \url{http://arxiv.org/abs/1511.04328}, (2015).

\bibitem{EPO2}
 V.~N.~Gourev, S.~G.~Manolev, V.~G.~Yordanov, and T.~M.~Mishonov,
 \textit{Measuring Plank constant with colour LEDs and compact disk},
 arXiv:1602.06114 [physics.ed-ph], \url{http://arxiv.org/abs/1602.06114}, (2015).

\bibitem{EPO3}
 S.~G.~Manolev, V.~G.~Yordanov, N.~N.~Tomchev, and T.~M.~Mishonov,
 \textit{Volt-Ampere characteristic of "black box" with a negative resistance},
 arXiv:1602.08090 [physics.ed-ph], \url{http://arxiv.org/abs/1602.08090}, (2015).

\bibitem{Maxwell}
	J.~C.~Maxwell, \textit{On a Method of Making a Direct Comparison of Electrostatic with Electromagnetic Force;
		with a Note on the Electromagnetic Theory of Light}, Phil. Trans. \textbf{CLVIII}, June 18 (1868).
\bibitem{Berkeley}
	E.~M.~Purcell, \textit{Electricity and Magnetism}, Berkeley Physics Course \textbf{Vol. 2} (McGraw-Hill, New York, 1963), problem~7.25, Fig.~7.42.
\bibitem{MIT}
	MIT~OCW, \textit{Course~8.02T},
	\url{http://ocw.mit.edu/courses/physics/8-02t-electricity-and-magnetism-spring-2005/labs/}, Experiments 2 and 8, (2005).
\bibitem{BG}
	V.~N.~Gourev, V.~G.~Yordanov and T.~M.~Mishonov,
	\textit{Measurement of the speed of light with an analytic scale} (in Bulgarian),
	 \url{http://optics.phys.uni-sofia.bg/disk_CONGRESS/html/pdf/S1012.pdf},
	2nd Bulgarian Nat. Congress on Phys. Sciences,
 	Sofia, September 25-29, (2013).
\bibitem{Breziche}
	Gimnazia Brezice, \textit{Determination of $\varepsilon_0$ as a maturity exam for gymnasium (high school) in Brezice Slovenia} (in Slovenian), \url{http://www2.arnes.si/~bivsic/fizika/vaje/laboratorijske_vaje_matura.pdf}, Laboratorijske vaje za maturo, \url{gimnazija.brezice@guest.arnes.si}, (2014).

\end{thebibliography}
\end{document}